\documentclass{aa}
\usepackage{txfonts}
\usepackage{graphicx}
\usepackage{epsfig}

\usepackage{natbib}
\bibpunct{(}{)}{;}{a}{}{,} 
\newcommand{\refeq}[1]{Eq.\,(\ref{#1})}
\newcommand{\reffig}[1]{Fig.\,\ref{#1}}
\newcommand{\reftab}[1]{Tab.\,\ref{#1}}
\newcommand{\refsec}[1]{Sect.\,\ref{#1}}
\begin{document}

\title{Prompt high-energy 
emission from gamma-ray bursts in the internal shock model}
\author{
\v Zeljka Bo\v snjak\inst{1} \and Fr\'{e}d\'{e}ric Daigne\inst{1,2} \and
Guillaume Dubus\inst{3,1}}

\institute{Institut d'Astrophysique de Paris, UMR 7095
Universit\'{e} Pierre et Marie Curie-Paris 6 -- CNRS, 
98 bis boulevard Arago, 75014 Paris, France 
\and
Institut Universitaire de France
\and
Laboratoire d'Astrophysique de Grenoble, UMR 5571 Universit\'{e} Joseph
Fourier -- CNRS, BP 53, 38041 Grenoble, France}

\offprints{F. Daigne (\texttt{daigne@iap.fr})}

\date{}
\titlerunning{Prompt HE emission from GRBs in the internal
shock model}
\authorrunning{Bo\v snjak, Daigne \& Dubus}

\abstract  
{Gamma-ray bursts (GRB) are powerful, short duration events with a
spectral luminosity peaking in the keV - MeV ({BATSE})
range . The prompt emission is thought to arise from 
 electrons accelerated in internal shocks propagating within a highly
relativistic outflow.}
{The launch of \textit{Fermi} offers the prospect of observations with unprecedented sensitivity in high-energy (HE, $>$100\ MeV) gamma-rays. The aim is to explore the predictions for HE emission from internal shocks, taking into account both dynamical and radiative aspects, and to deduce how HE observations constrain the properties of the relativistic outflow.}
{The prompt GRB emission is modelled by combining a time-dependent
radiative code, solving for the electron and photon distributions, with
a dynamical code giving the evolution of the 
physical conditions in the shocked regions of the outflow.
 Synthetic lightcurves and spectra are generated and compared to observations.}
{The HE emission deviates significantly from analytical estimates, which
tend to overpredict the IC component, when the time dependence and full
cross-sections are included. The exploration of the parameter space
favors the case where the dominant process in the {BATSE} range
is synchrotron emission. The HE component becomes stronger for weaker
magnetic fields. The HE lightcurve 
can display a prolonged pulse duration due to IC emission, or even a delayed peak
compared to the {BATSE} range.
Alternatively, having dominant IC
emission in the {BATSE} range requires most electrons to be
accelerated into a steep power-law distribution and implies strong
second order IC scattering. In this case, the {BATSE} and HE
lightcurves are very similar. }
{The combined dynamical and radiative approach allows a firm
appraisal
of GRB HE prompt emission. A diagnostic procedure is
presented to identify from observations the dominant emission process
and derive constrains on the bulk Lorentz factor, particle density and
magnetic field of the outflow.} 
\keywords{gamma-rays: bursts; shock-waves; radiation mechanisms: non-thermal}

\maketitle

\section{Introduction}
The forthcoming first results of the \textit{Fermi gamma-ray space
telescope} 
call for a
detailed
study of the high energy (above 100 MeV) gamma--ray
burst (GRB) emission. Current observational information on very
high-energy gamma--rays emitted in a GRB date from the {EGRET}
({Energetic Gamma--Ray Experiment Telescope}) mission on
board the \textit{CGRO} (\textit{Compton Gamma Ray Observatory}). It detected
high energy photons from a handful of GRBs; the most energetic (18 GeV)
photon was detected in the case of GRB 940217 \citep{hurley:94}. \citet{gonzalez:03} reported the observation of a bright high-energy component in GRB
941017, showing a strong temporal evolution, distinct from the low-energy
($<$2 MeV) component. The inspection of the sample of gamma--ray
bursts that were observed both by 
{EGRET} and {BATSE} ({Burst and Transient Source Experiment})
indicates that these bursts were among the brightest ones detected by {BATSE}
\citep[e.g.][]{baring:06};
as {BATSE} trigger was sensitive in the
lower energy range, there could be a population of bursts  with high
energy photons that did not trigger {BATSE}
 \citep{jones:96}. \citet{kaneko:08} reported the 
spectral analysis of combined {BATSE} and {EGRET}
data for 15 bright GRBs in energy range $\sim$30 keV-200 MeV,
emphasizing the importance of such broadband spectral analysis in
constraining the high-energy spectral indices and break energies of GRBs
that have significant MeV emission. More recently
\citet{giuliani:08}
reported observations of GRB 080514B by \textit{AGILE} showing
some evidence that the emission above 30 MeV extends for a longer
duration
than the emission observed at lower energies.
 Evidence of even higher (TeV) energy emission from GRBs was
 reported from ground-based experiments, based on the detection of
 extensive air showers produced by high energy photons propagating in the
 atmosphere \citep{atkins:00}.\\

The observation of high energy spectral components in GRBs can provide 
strong constraints on
present models for the GRB prompt phase. 
GRBs are believed to be produced by ultra-relativistic
($\Gamma \ga 100$)
outflows ejected from a newly formed
compact stellar mass source. The prompt gamma-ray emission is usually
interpreted as radiation from accelerated electrons in shock waves that
propagate within the outflow \citep{rees:94}. Such internal shocks can form if the
ejection process by the central source is highly variable. A high
energy spectral component is expected
within this
framework \citep[see e.g.][]{papathanassiou:96,sari:97,pilla:98,guetta:03,peer:04,razzaque:04,asano:07,gupta:07,galli:08,fan:08,ando:08}. The typical GRB spectrum in the low gamma-ray range, as observed for
instance by {BATSE}, is a smoothly connected broken power
law with a break energy in the range 0.1 - 1
MeV. This component can be
directly produced by
synchrotron radiation from the shock accelerated electrons,
or by inverse Compton scatterings of low-energy synchrotron
photons by the relativistic electrons.
Thus observations of the GRB spectrum extending to very high energy
emission (GeV ranges) can be expected when the keV-MeV photons
are inverse Compton scattered (provided that the $\gamma\gamma$ opacity
in the source is low). Depending on the relevant parameters, the flux
of the high energy component can be even comparable to the prompt GRB
gamma--ray flux in {BATSE} energy range. \\

Significant observational progress is expected with the
launch of \textit{Fermi}, whose two instruments, {GBM} (\textit{GLAST} burst monitor) and
{LAT} (Large Array Telescope) will allow the observation of GRBs in an unprecedented
spectral range from 8 keV to 10 GeV or above \citep{gehrels:99}. The {LAT} has
a large field of view ($\sim 2\ \mathrm{sr}$), is about
10 times more sensitive than {EGRET} and has a very short dead
time of $\sim 100\,\mathrm{\mu s}$ (compared to 100 ms for {EGRET}). \textit{Fermi}
should therefore detect 100 to 200 GRBs per year ({GBM}+{LAT}), with an
appreciable number of them being bright enough above 100 MeV to allow a
good characterization of their temporal and spectral properties in the
high-energy gamma-ray range.\\

This paper presents a detailed calculation of the GRB prompt emission in the context of the internal shock
model, focussing on the high energy (above 100 MeV) range. The emission
in the shocked region is computed using a radiative code that was
developed to solve simultaneously the time evolution of the electron
and photon distribution, which is a significant improvement compared to
studies based on an analytical estimate of the spectrum. This  radiative
calculation is less detailed than in previous studies \citep{peer:05,asano:07}
as it does not include components emitted by shock-accelerated protons
or by electron--positron pairs created from $\gamma\gamma$
annihilation. However it includes all the relevant processes for the
emission from shock-accelerated electrons, whose contribution is
expected to be dominant. In addition, this radiative calculation is combined for the first time with a detailed dynamical simulation of
the internal shock phase, which allows us not only to estimate the spectrum
of the prompt GRB emission, but to generate full synthetic GRBs with
lightcurves and spectra. This approach is described in
\refsec{sec:method}. The
effect of the parameters describing the physical conditions in the
shocked medium on the shape of
the emitted spectrum in the comoving frame is shown in
\refsec{sec:spec_com}. 
The
parameter space of the internal shock model is explored in a systematic
way in
\refsec{sec:parameterspace}, which allows the identification of the different classes of
high-energy spectra that can be expected. We show how \textit{Fermi}
data will allow us to diagnose the dominant radiative process (synchrotron
radiation vs inverse Compton scatterings), the physical conditions in
the shocked medium (electron distribution, magnetic field) and the
properties of the relativistic outflow (Lorentz factor and injected
kinetic power). Finally, \refsec{sec:completegrbs} describes examples of
synthetic bursts (lightcurves and spectra) and
discuss how the comparison between the {LAT} and the {GBM}
lightcurves and the observed spectral evolution in \textit{Fermi} bursts
are also powerful tools to better constrain the physical processes at
work in GRBs.
\refsec{sec:conclusions} summarizes the results of this study.

\section{Internal shocks: dynamics and radiative processes}
\label{sec:method}
We assume that a relativistic
outflow is ejected by the central source of the gamma-ray burst, and
that,
due to initial variability in the distribution of the Lorentz factor,
shock waves form and propagate within this outflow \citep[internal shocks,][]{rees:94}. A fraction of the kinetic energy
which is dissipated in the shock waves is radiated and produces the
observed prompt GRB. Here, we focus on the most discussed version of the
internal shock model, where the radiation is due to shock-accelerated
electrons in optically thin conditions. It has been suggested that shock accelerated protons could
also contribute to the high-energy emission \citep{razzaque:04,asano:07,asano:08}, or that the
emission could occur in optically thick regions leading to quasi-thermal
comptonization \citep{ghisellini:99,meszaros:00,peer:04,giannios:07}, or that the
dominant process is not related to shock-accelerated electrons but
rather to decaying pions \citep{paczynski:94}. These alternative
possibilities are not considered in this paper.\\

In order to follow the time evolution of the photon spectrum
emerging from the relativistic outflow during the internal shock phase,
several steps are needed:
\begin{enumerate}
\item The dynamics of the internal shock phase must be
      followed
to determine the physical conditions behind each shock wave;
\item In the shocked medium, electrons are accelerated and the
      magnetic field is amplified. The emitted photon spectrum has to be
      computed from the time-dependent evolution of the
      relativistic electrons. This evolution is governed by several radiative processes that are in competition
      with the adiabatic
      cooling due to the spherical expansion.
\item From the evolution of the emission in the comoving frame of the
      shocked material, one can deduce the observed prompt GRB lightcurve and
      spectrum.   
\end{enumerate}
Some aspects of this project have already been studied
by several authors, who focus on the second step (radiative
processes in the comoving frame) after assuming a typical collision
between two relativistic shells for the first step. This has been done
either using an approximate analytical or semi--analytical estimate of
the spectrum
\citep[e.g.][]{papathanassiou:96,guetta:03,gupta:07,galli:08,fan:08,ando:08} or
a detailed radiative code \citep{peer:04,asano:07}. Such studies allow
to discuss 
 different emission mechanisms of high energy photon production
during internal shocks 
and to derive the expected high energy photon spectrum from
one single shocked relativistic shell. However, they cannot produce full
lightcurves and time-evolving spectra, and
evaluate
the role of
the dynamics of internal shocks in the observed spectral evolution. 
In this work we attempt to improve this approach by combining a complete
model for the dynamics of the internal shocks with a detailed
calculation of the relevant radiative processes occurring in the shocked
medium. This allows us for the first time to obtain the time evolution
of the high-energy gamma-ray emission in a GRB. The procedure we have
adopted in described in the present section.

\subsection{Dynamical evolution during the internal shock phase}
\label{sec:dynamics}
The dynamics of internal shocks within a relativistic outflow has been
described in \citet{kobayashi:97} in the case where the central engine
is emitting a discrete number of shells, separated by short periods
without any ejection. In this scenario, each pulse observed in the GRB
lightcurve is due to a collision between two shells. One potential
problem with this approach is
that the pulse shape in the decay phase is dominated by the so-called
curvature effect, i.e. the spreading of the arrival time of photons
emitted simultaneously on a curved surface. Such a decay is too fast
compared to observations \citep[see e.g.][]{soderberg:01}. In this
paper, the dynamics of internal shocks is rather computed using the model
developed by \citet{daigne:98}, where the relativistic ejection is now 
considered as a continuous process. Instead of collisions between
discrete shells, internal shocks are in this case shock waves
propagating within the outflow. In the observed lightcurve, the shape of
pulses in their decay phase is then determined by the hydrodynamical timescale associated with the
propagation of the shock waves, rather than the curvature
effect (except at the very end of this dynamical phase). Slow pulse decays can easily be obtained, which
greatly improves the agreement with observations \citep{daigne:03}.\\
\begin{figure*}
\centerline{\psfig{file=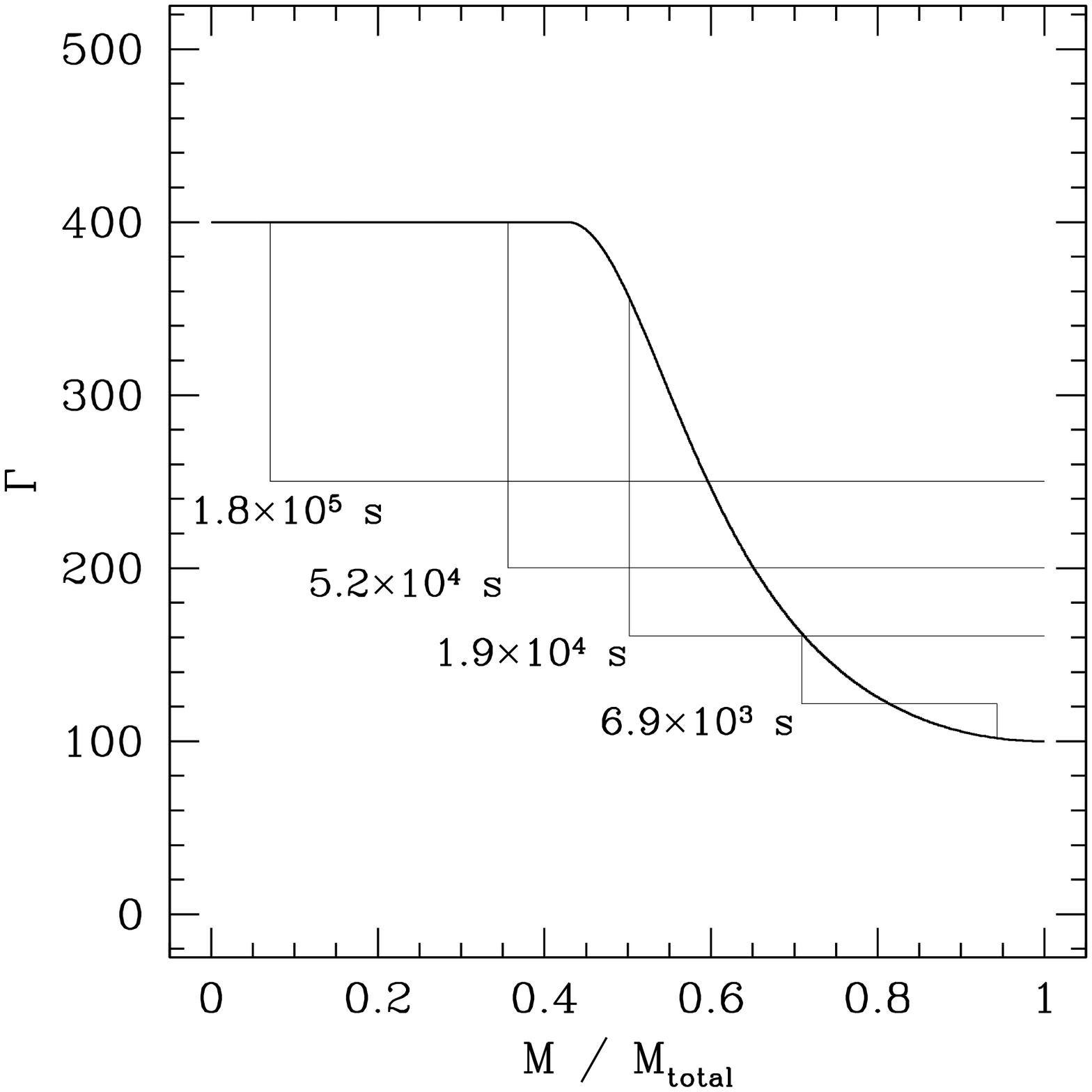,width=0.49\textwidth}\psfig{file=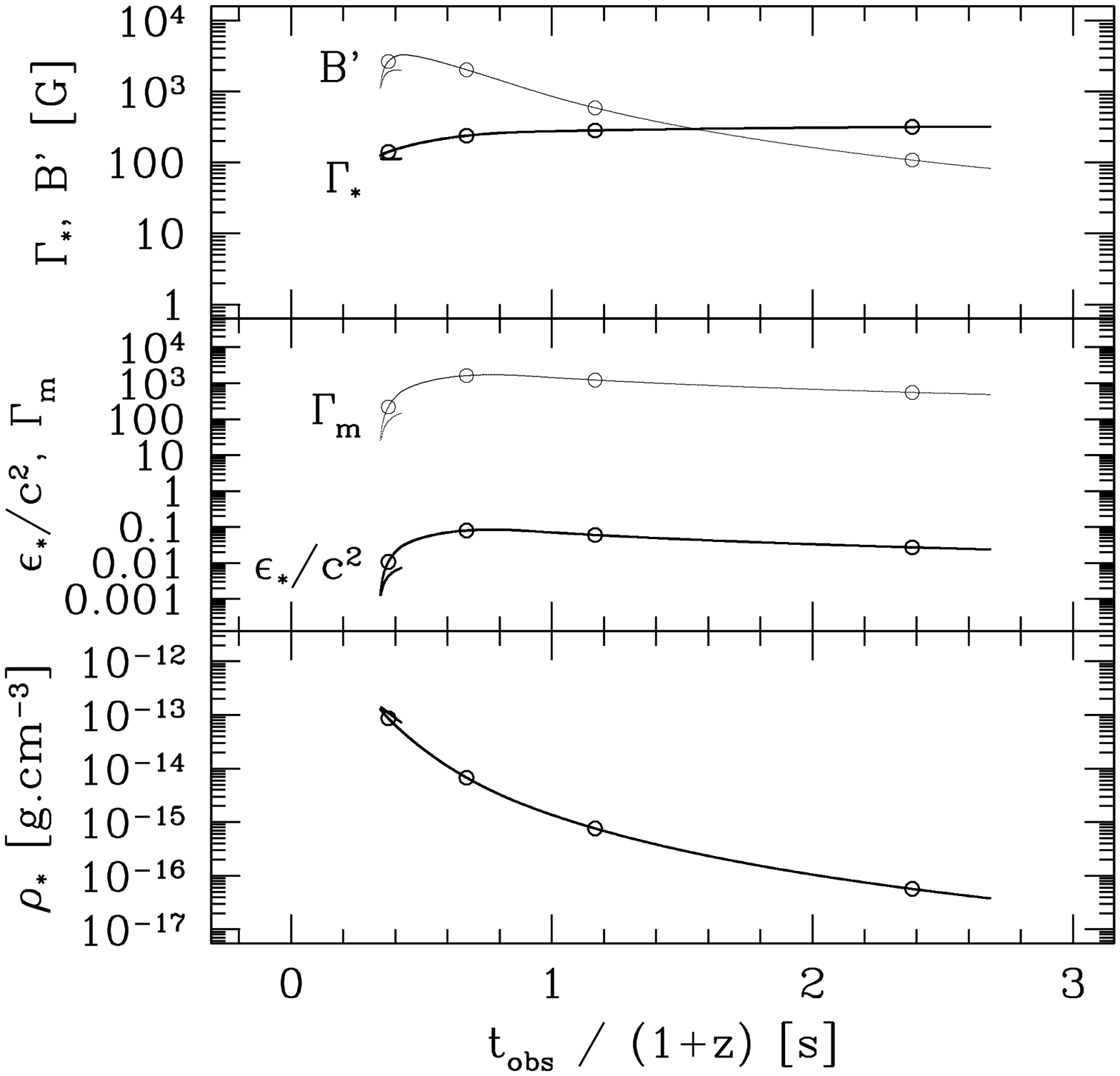,width=0.49\textwidth}}
\caption{\textbf{Dynamics of internal shocks: an example.}
 \textit{Left.} Evolution of the distribution
 of the Lorentz factor in the relativistic outflow. In this example, the
 initial distribution (thick solid line)
 corresponds to a case where material has been ejected for $t_\mathrm{w}=$ 2 s by the
 central source, with a Lorentz factor $\Gamma(t)$ increasing from 100 to 400. Two
 internal shock waves form when the faster part catches up with
 the slower one, as it is shown by the evolution of the distribution of the Lorentz
 factor with  time
 (thin solid lines). The dynamics is computed assuming an injected
 kinetic power
 $\dot{E}=5\times 10^{52}\ \mathrm{erg~s^{-1}}$. \textit{Right.} Corresponding physical conditions
 ($\Gamma_{*}$, $\rho_{*}$ and $\epsilon_{*}$) in the shocked
 material (see text), plotted as a function of the photon arrival time. 
For each curve, the two branches correspond to the two shock
 waves that form in the outflow. Two additional quantities are also
 plotted, $\Gamma_\mathrm{m}$ and $B'$, assuming
 $\epsilon_\mathrm{e}=1/3$, $\zeta=10^{-2}$, $p=2.5$ and
 $\epsilon_\mathrm{B}=1/3$. Circles indicate the physical conditions at
 times $t=6.9\times 10^{3}$, $1.9\times 10^{4}$, $5.2\times 10^{4}$ and
 $1.8\times 10^{5}\ \mathrm{s}$, for which the Lorentz factor distribution
 in the outflow is plotted in the left panel.
Notice that the shock propagating
 forward reaches rapidly the front edge so that, in this example, the
 emission will be dominated by one shock wave only. This example will
 lead to  a single pulse burst. In practice, the initial Lorentz factor
 considered here has to be seen as a building block for more realistic
 distributions leading to multi-pulse lightcurves (see \refsec{sec:multipulses}).}
\label{fig:exampledyn}
\end{figure*}

The dynamics during the internal shock phase is entirely determined from
the following parameters\,: the total duration $t_\mathrm{w}$ of the
relativistic ejection and the history of the Lorentz factor $\Gamma(t)$
and of
the injected kinetic power $\dot{E}(t)$ during this ejection. In
practice, the outflow is described  as a
series of shells emitted regularly over a timescale $\Delta t \ll
t_\mathrm{w}$, so that the number of shells is much larger that the
number of pulses in the lightcurve. These shells interact only by direct
collisions, so that the propagation of a shock wave is discretized by a
succession of shocks between shells. The details of the implementation
of this model
are described in \citet{daigne:98}. This method has been validated by a
comparison with the results of a 1D Lagrangian relativistic hydrocode
\citep{daigne:00}. Relativistic hydrodynamical simulations of internal shocks
have also been performed by \citet{mimica:04} in the context of blazars
and by \citet{mimica:07} in the context of GRBs. The authors discuss the
efficiency of the conversion of kinetic energy into radiation, and
especially
the impact of the possible magnetization of the outflow, which is not
considered
in the present paper.
The output of a simulation of the internal shock dynamics is the time 
evolution of the physical conditions  in the shocked
medium behind each shock wave (comoving mass density $\rho_{*}$,
comoving specific energy density $\epsilon_{*}$, and Lorentz factor $\Gamma_{*}$). This is illustrated in a simple example shown
in \reffig{fig:exampledyn}, where the Lorentz factor distribution in
the outflow is plotted at different times $t$, and the physical
conditions in the shocked medium are plotted as a function of $
t_\mathrm{obs}/(1+z)=t-R/c$ (arrival time in the observer frame of photons emitted along
the line of sight at radius $R$ and time $t$). \\

To estimate the typical radius and shock conditions in internal shocks,
a simple ``two shells'' model is often used \citep[see
e.g.][]{rees:94,barraud:05,daigne:07,kumar:08}. We consider the ejection of two equal mass relativistic shells
with Lorentz factor $\Gamma_{1}$ and $\Gamma_{2}$ from
the central source. Shell 1 is ejected first and shell 2 after shell 1,
with a delay $\tau$. If the contrast $\kappa=\Gamma_{2}/\Gamma_{1}$ is
greater than unity, an internal shock will occur at a radius
\begin{equation}
R_\mathrm{is} \simeq \frac{8\kappa^{2}}{(\kappa-1)(\kappa+1)^{3}}\bar{\Gamma}^{2}c\tau\ ,
\label{eq:isradius}
\end{equation}
where the average Lorentz factor is
$\bar{\Gamma}=(\Gamma_{1}+\Gamma_{2})/2$. The fraction of the kinetic
energy of the shells which is dissipated in the collision is
\begin{equation}
f_\mathrm{dyn}\simeq \frac{\left(\sqrt{\kappa}-1\right)^{2}}{\kappa+1}\ .
\end{equation}
Then, if the injected kinetic power during the relativistic
ejection phase is $\dot{E}$, the Lorentz factor, comoving mass density and
comoving specific internal energy density in the shocked material are
given by
\begin{eqnarray}
\Gamma_{*}   & \simeq & \frac{2\sqrt{\kappa}}{1+\kappa}\bar{\Gamma}\ ,\nonumber\\
\rho_{*}     & \simeq & \frac{\dot{E}}{4\pi R_\mathrm{is}^{2}\Gamma_{*}^{2}c^{3}}\ ,\nonumber\\
\epsilon_{*} & \simeq & \frac{\left(\sqrt{\kappa}-1\right)^{2}}{2\sqrt{\kappa}}c^{2}\ .
\label{eq:iscom}
\end{eqnarray}
These simple scaling laws will be used to explore the parameter space of
the internal shock
model in the next section.\\

Once the dynamics of the internal shock phase is well understood and the
physical conditions in the shocked material are known, more assumptions
are necessary to compute the emission. This is described in the next subsection.

\subsection{Physical conditions in the shocked medium}

The physics of the acceleration of particles in relativistic shock waves, as well as the amplification of the
magnetic field, is far from being fully understood. It is therefore impossible
in our state of knowledge to directly estimate the electron distribution
and  the magnetic field in the shocked medium from $\Gamma_{*}$,
$\rho_{*}$ and $\epsilon_{*}$ using first principles. Therefore, the
microphysics related to these processes is usually parameterized in a
very simple way, which is adopted in the present  paper: (i) it is assumed that a
fraction $\epsilon_\mathrm{e}$ of the dissipated energy is injected in a
fraction $\zeta$ of the ambient electrons that are accelerated to
relativistic energies, with a power-law distribution of slope $-p$. Note
that most GRB studies (prompt and afterglow emission modelling) are restricted to the case $\zeta=1$ (all electrons
are accelerated) but numerical simulations of particle acceleration in
relativistic shocks suggest that it may not be the case \citep[see e.g.][]{bykov:96,eichler:05,spitkovsky:08}; (ii) it is assumed that a
fraction $\epsilon_\mathrm{B}$ of the dissipated energy is injected in
the magnetic field. We do not investigate in this paper an alternative scenario,
where the magnetic field is dominated by a large-scale component
anchored in the central source \citep[see e.g.][]{spruit:01}. With these four additional parameters
($\epsilon_\mathrm{e}$, $\zeta$, $p$ and $\epsilon_\mathrm{B}$),  the
number density of non-thermal electrons can be computed
\begin{equation}
n^\mathrm{acc}_\mathrm{e} \simeq \zeta\frac{\rho_{*}}{m_\mathrm{p}}\ ,
\end{equation}
as well as their initial distribution 
\begin{equation}
n\left(\gamma\right) \simeq (p-1)\frac{n^\mathrm{acc}_\mathrm{e}}{\Gamma_\mathrm{m}}\left(\frac{\gamma}{\Gamma_\mathrm{m}}\right)^{-p}\ \mathrm{for}\ \gamma\ge\Gamma_\mathrm{m}
\label{eq:epowerlaw}
\end{equation}
with
\begin{equation}
\Gamma_\mathrm{m} \simeq \frac{p-2}{p-1}\frac{\epsilon_\mathrm{e}\ \rho_{*}\epsilon_{*}}{n^\mathrm{acc}_\mathrm{e}m_\mathrm{e}c^{2}}\simeq \frac{p-2}{p-1}\frac{\epsilon_\mathrm{e}}{\zeta}\frac{m_\mathrm{p}}{m_\mathrm{e}}\frac{\epsilon_{*}}{c^{2}}\ .
\label{eq:gm}
\end{equation}
The magnetic field in the comoving frame of the shocked material is
given by
\begin{equation}
B' \simeq \sqrt{8\pi \epsilon_\mathrm{B}\ \rho_{*} \epsilon_{*}}\ .
\end{equation}
The evolution of $\Gamma_\mathrm{m}$ and $B'$ is plotted for our example
in \reffig{fig:exampledyn}.\\

In practice, it is assumed that the relativistic electron distribution
extends up to a maximum Lorentz
factor $\Gamma_\mathrm{M}$, defined  as the Lorentz factor where
the acceleration timescale becomes comparable to the minimum of the
radiative timescale and the escape timescale (see below). This
corresponds to the most efficient acceleration that can be expected.
In the
comoving frame of the shocked region, the acceleration
timescale of an electron with Lorentz factor $\gamma$
is estimated as $R'_\mathrm{L}(\gamma)/c$, where 
\begin{equation}
R'_\mathrm{L}(\gamma) = \frac{\gamma m_\mathrm{e} c^{2}}{e B'}
\end{equation}
 is the Larmor
radius. This leads to
\begin{equation}
\Gamma_\mathrm{M} = \min{\left(\left(\frac{6\pi e}{\sigma_\mathrm{T}B'}\right)^{1/2};\frac{e B' t'_\mathrm{ex}}{m_\mathrm{e} c}\right)}\ ,
\label{eq:gmax}
\end{equation}
where the radiative timescale is taken to be equal to the synchrotron
timescale (\refeq{eq:tsyn} below) and the escape timescale is
identified with the dynamical timescale $t'_\mathrm{ex}$
(\refeq{eq:tex} below).
Note that when inverse Compton losses are important, this expression
overestimates the maximum Lorentz factor $\Gamma_\mathrm{M}$. This is further
discussed later on.

\subsection{Emission in the comoving frame}
\label{sec:radiation}
\paragraph{Timescales.} Two timescales are necessary to characterize the physics in the shocked
region: (i) the dynamical timescale 
\begin{equation}
t'_\mathrm{ex} \simeq \frac{R}{\Gamma_{*} c}\ ,
\label{eq:tex}
\end{equation} 
which is the
typical timescale associated with the adiabatic cooling due to the
spherical expansion; and (ii) the radiative timescale $t'_\mathrm{rad}$,
defined as the timescale necessary for the relativistic electrons to
radiate most of their energy. As described in \citet{sari:98}, electrons
with $t'_\mathrm{rad}\ll t'_\mathrm{ex}$ are in ``fast cooling'' regime
and will radiate efficiently, whereas electrons with $t'_\mathrm{rad}\gg
t'_\mathrm{ex}$ are in ``slow cooling'' regime and will loose most of
their energy via the adiabatic cooling. In internal shocks, the short
variability timescale observed in the lightcurves imposes that all
electrons are in fast cooling regime \citep{rees:94,sari:96,kobayashi:97}. This is also probably required by
pure energetic considerations, as the huge
gamma-ray luminosities observed in GRBs are very difficult to understand
if electrons are not radiating efficiently. From a numerical point of
view, the advantage of being in fast cooling regime is that the emission
is produced over a short timescale: relativistic electrons accelerated
in one collision will radiate most of their energy before the next
collision occurs. This allows to compute the emission in an independent
way: for each dyn\-amical timestep (duration $\sim t'_\mathrm{ex}$), the
radiation in the shocked region is computed assuming that the dynamical
quantities (e.g. the density) do not vary.

\paragraph{Geometry.}
The shocked region is a shell with radius $R$, opening angle
$\Delta\theta$ (equal to the opening angle of the outflow, that can be
considered as constant in the internal shock phase, the lateral
expansion becoming efficient only when the outflow has notably
decelerated) and comoving width $\Delta'\sim c t'_\mathrm{ex}$. During
the dynamical timescale $t'_\mathrm{ex}$, the emission in the comoving
frame can be computed assuming constant dynamical quantities. 
The causally
connected region during this duration has a size $c t'_\mathrm{ex}$
which is small compared to the lateral size of the shell
$R\Delta\theta$, as long as $\Delta\theta\gg
1/\bar{\Gamma}$. In the comoving frame of the shocked region, one can
therefore neglect the curvature and consider a infinite plane layer with
width $c t'_\mathrm{ex}$.\\

As electrons are in fast cooling regime with $t'_\mathrm{rad}\ll
t'_\mathrm{ex}$, most of the evolution occurs on a short timescale,
corresponding to a causally connected region of size $c t'_\mathrm{rad}$
much smaller than the physical width of the region. Therefore, it is
justified to assume that, if the shell is initially homogeneous, it
will remain so for most of the evolution: the density distribution of
electrons will depend on time, but not on the position in the shocked
region. The same will happen for the photon distribution, which will
appear as isotropic everywhere in the shocked region. This is of course
not strictly valid within a distance $ct'_\mathrm{rad}$ from the edge of
the shell, but the corresponding volume is negligible, as
$t'_\mathrm{rad}\ll t'_\mathrm{ex}$.

\paragraph{The photon field}
\label{sec:photonfield}
At time $t'=0$, just after the collision, when the particle acceleration
and the amplification of the magnetic field are achieved (it is assumed that these
processes operate on timescales which are short compared to the
radiative and the dynamical timescales), the electron distribution is
given by \refeq{eq:epowerlaw} and the photon density is
zero. This is justified as all electrons that were shock-accelerated earlier have
already cooled.\\

The photon density distribution at time $t'$ at a given position in the shocked region is
given by
\begin{equation}
n_{\nu'}(t')=\frac{4\pi}{c} \frac{I_{\nu'}(t')}{h\nu'}\ ,
\end{equation}
due to the isotropy of the photon field (see above). If 
absorption is neglected at this stage of the discussion,
the specific
intensity $I_{\nu'}(t')$ is built by integration of
\begin{equation}
\frac{dI_{\nu'}}{ds'}=j_{\nu'}
\end{equation}
along a ray, from $s'=0$ to $s'=ct'$ (where $s'=0$ is the position where
$n_{\nu'}$ is computed), due to the finite speed of
light. Assuming an isotropic emission by electrons, this leads to
\begin{eqnarray}
I_{\nu'} & = & \frac{1}{4\pi}\int_{0}^{c t'} ds' P_{\nu'}\left(s',\tilde{t}'=t'-\frac{s'}{c}\right)\nonumber\\
         & = & \frac{c}{4\pi}\int_{0}^{t'}\ d\tilde{t}' P_{\nu'}(\tilde{t}')\ ,
\end{eqnarray}
where  the homogeneity of the shock region is taken into account (see
above). Finally, the photon density distribution is given by
\begin{equation}
n_{\nu'}(t') = \int_{0}^{t'}d\tilde{t}' \frac{P_{\nu'}(\tilde{t}')}{h\nu'}\ ,
\end{equation}
where it appears clearly that the local photon field is built by
accumulating photons coming from a growing region of size $c\tilde{t}'$ and
therefore depends on the whole
history of the emission between $\tilde{t}'=0$
and $\tilde{t}'=t'$. In the next paragraph this equation is expanded by
giving explicitly the emission processes that are considered in the
present study and including the absorption
processes that were neglected in this paragraph.

\paragraph{Radiative processes.}
Many radiative processes can operate in the shocked medium. In this
paper, we focus on the processes that are expected to be dominant if the
radiation is mainly produced by electrons, i.e. we do not include
contributions associated to a possible population of relativistic
protons accelerated in the shock. Such a component is included in the
calculations made by \citet{asano:07} for a typical shock. Their results
show that (i) for most parameters, the proton contribution is
negligible, especially below a few GeV; (ii) it is only when
$\epsilon_\mathrm{B}\gg \epsilon_\mathrm{e}$, i.e. when most of the
dissipated energy is injected in the magnetic field and in protons,
rather than in electrons, that a non negligible proton component
emerges.\\ 

Accelerated relativistic electrons in the
amplified magnetic field will radiate via the synchrotron process. These
synchrotron photons can be scattered to higher energies by relativistic
electrons (inverse Compton). At low energy, they can also be absorbed
(synchrotron-self absorption). At high energies, photon--photon
annihilation can occur, producing electron-positron pairs. The
corresponding pairs could contribute to the radiation, but this
contribution is not considered in the present paper, as we limit our
studies to cases where the production of pairs is weak (see next
section). We did not
consider in this study the case of the ``jitter radiation''
\citep{medvedev:00,medvedev:08} that is an alternative to the standard
synchrotron radiation.\\

Based on the timescales and the geometry discussed above, we have
implemented a radiative code to solve the evolution of electrons and
photons in the comoving frame of the shocked medium during a dynamical
timestep. Two equations are solved, one for the
evolution of the comoving electron density distribution $n\left(\gamma,t'\right)$\,:
\begin{equation}
\frac{\partial n}{\partial t'} =  -\frac{\partial}{\partial\gamma}\left[\left.\frac{d\gamma}{dt'}\right|_{\mathrm{syn+ic+ad}} n\left(\gamma,t'\right) \right]
\label{eq:electrons}
\end{equation}
and one for the evolution of the comoving photon density distribution
$n_{\nu'}(t')$\,:
\begin{eqnarray}
\frac{\partial n_{\nu'}}{\partial t'} & = & \int d\gamma\ n\left(\gamma,t'\right) \frac{P_{\nu'}^\mathrm{syn+ic}(\gamma)}{h\nu'}\nonumber\\
& &  - c n_{\nu'}(t')\int d\gamma\ n\left(\gamma,t'\right)\sigma_\mathrm{sa}\left(\gamma,\nu'\right)\nonumber\\
& & - c n_{\nu'}(t')\int d\tilde{\nu}'\ n_{\tilde{\nu}'}(t') \sigma_{\gamma\gamma}\left(\nu',\tilde{\nu}'\right)\ .
\label{eq:photons}
\end{eqnarray}
The indexes $\mathrm{syn}$, $\mathrm{ic}$, $\mathrm{ad}$, $\mathrm{sa}$
and $\mathrm{\gamma\gamma}$ stand respectively for the following
processes: synchrotron radiation, inverse Compton scattering,
adiabatic cooling, synchrotron self-absorption and photon--photon
annihilation.\\ 

The expressions of the different terms appearing in
Eqs.~(\ref{eq:electrons}) and~(\ref{eq:photons}) are listed in
appendix~\ref{sec:radproc} and the
numerical method to solve this set of equations is described in
appendix~\ref{sec:mnum}. 
The adiabatic losses are estimated by
$\left.d\gamma/dt'\right|_\mathrm{ad}=-\gamma/t'_\mathrm{ex}$. The
synchrotron radiation is computed exactly, assuming an isotropic
distribution of the pitch angle $\alpha$ between the electron velocity
and the magnetic field. The synchrotron self-absorption is also computed
using the exact cross-section \citep[see e.g.][]{rybicki:79}. Note that
the corresponding heating
term at low energy is neglected in
\refeq{eq:electrons}.
 Inverse Compton scatterings are computed using the
approximate kernel derived by \citet{jones:68}, which is
a  very good approximation, even in the Klein-Nishina regime. Note that \refeq{eq:photons}
does not include the loss term at low frequency corresponding to the
source term at high energy. This is because the Thomson optical depth is
always low in our case (see next section). We do not examine
situations where comptonization could occur. Finally, the
full cross-section for gamma-gamma annihilation is used, assuming an isotropic
photon field \citep{gould:67}. As mentioned above, the present version of the code does not include the
associated pair creation term, so that we limit the study to cases where
it is negligible (see next section).\\
\begin{figure*}
\centerline{\psfig{file=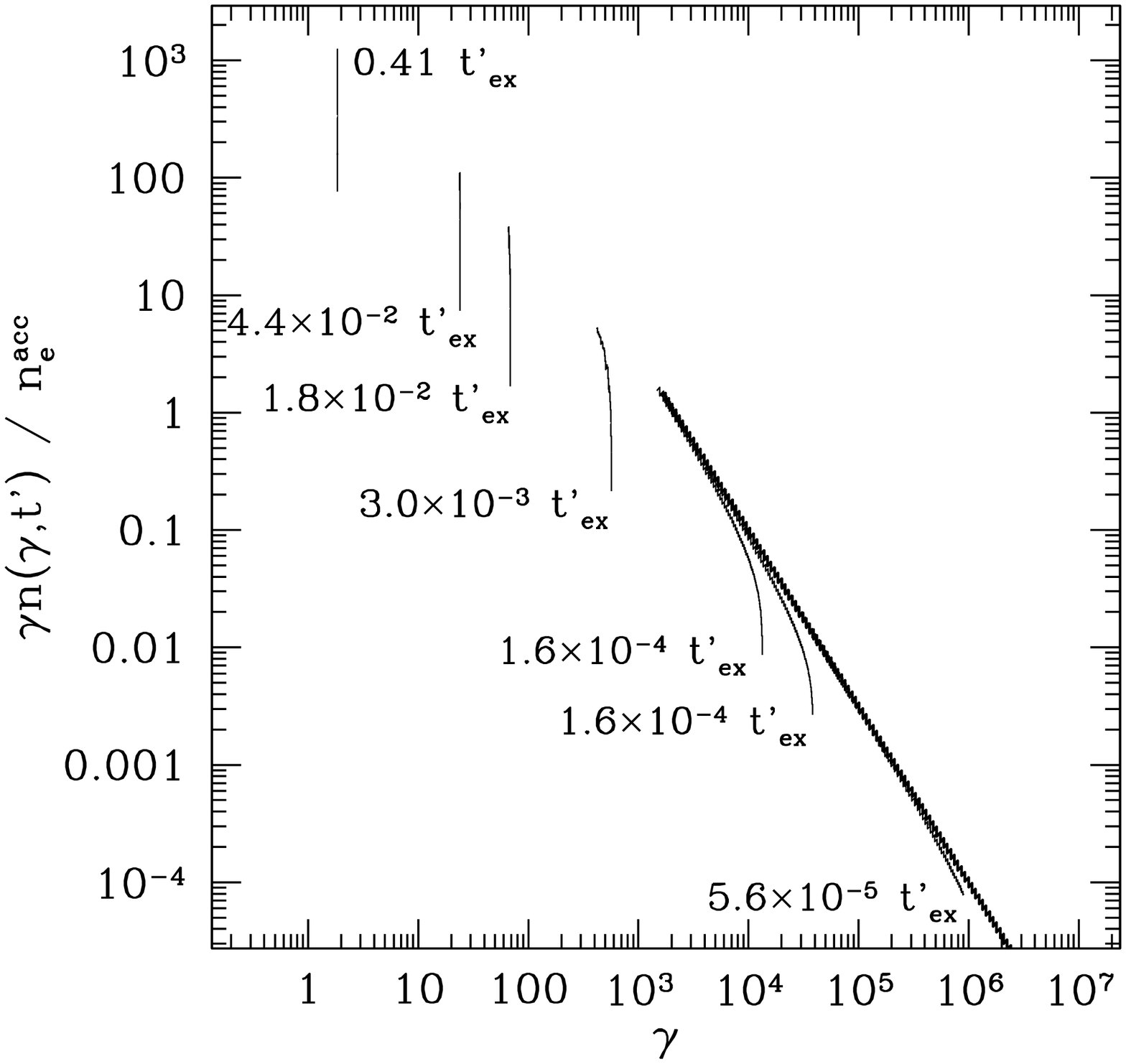,width=0.49\textwidth}\psfig{file=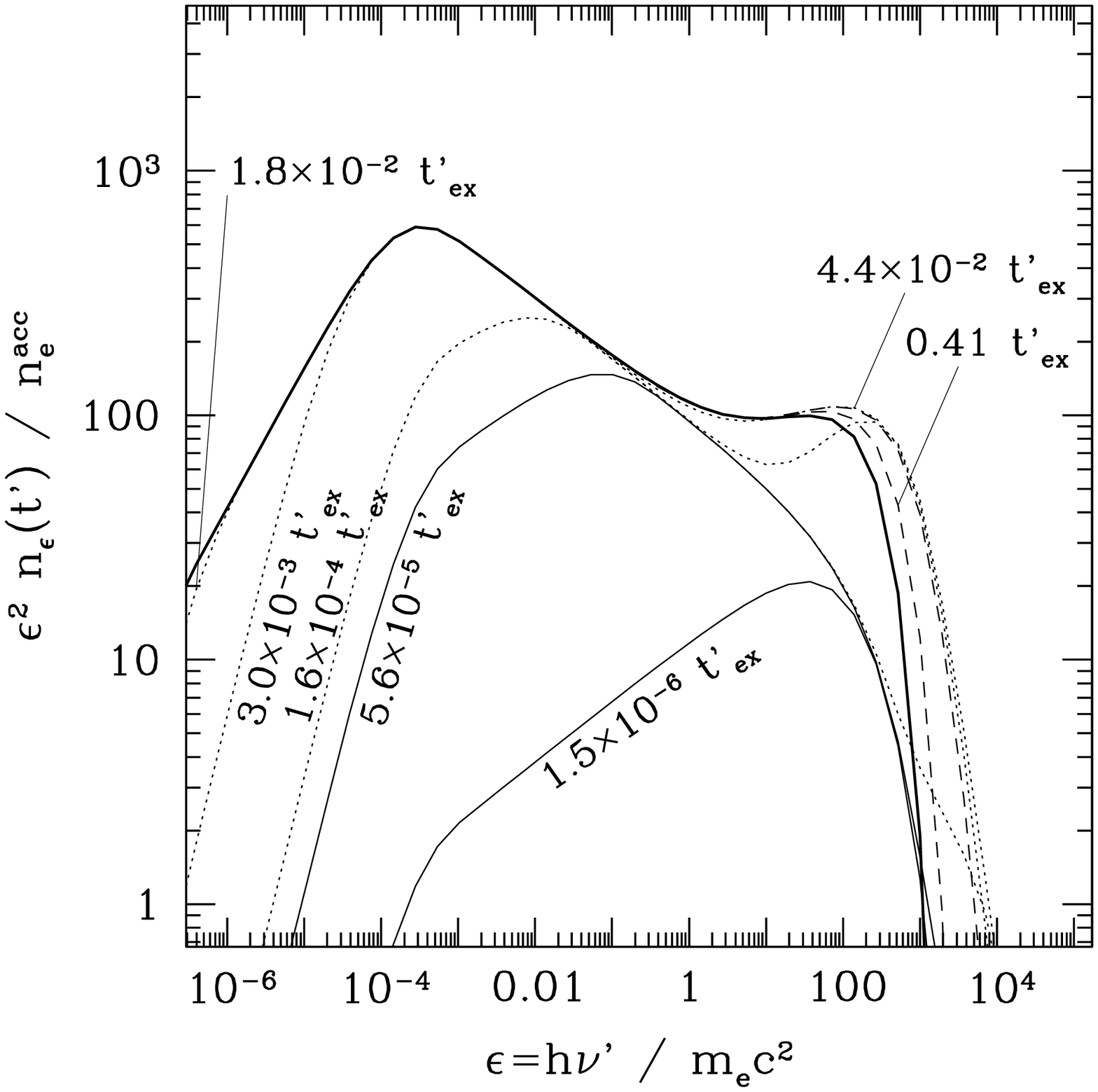,width=0.49\textwidth}}
\caption{\textbf{Emission in the comoving frame: an example.} This
 figure shows the result of the radiative calculation in the comoving
 frame of the shocked material, at time $1.9\times 10^{4}\ \mathrm{s}$ in the example shown in
 \reffig{fig:exampledyn}. \textit{Left.} Evolution of the
 relativistic electron population. The initial distribution is plotted
 with a thick line. \textit{Right.} Evolution of the photon spectrum 
 $\left.\nu'\right.^{2} n_{\nu'}$. The final spectrum at $t'=t'_\mathrm{ex}$
 is plotted with a thick line. Several phases are well identified: the
 synchrotron spectrum is built very early (thin solid line). Once these
 seed photons are present,  inverse Compton scatterings produce high
 energy emission (dotted line). At late times, high-energy photons
 annihilate which reduces the intensity of the second peak in the
 spectrum (dashed line). The microphysics parameters are
 $\epsilon_\mathrm{e}=1/3$, $\zeta=10^{-2}$, $p=2.5$ and
 $\epsilon_\mathrm{B}=1/3$ so that the initial Thomson optical depth is
 $\tau_\mathrm{T}^\mathrm{acc}\simeq 6\times 10^{-5}$ and the critical
 Lorentz factor equals $\Gamma_\mathrm{c}\simeq 2.4$  (see text).}
\label{fig:examplecom}
\end{figure*}

Following \citet{sari:98}, it is convenient to define $\Gamma_\mathrm{c}$ as the Lorentz
factor of electrons whose synchrotron timescale 
\begin{equation}
t_\mathrm{syn}'\left(\gamma\right) = \frac{6\pi m_\mathrm{e} c}{\sigma_\mathrm{T} \left.B'\right.^{2} \gamma}
\label{eq:tsyn}
\end{equation}
is equal to the
adiabatic cooling timescale, i.e.
\begin{equation}
\Gamma_\mathrm{c} = \frac{6\pi m_\mathrm{e} c}{\sigma_\mathrm{T} \left.B'\right.^{2} t'_\mathrm{ex}}\ .
\label{eq:gc}
\end{equation}
When the synchrotron process is dominant, electrons with $\gamma >
\Gamma_\mathrm{c}$ are in fast cooling regime. When inverse Compton scatterings
become efficient, the effective transition between slow and fast
cooling occurs at a Lorentz factor lower than $\Gamma_\mathrm{c}$ as
the radiative timescale becomes shorter than the synchrotron
timescale.\\

The solution at $t'=t'_\mathrm{ex}$ of system of
Eqs.~(\ref{eq:electrons}) and~(\ref{eq:photons}) is entirely determined
by the expansion timescale $t'_\mathrm{ex}$,
 the shape of the initial electron distribution (i.e. mainly
$\Gamma_\mathrm{m}$ and $p$), the relativistic electron density
$n_\mathrm{e}^\mathrm{acc}$ and the magnetic field $B'$. Rather than
using these two last quantities, it is convenient to consider
alternatively the critical Lorentz factor
$\Gamma_\mathrm{c}$ and  the initial Thomson optical depth associated to
relativistic electrons 
\begin{equation}
\tau^\mathrm{acc}_\mathrm{T}=\sigma_\mathrm{T}
n^\mathrm{acc}_\mathrm{e} c t'_\mathrm{ex}\ .
\label{eq:tauacc}
\end{equation}

The radiative calculation has to be made for each collision occurring in
the dynamical phase, i.e. at each instant along the propagation of a shock
wave within the relativistic outflow. \reffig{fig:examplecom}
shows one of these elementary calculations. This case has been selected
as the effect of each process is clearly identified (see caption of the
figure). Possible
additional effects (scatterings or absorption) between photons emitted
in a shocked region and electrons or photons present in another
shocked region, which could affect the high-energy spectrum
\citep{gruzinov:00}, are not considered in the present paper but will be
investigated in the future. 
We also ignore the effects of triplet pair production which can occur
when electrons of very high energies
encounter soft photons : the cross-section for this process becomes
larger than the inverse Compton cross-section in the deep Klein-Nishina regime, for
$\gamma \epsilon \ga 250$, where $\gamma$ is the electron Lorentz factor
and $\epsilon=h\nu/m_{e}c^{2}$
the soft photon energy \citep{mastichiadis:91}. Finally,
we do not include the possible
interaction of the prompt gamma-rays emitted from internal shocks and
the circumburst environment, that could also lead to an additional early
high-energy component \citep{beloborodov:02,beloborodov:05}.

\subsection{Observed flux}
Once the emission in the comoving frame is computed at each instant
along the propagation of internal shocks within the relativistic
outflow, the observed flux as a function of time is computed by summing
up the contributions of all shock waves, taking into account\,: (i) the
relativistic effects (Lorentz transformation from the comoving frame of
the shocked region to a fixed frame); (ii) the curvature of the emitting
surface; (iii) the cosmological effects due to the redshift of the GRB
source. The two first points require an integration over
equal-arrival times surfaces, that is carried out following equations given in \citet{woods:99}.
Any absorption in the gamma-ray range due to pair creation on
the extragalactic background light is neglected. This would be
important, depending on the redshift, above $\sim 10\,\mathrm{GeV}$.
Examples of synthetic lightcurves and spectra produced following
the complete procedure described in this section are presented in
\refsec{sec:completegrbs}. 

\section{The emitted spectrum in the comoving frame}
\label{sec:spec_com}
As described in \refsec{sec:method}, the emitted spectrum in the
comoving frame of the shocked material is entirely determined by four
parameters:
(i) the magnetic field $B'$; (ii) the adiabatic cooling timescale
$t_{\mathrm{ex}}'$; (iii) the relativistic electron density
$n_\mathrm{e}^\mathrm{acc}$, and (iv) the shape of the initial
distribution of the Lorentz factor of accelerated electrons, i.e. the
slope $p$ and the minimum Lorentz factor $\Gamma_\mathrm{m}$ for a
power-law distribution. A clear insight in the way that every of these
parameters affects the radiative processes is necessary to
anticipate the characteristics of the photon spectrum resulting from the
shock-accelerated electrons. \\

The final observed photon spectrum comprises the contributions of all
the photons emerging from the collisions occurring during the evolution
of the relativistic outflow. We focus first
on the radiative
processes and the photon spectrum occurring after a single collision only
and will describe later (\S~\ref{sec:completegrbs}) the complete GRB lightcurve and spectrum.\\

We have carried out spectral calculations corresponding to a large
exploration of the parameter space
$\left(B',t'_\mathrm{ex},n_\mathrm{e}^\mathrm{acc},\Gamma_\mathrm{m}\right)$
describing the physical conditions
in the shocked material,
assuming  a fixed electron slope $p=2.5$. We computed 2744 spectra
corresponding to\,: (i) 7 values of
the magnetic field $\log{\left(B'/1\ \mathrm{G}\right)}=1$, $1.5$, $2$, $2.5$, $3$, $3.5$ and $4$;
(ii) 7 values of the dynamical timescale
$\log{\left(t'_\mathrm{ex}/1\ \mathrm{s}\right)}=0$, $0.5$, $1$, $1.5$,
$2$, $2.5$ and $3$;
(iii) 7 values of the electron density
$\log{\left(n_\mathrm{e}^\mathrm{acc}/1\ \mathrm{cm^{-3}}\right)}=4$,
$5$, $6$, $7$, $8$, $9$ and $10$;
(iv) and 8 values of the minimum electron
Lorentz factor $\log{\Gamma_\mathrm{m}}=1.5$, $2$, $2.5$, $3$, $3.5$,
$4$, $4.5$ and $5$.

\begin{figure}
\begin{center}
\includegraphics[width=0.49\textwidth]{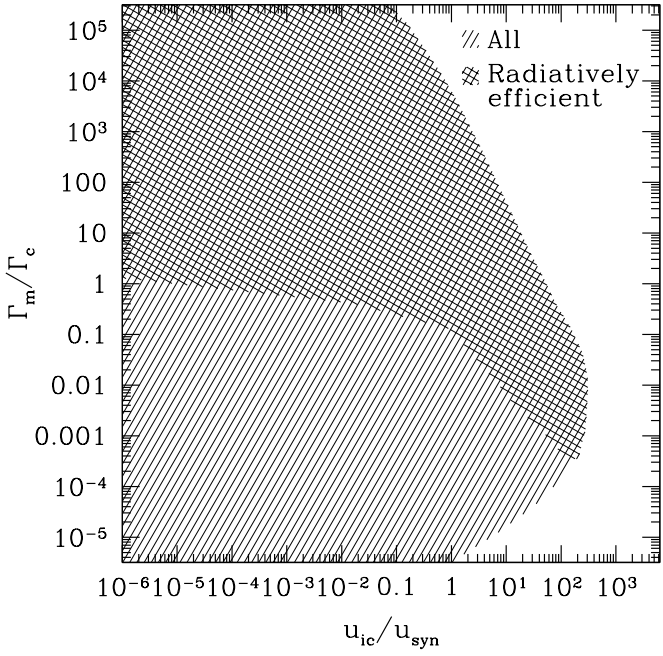} 
\end{center}
\caption{\textbf{Radiative efficiency.} The explored region of the
 parameter space of internal shocks is shown in the plane
 $u_\mathrm{ic}/u_\mathrm{syn}$ versus
 $\Gamma_\mathrm{m}/\Gamma_\mathrm{c}$ (shaded region labeled
 'All'). The x-axis quantity measures the efficiency of inverse Compton
 scatterings and the y-axis compares the synchrotron and the
 adiabatic cooling timescales. Radiatively efficient cases, i.e. cases
 where most of the energy initially injected in relativistic electrons
 is radiated (see text) are indicated (shaded region labeled
 'Rad. efficient'). Inverse Compton scatterings reduce the effective
 radiative timescale of relativistic electrons. Therefore models 
with
large
 $u_\mathrm{ic}/u_\mathrm{syn}$
can be efficient even for 
 low $\Gamma_\mathrm{m}/\Gamma_\mathrm{c}$. For negligible
 inverse Compton emission, the radiatively efficient region corresponds
 to the usual synchrotron fast cooling regime.}
\label{fig:efficiency} 
\end{figure}

\subsection{Radiative efficiency and transparency}
In 
the shocked medium, the
evolution of the relativistic electron distribution is governed by
several radiative processes that are in competition with the adiabatic
cooling due to the spherical expansion. The efficiency of converting the
energy deposited in relativistic electrons in radiation
depends strongly on the relative magnitudes of the radiative cooling
timescale $t_{\mathrm{rad}}'$ of relativistic electrons and the
adiabatic cooling timescale of expanding shell $t_{\mathrm{ex}}'$ (see
\refsec{sec:radiation}). The observed short timescale variability
as well as the high isotropic equivalent energy radiated in gamma-rays
imply that electrons are radiating efficiently, i.e. that
$t'_\mathrm{rad} < t'_\mathrm{ex}$, the so-called fast-cooling regime \citep{sari:98}.
Therefore, in the following, we have only considered the region of the
parameter space where the radiative efficiency is high, i.e.
$u_\mathrm{rad}/u_\mathrm{e}^\mathrm{acc}>0.5$, where
$$
u_\mathrm{e}^\mathrm{acc} = \int d\gamma\ n\left(\gamma,t'=0\right)\gamma
m_\mathrm{e} c^{2}
$$
is the initial energy density in
relativistic electrons and 
$$
u_\mathrm{rad}=\int d\nu'\ n_{\nu'}\left(t'=t'_\mathrm{ex}\right)
h\nu'
$$
is the final energy density
contained in the radiated photons. \reffig{fig:efficiency} shows
this region in the plane
$\Gamma_\mathrm{m}/\Gamma_\mathrm{c}$--$u_\mathrm{ic}/u_\mathrm{syn}$,
where $u_\mathrm{syn}$ (resp. $u_\mathrm{ic}$) is the component of
$u_\mathrm{rad}$ corresponding to synchrotron emission (resp. inverse
Compton
emission). Clearly, when inverse Compton scatterings are inefficient,
the electron radiative timescale is the synchrotron timescale and
the efficiency condition is equivalent to
$\Gamma_\mathrm{m}>\Gamma_\mathrm{c}$ \citep{sari:98}. However, inverse
Compton scatterings reduce the effective electron radiative timescale
and models with $\Gamma_\mathrm{m}<\Gamma_\mathrm{c}$ can still be
efficient if $u_\mathrm{ic} > u_\mathrm{syn}$. In these models, the
Lorentz factor $\tilde{\Gamma}_\mathrm{c}$ of electrons whose radiative
timescale is equal to $t'_\mathrm{ex}$ can be much lower than
$\Gamma_\mathrm{c}$ (defined from the synchrotron timescale only).\\

As we do not consider scenarios where the emitting region is optically
thick \citep[for instance a comptonized spectrum, see e.g.][]{ghisellini:99,meszaros:00,peer:04}, we also limit the
discussion to the region of the parameter space where the medium is optically
thin for Thomson scatterings, i.e. $\tau_\mathrm{T}=\sigma_\mathrm{T}
n_\mathrm{e} c t'_\mathrm{ex} < 0.1$, $n_\mathrm{e}$ being the total density
of electrons (relativistic or not). Using the two-shells model, this condition leads
to a minimum value for the Lorentz factor:
\begin{equation}
\bar{\Gamma} > 1.58\ \left(\frac{\left(\kappa-1\right)\left(\kappa+1\right)^{3}\left(\kappa^{2}+1\right)^{3}}{64\kappa^{7/2}}\ \frac{\sigma_\mathrm{T}\dot{E}}{4\pi m_\mathrm{p}c^{4}\tau}\right)^{1/5}\ .
\end{equation}
This minimum Lorentz factor of the outflow is of the order of $100-200$,
for $\tau\simeq 1\mathrm{s}$, $\dot{E}\simeq 10^{52}\
\mathrm{erg~s^{-1}}$ and $\kappa\simeq 2-4$. 
However, an additional effect must be taken into
account:
high energy photons can annihilate
into $e^{\pm}$ pairs, and the corresponding new leptons will increase
the Thomson optical depth. Therefore the true transparency condition that
we impose is
\begin{equation}
\tau_\mathrm{T}^\mathrm{tot}=\tau_\mathrm{T}\left(1+\zeta n_\mathrm{\pm}/n_\mathrm{e}^\mathrm{acc} \right)< 0.1\ ,
\end{equation} 
\label{sec:transparency}
where $n_{\pm}$ is the final density of leptons produced by pair
annihilation. This will increase the minimum value of the Lorentz factor
derived above. Compared to analytical estimates of the minimum Lorentz
factor \citep[see e.g.][]{lithwick:01}, we use here a precise estimate
of the pair production factor $n_\mathrm{\pm}/n_\mathrm{e}^\mathrm{acc}$
which is a byproduct of our radiative calculation.
\citet{lithwick:01} have also considered a
third transparency condition to derive a minimum value for the Lorentz
factor from GRB observations: the absence in the MeV spectrum of a
cutoff due to $\gamma\gamma$ annihilation. We will discuss this
condition in \refsec{sec:cutoff}, as it is expected that this cutoff
could be observed by \textit{Fermi} in some GRBs in the future.
A consequence of our transparency condition is that all the cases
presented in this paper correspond to
situations where the fraction of energy (initially in high energy photons) which is
deposited in pairs is small.
This justifies that the emission of these
leptons is neglected in our present calculations.\\

Except for the two limitations (efficiency and
transparency), all parameters are a priori acceptable. Indeed, models
for the central engine of gamma-ray bursts are not in a state where a
distribution function can be provided for the injected kinetic power or
the initial Lorentz factor in the outflow. Even the expected range of each
quantity is highly uncertain.
Therefore, the most promising way to estimate such
physical quantities is to apply a detailed spectral model as described in
this paper to recover the internal shock parameters that can reproduce
observed lightcurves and spectra. When the low energy gamma-ray range only
(e.g. {BATSE} data) is used, there is a large degeneracy. Hopefully,
observations in the high-energy gamma-ray range (\textit{Fermi} data) will
improve this situation. For this reason, we study in this section how the broad
spectral shape is affected by each parameter of the model. Before this,
we recall the main scaling laws that are expected from analytical
considerations, and check their validity with our detailed calculation.\\
\subsection{Analytical estimates}
\paragraph{Synchrotron component.}
The dimensionless photon frequency in the comoving frame is defined by
$\epsilon=h\nu'/m_\mathrm{e}c^{2}$. The exact solution of
Eqs.~(\ref{eq:electrons}) and~(\ref{eq:photons}) can be obtained when only
synchrotron radiation and adiabatic cooling are included. The time-averaged (over $t'_\mathrm{ex}$) electron
distribution $\bar{n}(\gamma)$ is very close to a broken power-law:
\begin{eqnarray}
\frac{\bar{n}(\gamma)}{n_\mathrm{e}^\mathrm{acc}} & \simeq & \left\lbrace
\begin{array}{cl}
\frac{1}{\Gamma_\mathrm{c}} 
\left\lbrace\begin{array}{cl}
\left(\frac{\gamma}{\Gamma_\mathrm{c}}\right)^{-2} & \mathrm{if}\ \Gamma_\mathrm{c}<\gamma < \Gamma_\mathrm{m}\ ,\\
\left(\frac{\Gamma_\mathrm{m}}{\Gamma_\mathrm{c}}\right)^{-2}
\left(\frac{\gamma}{\Gamma_\mathrm{m}}\right)^{-(p+1)} & \mathrm{if}\ \gamma > \Gamma_\mathrm{m}\ .
\end{array}\right.
& \!\!\left(\Gamma_\mathrm{m}>\Gamma_\mathrm{c}\right),\\
\frac{p-1}{\Gamma_\mathrm{m}}
\left\lbrace\begin{array}{cl}
\left(\frac{\gamma}{\Gamma_\mathrm{m}}\right)^{-p} & \mathrm{if}\ \Gamma_\mathrm{m}<\gamma < \Gamma_\mathrm{c}\ ,\\
\left(\frac{\Gamma_\mathrm{c}}{\Gamma_\mathrm{m}}\right)^{-p}
\left(\frac{\gamma}{\Gamma_\mathrm{c}}\right)^{-(p+1)} & \mathrm{if}\ \gamma > \Gamma_\mathrm{c}\ .
\end{array}\right.
 & \!\!\left(\Gamma_\mathrm{m}<\Gamma_\mathrm{c}\right).\\
\end{array}
\right.\nonumber\\
\label{eq:nbarge}
\end{eqnarray}
An accurate approximation of
the corresponding photon spectrum $\epsilon^{2}\left.n_{\epsilon}\right|_\mathrm{syn}$ is given by a
broken powerlaw shape \citep{sari:98}:
\begin{equation}
\frac{\left.\epsilon^{2} n_{\epsilon}\right|_\mathrm{syn}}{n_\mathrm{e}^\mathrm{acc}\Gamma_\mathrm{m} m_\mathrm{e}c^{2}} \simeq 
\left\lbrace
\begin{array}{lll}
\left(\frac{\epsilon_\mathrm{c}}{\epsilon_\mathrm{m}}\right)^{1/2} &
 \left(\frac{\epsilon}{\epsilon_\mathrm{c}}\right)^{4/3} & \mathrm{if}\
 \epsilon < \epsilon_\mathrm{c}\ ,\\
& \left(\frac{\epsilon}{\epsilon_\mathrm{m}}\right)^{1/2} & \mathrm{if}\
 \epsilon_\mathrm{c} < \epsilon < \epsilon_\mathrm{m}\ ,\\
& \left(\frac{\epsilon}{\epsilon_\mathrm{m}}\right)^{-\frac{p-2}{2}} &
 \mathrm{if}\ \epsilon > \epsilon_\mathrm{m}\ ,
\end{array}
\right.
\label{eq:synfast}
\end{equation}
in the synchrotron fast cooling regime ($\Gamma_\mathrm{m} >
\Gamma_\mathrm{c}$) and 
\begin{equation}
\frac{\left.\epsilon^{2} n_{\epsilon}\right|_\mathrm{syn}}{n_\mathrm{e}^\mathrm{acc}\Gamma_\mathrm{m} m_\mathrm{e}c^{2}} \simeq 
\left(\frac{\Gamma_\mathrm{m}}{\Gamma_\mathrm{c}}\right)^{p-2} 
\left\lbrace
\begin{array}{lll}
\left(\frac{\epsilon_\mathrm{m}}{\epsilon_\mathrm{c}}\right)^{\frac{3-p}{2}} & \left(\frac{\epsilon}{\epsilon_\mathrm{m}}\right)^{4/3} & \mathrm{if}\ \epsilon < \epsilon_\mathrm{m}\ ,\\
& \left(\frac{\epsilon}{\epsilon_\mathrm{c}}\right)^{\frac{3-p}{2}} & \mathrm{if}\ \epsilon_\mathrm{m} < \epsilon < \epsilon_\mathrm{c}\ ,\\
& \left(\frac{\epsilon}{\epsilon_\mathrm{c}}\right)^{-\frac{p-2}{2}} & \mathrm{if}\ \epsilon > \epsilon_\mathrm{c}\ ,
\end{array}
\right.
\label{eq:synslow}
\end{equation}
in the synchrotron slow cooling regime ($\Gamma_\mathrm{m} < \Gamma_\mathrm{c}$).
In these expressions,  $\epsilon_\mathrm{m}$ (resp. $\epsilon_\mathrm{c}$) is the
synchrotron frequency of electrons at $\Gamma_\mathrm{m}$
(resp. $\Gamma_\mathrm{c}$). It appears clearly that the first case
(synchrotron fast cooling regime) is efficient as most of the initial
energy deposited in relativistic electrons is radiated. In the second
case (synchrotron slow cooling regime), only a small fraction of the
electron energy is 
radiated. The peak of the
synchrotron emission in $\epsilon^{2}n_{\epsilon}$ (equivalent to $\nu
F_{\nu}$) is located at frequency
$\epsilon_\mathrm{p,syn}=\max{\left\lbrace\epsilon_\mathrm{m};\epsilon_\mathrm{c}\right\rbrace}$ for $2<p<3$
($\epsilon_\mathrm{m}$ if $p>3$), i.e. from
\refeq{eq:nusyn} at energy
\begin{eqnarray}
h \nu'_\mathrm{p,syn}
& \simeq &
0.17\ \mathrm{eV}\
\left(\frac{B'}{1000\ \mathrm{G}}\right) \left(\frac{\Gamma_\mathrm{p}}{100}\right)^{2}\nonumber\\
\mathrm{with}\ \Gamma_\mathrm{p} & = & \left\lbrace\begin{array}{cl}
\max{\left\lbrace\Gamma_\mathrm{m};\Gamma_\mathrm{c}\right\rbrace} & \mathrm{if}\ 2<p<3\\
\Gamma_\mathrm{m}                                      & \mathrm{if}\ p>3
\end{array}\right. \ .
\label{eq:epsyn}
\end{eqnarray}
Our estimate of the maximum Lorentz factor of relativistic electrons
(\refeq{eq:gmax}) leads to a cutoff in the synchrotron component at frequency
\begin{equation}
h\nu'_\mathrm{M,syn} \simeq \min{\left\lbrace230\ \mathrm{MeV}\ ;\ 5300\ \mathrm{TeV}\ \left(\frac{B'}{1000\ \mathrm{G}}\right)^{3}\left(\frac{t'_\mathrm{ex}}{1\ \mathrm{s}}\right)^{2} \right\rbrace}
\ .
\end{equation}
Except for very low magnetic fields and very short dynamical timescales,
it is always the first limit, when the acceleration timescale becomes
of the order of the synchrotron timescale, that dominates.\\

\begin{figure*}[t!]
\begin{center}
\includegraphics*[width=0.3694\textwidth,viewport=0cm 0cm 20.2cm
 20.2cm]{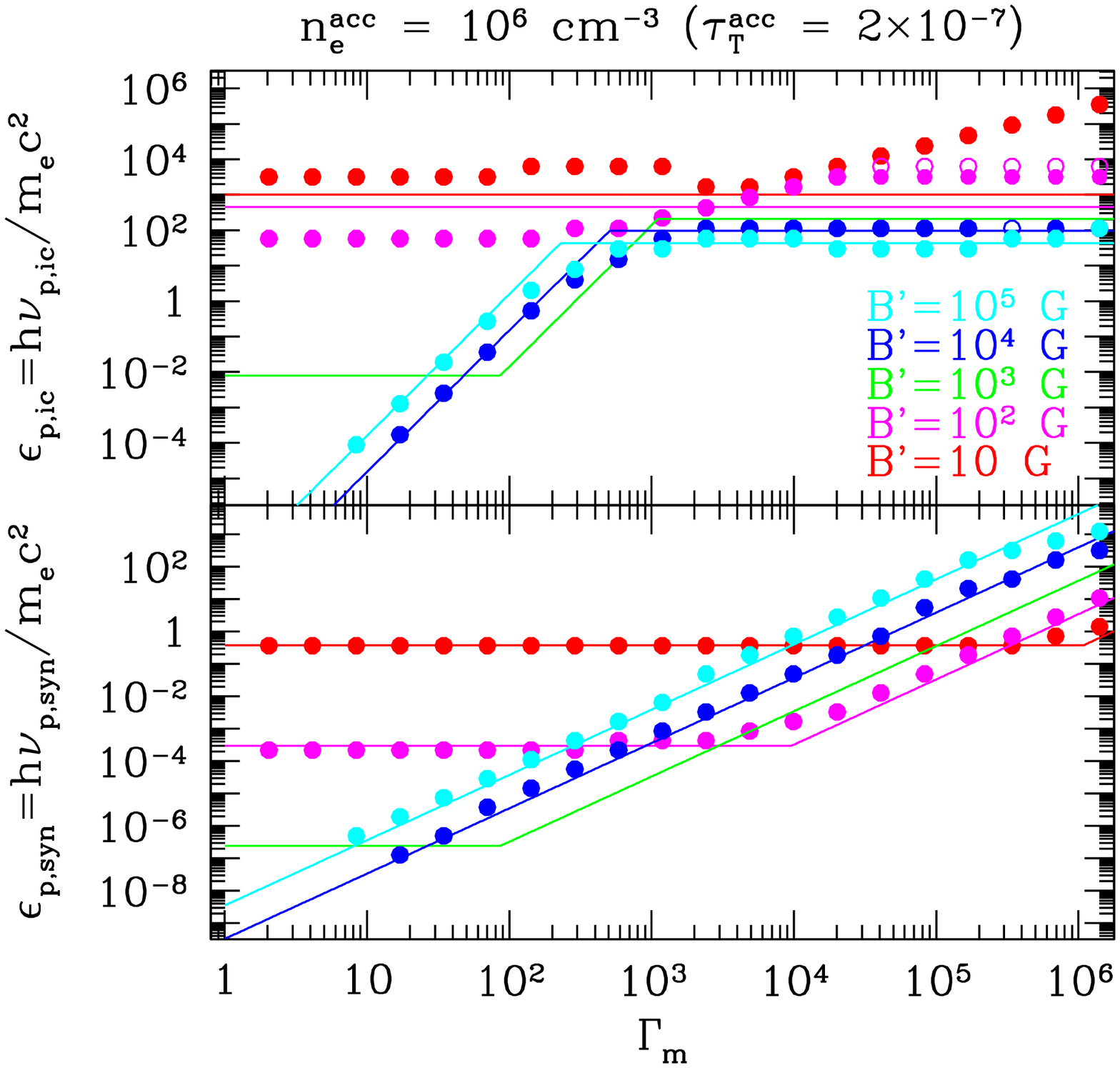}
\includegraphics*[width=0.3054\textwidth,viewport=3.5cm 0cm 20.2cm
 20.2cm]{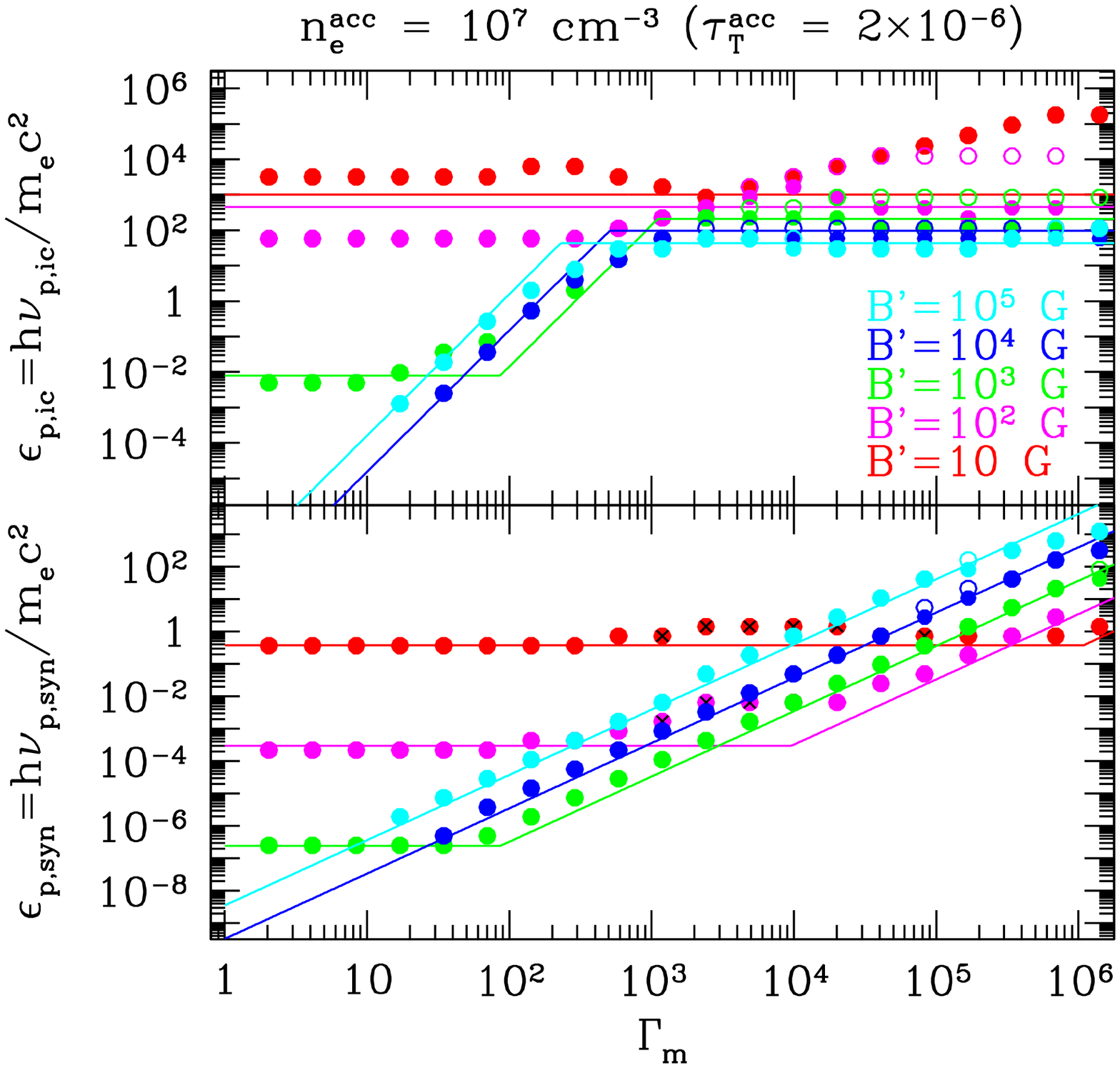}
\includegraphics*[width=0.3054\textwidth,viewport=3.5cm 0cm 20.2cm 20.2cm]{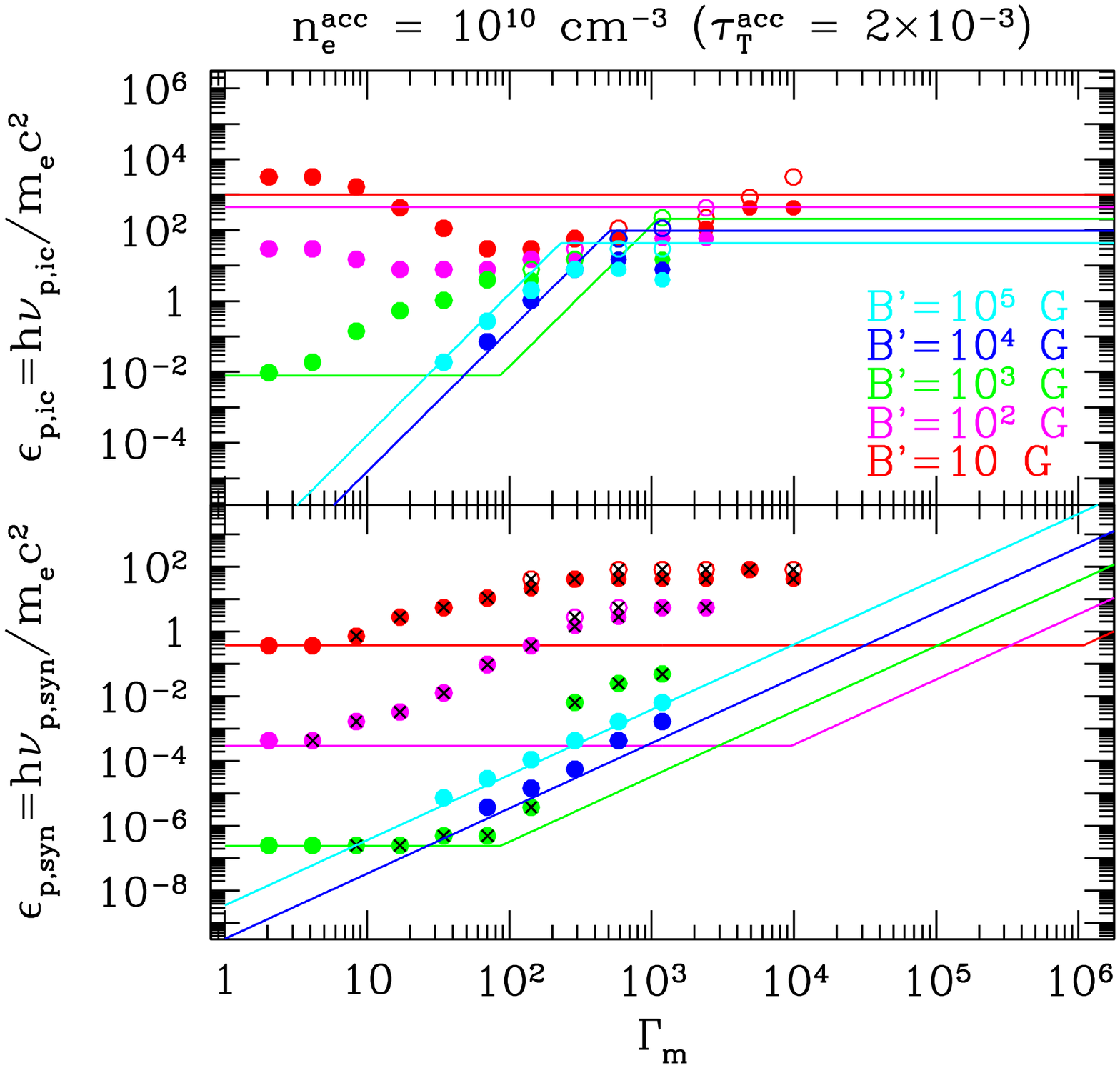}
\end{center}
\caption{\textbf{Synchrotron and inverse Compton peaks: comparison with
 analytical estimates.} The photon peak energy of the synchrotron
 (bottom) and inverse Compton (top) components is plotted as a function
 of the initial minimum Lorentz factor of the accelerated electrons
 $\Gamma_\mathrm{m}$, for five different values of the magnetic field
 (color coded) and increasing values of the accelerated electron density
 $n_\mathrm{e}^\mathrm{acc}$ in the three panels. The dynamical
 timescale is fixed to $t'_\mathrm{ex}=10\ \mathrm{s}$. Values
 obtained from our radiative code are indicated with open (resp. filled) circles
 corresponding to calculations without (resp. with) $\gamma\gamma$
 annihilation. In the bottom panels, when the circle is filled with a black
 cross, the inverse Compton process is dominant compared to the
 synchrotron radiation. These results are compared to the analytical
 estimates (solid lines) discussed in the text: \refeq{eq:epsyn} for
 the synchrotron component and Eqs.~(\ref{eq:epic}) renormalized by a factor
 $10$,~(\ref{eq:ekn}) renormalized by a factor $0.1$ and~(\ref{eq:emic})
 for the inverse Compton component. Note that, in the bottom panels, the transition
 between the horizontal and rising branches of the analytical estimate
 for $\epsilon_\mathrm{p,syn}$ corresponds to the
 transition between the synchrotron slow and fast cooling regimes.}
\label{fig:estim}
\end{figure*}

\begin{figure*}
\begin{center}
\includegraphics*[width=0.3694\textwidth,viewport=0cm 0cm 20.2cm
 20.2cm]{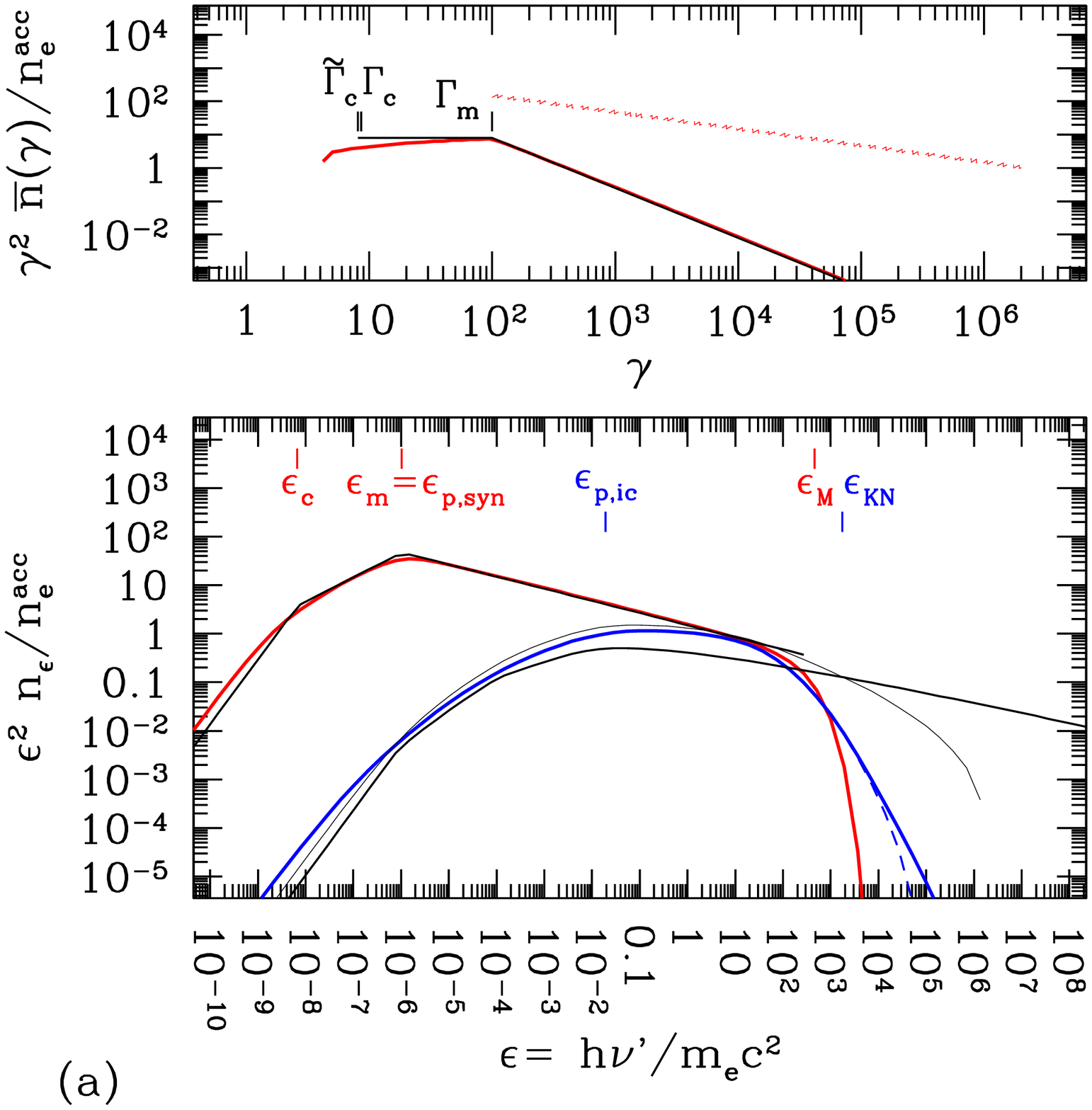}
\includegraphics*[width=0.3054\textwidth,viewport=3.5cm 0cm 20.2cm
 20.2cm]{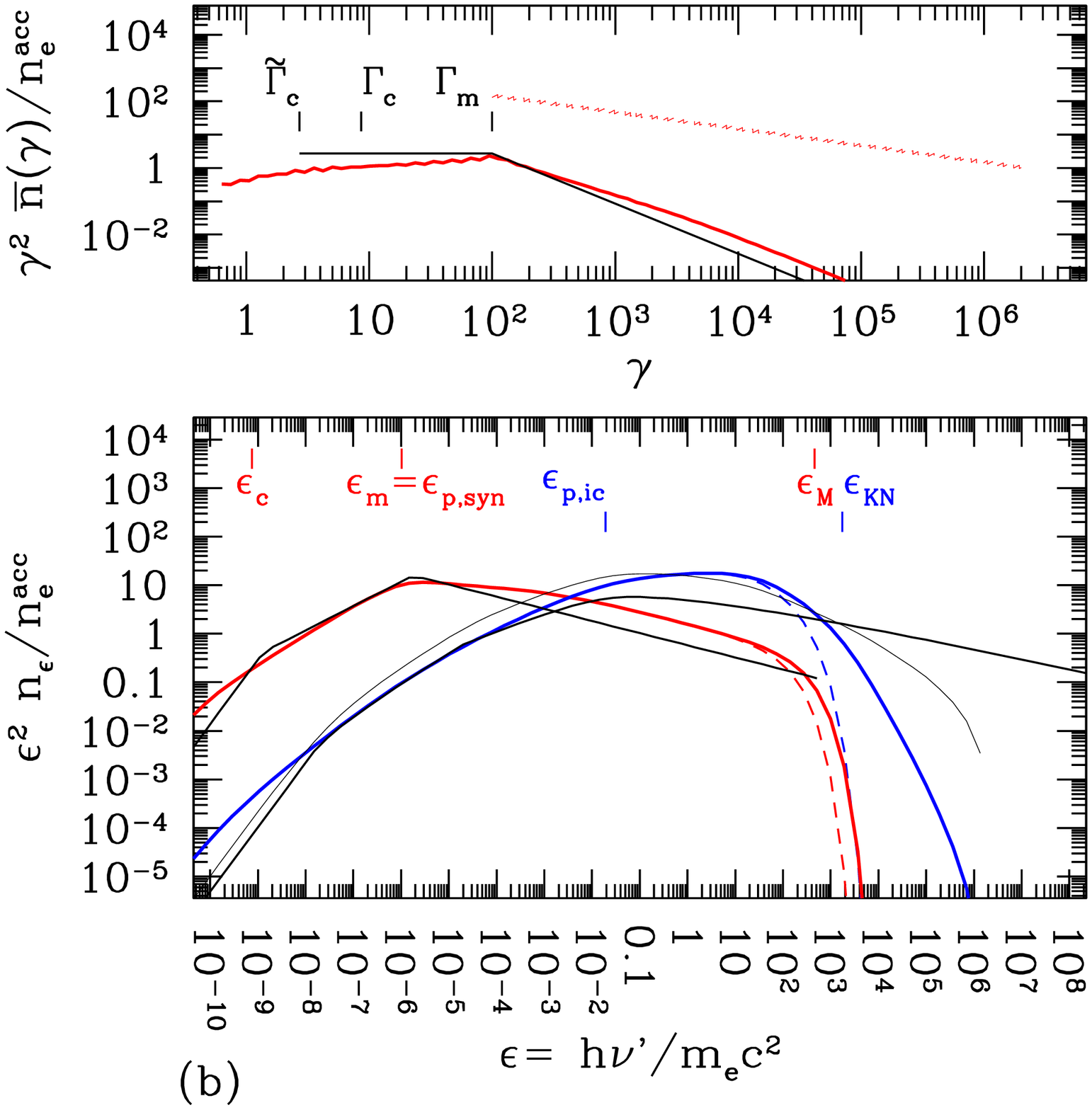}
\includegraphics*[width=0.3054\textwidth,viewport=3.5cm 0cm 20.2cm
 20.2cm]{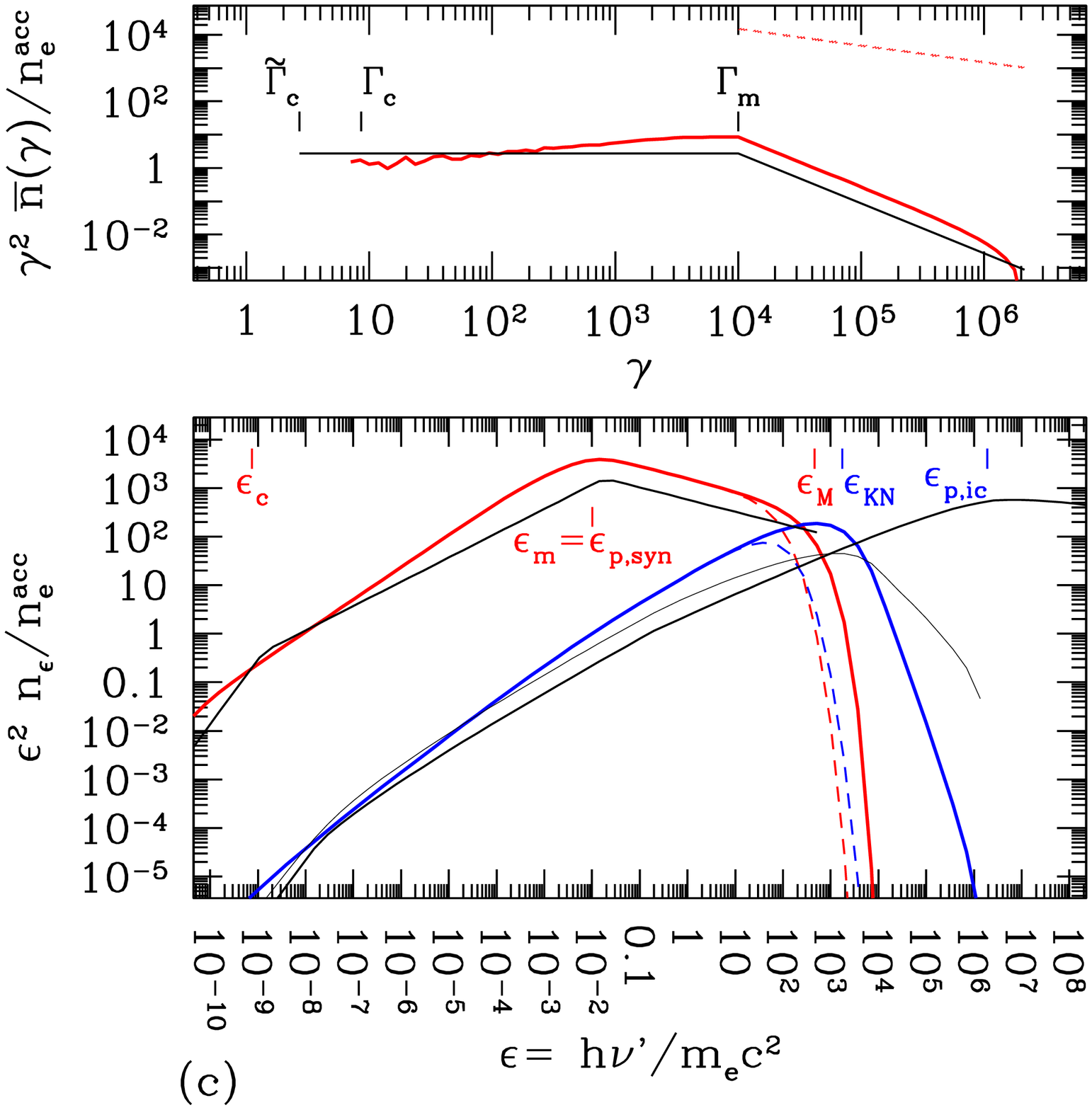}
\end{center}
\caption{\textbf{Synchrotron and Inverse Compton spectral components: comparison
 with analytical estimates.} The time-averaged electron distribution and
 the final photon spectrum are shown for 3 different cases: (a)
 $\Gamma_\mathrm{m}=10^{2}$ and $n_\mathrm{e}^\mathrm{acc}=10^{8}\ \mathrm{cm^{-3}}$;
 (b)  $\Gamma_\mathrm{m}=10^{2}$ and $n_\mathrm{e}^\mathrm{acc}=10^{10}\ \mathrm{cm^{-3}}$; 
(c)  $\Gamma_\mathrm{m}=10^{4}$ and $n_\mathrm{e}^\mathrm{acc}=10^{8}\ \mathrm{cm^{-3}}$.
In all cases, $p=2.5$, $B'=3000\ \mathrm{G}$ and $t'_\mathrm{ex}=10\ \mathrm{s}$. Synchrotron self-absorption and $\gamma\gamma$ annihilation are not included in the calculation. The time-averaged
 electron distribution obtained with our radiative code is plotted in
 the top panel with a red line and
 compared to the analytical prediction in black. 
The characteristic
 electron Lorentz factors discussed in the text are indicated with vertical bars. The initial
 distribution is also plotted with a dotted red line. The
 synchrotron (resp. inverse Compton) component of the final photon
 spectrum is plotted in the bottom panel with a red (resp. blue) line and is
 compared to the analytical estimate in black 
(thick line:
 analytical estimate by \citet{sari:01}; thin line: semi-analytical
 estimate described in the text, which includes Klein-Nishina corrections). For indication, the
 result of the numerical calculation including $\gamma\gamma$
 annihilation is also shown in dashed red and blue lines. The
 characteristic photon frequencies discussed in the text are indicated
 with vertical bars.
Klein-Nishina corrections are negligible (except at high energy) in case (a) and (b). Synchrotron radiation is dominant in cases (a) and (c),
 whereas inverse Compton scatterings dominate in case (b).}
\label{fig:testic}
\end{figure*}

As shown in \reffig{fig:estim} (bottom panels) and
\reffig{fig:testic} (case (a)), the numerical results of our radiative code when synchrotron
radiation is dominant are in an excellent
agreement with these analytical estimates, showing the good accuracy of
the synchrotron spectrum described by \citet{sari:98}.\\

The timescale associated with the synchrotron self-absorption at
frequency $\nu'$ can be estimated by
\begin{equation}
t'_\mathrm{a}\left(\nu'\right) \simeq \frac{8\pi m_\mathrm{e}\left.\nu'\right.^{2}}{c}
\left(\int\frac{d\gamma}{\gamma}\bar{n}\left(\gamma\right)P^\mathrm{syn}_{\nu'}(\gamma)\left[2-\frac{d\ln{\bar{n}}}{d\ln{\gamma}}\right]\right)^{-1}\ .
\end{equation}
From \refeq{eq:nbarge}, one gets in the synchrotron fast cooling regime
\begin{equation}
\frac{t'_\mathrm{a}\left(\nu'\right)}{t'_\mathrm{ex}} \simeq \frac{4\pi \left.\nu'_\mathrm{c}\right.^{3}}{n_\mathrm{e}^\mathrm{acc} c^{3}}
\left\lbrace\begin{array}{cl}
\left(\nu'/\nu'_\mathrm{c}\right)^{5/3} & \mathrm{if}\ \nu'<\nu'_\mathrm{c}\ ,\\
\left(\nu'/\nu'_\mathrm{c}\right)^{3}   & \mathrm{if}\ \nu'_\mathrm{c}<\nu'<\nu'_\mathrm{m}\ ,\\
\left(\nu'_\mathrm{m}/\nu'_\mathrm{c}\right)^{3}
\left(\nu'/\nu'_\mathrm{m}\right)^{\frac{p+5}{2}}   & \mathrm{if}\ \nu'>\nu'_\mathrm{m}\ ,\\
\end{array}\right.
\end{equation}
and in the synchrotron slow cooling regime
\begin{eqnarray}
\frac{t'_\mathrm{a}\left(\nu'\right)}{t'_\mathrm{ex}} & \simeq &
\frac{4\pi \left.\nu'_\mathrm{m}\right.^{3}}{n_\mathrm{e}^\mathrm{acc} c^{3}}
\frac{\Gamma_\mathrm{c}}{\Gamma_\mathrm{m}}
\left\lbrace\begin{array}{cl}
\left(\nu'/\nu'_\mathrm{m}\right)^{5/3}     & \mathrm{if}\ \nu'<\nu'_\mathrm{m}\ ,\\
\left(\nu'/\nu'_\mathrm{m}\right)^{\frac{p+4}{2}} & \mathrm{if}\ \nu'_\mathrm{m}<\nu'<\nu'_\mathrm{c}\ ,\\
\left(\nu'_\mathrm{c}/\nu'_\mathrm{m}\right)^{\frac{p+4}{2}}
\left(\nu'/\nu'_\mathrm{c}\right)^{\frac{p+5}{2}}   & \mathrm{if}\ \nu'>\nu'_\mathrm{c}\ .\\
\end{array}\right.\nonumber\\
\end{eqnarray}
At high frequency, this timescale is very long and the synchrotron
self-absorption process is negligible. It will only affect the spectrum below
frequency $\nu'_\mathrm{a}$ for which
$t'_\mathrm{a}(\nu'_\mathrm{a})\simeq t'_\mathrm{ex}$. One can 
show that below $\nu'_\mathrm{a}$, the predicted slope of the absorbed
spectrum is $\epsilon^{2}n_{\epsilon}\propto \epsilon^{3}$ if $\nu'_\mathrm{a}<\nu'_\mathrm{c}$,
and $\epsilon^{2}n_{\epsilon}\propto \epsilon^{3.5}$ otherwise.
We checked with our
numerical code that the accuracy of these expressions is quite good as
long as the inverse Compton cooling is not dominant (otherwise the
true electron distribution differs from \refeq{eq:nbarge} used by
\citet{sari:98}, see below).

\paragraph{Inverse Compton component.}
If most scatterings between relativistic
electrons and synchrotron photons occur
in Thomson regime, the peak of the inverse
Compton component is expected at $\epsilon_\mathrm{ic,p}\simeq \Gamma_\mathrm{p}^{2}\epsilon_\mathrm{p,syn}$,
i.e.
\begin{equation}
h \nu'_\mathrm{p,ic}
\simeq 
1.7\ \mathrm{keV}\
\left(\frac{B'}{1000\ \mathrm{G}}\right) \left(\frac{\Gamma_\mathrm{p}}{100}\right)^{4}\ .
\label{eq:epic}
\end{equation}
The Thomson approximation is valid as long as $\Gamma_\mathrm{p}
\epsilon_\mathrm{p,syn} \ll 1$, i.e.
\begin{equation}
\Gamma_\mathrm{p} \ll \Gamma_\mathrm{KN} \simeq 3.0\times 10^{3}\ \left(\frac{B'}{1000\ \mathrm{G}}\right)^{-1/3}\ ,
\label{eq:gkn}
\end{equation}
which corresponds to
\begin{equation}
h\nu'_\mathrm{p,ic} \ll h\nu'_\mathrm{KN}\simeq \Gamma_\mathrm{KN}m_\mathrm{e}c^{2}\simeq 1.3\ \mathrm{GeV}\ \left(\frac{B'}{1000\ \mathrm{G}}\right)^{-1/3}\ .
\label{eq:ekn}
\end{equation}
A severe reduction of the high-energy spectrum should
always be expected above $\nu'_\mathrm{KN}$ in the comoving
frame due to Klein-Nishina corrections to the inverse Compton
cross-section. Even more, the maximum Lorentz factor of the relativistic
electrons given by \refeq{eq:gmax} leads to an absolute maximum energy
$\sim \Gamma_\mathrm{M} m_\mathrm{e}c^{2}$ for a scattered photon, i.e.
\begin{eqnarray}
h\nu'_\mathrm{M,ic} &  \simeq &
\min{\left\lbrace 1.9\ \mathrm{TeV} \left(\frac{B'}{1000\ \mathrm{G}}\right)^{-1/2}
\!\!\!\!\!\!\!
;\ 9000\ \mathrm{TeV} \left(\frac{B'}{1000\ \mathrm{G}}\right)\left(\frac{t'_\mathrm{ex}}{1\ \mathrm{s}}\right)\right\rbrace} \ .\nonumber\\
\label{eq:emic}
\end{eqnarray}
Again, except for very weak magnetic fields and very short dynamical
timescales, the maximum inverse Compton frequency is always given by the
first limit (acceleration limitation due to radiative losses). From these estimates, one can deduce that
the peak of the inverse Compton component should be found in all cases
at the frequency $\nu'\simeq \min{\left\lbrace\nu'_\mathrm{p,ic};\
\nu'_\mathrm{KN};\ \nu'_\mathrm{M,ic}\right\rbrace}$. \reffig{fig:estim}
(top panels) show that numerical results of our radiative code are in
a reasonable agreement with these theoretical predictions, as long as
inverse Compton scatterings are not the dominant cooling process for
electrons. The scaling in equation~(\ref{eq:epic}) appears to be correct
but the
normalization seems to be underestimated by a factor $\sim 10$. When
Klein-Nishina corrections become important, the peak of the inverse
Compton component appears typically  at frequency 
$\sim 0.1 \nu'_\mathrm{KN}$ compared to \refeq{eq:ekn}. This is not
surprising since the cross section for scatterings of a photon with frequency $\nu'$ by an
electron with Lorentz factor $\gamma$ shows non negligible Klein-Nishina
deviations well below the limit $h\nu'=\gamma m_\mathrm{e}c^{2}$. 
Significant deviations of one
order of magnitude from the simple estimate $0.1\nu'_\mathrm{KN}$ can be observed 
(see the cases with a weak magnetic field in \reffig{fig:estim}).
\\

When the Thomson regime is valid, the ratio of the
inverse Compton over the synchrotron power is given by the Compton
parameter, defined as the ratio of the energy density in photons
over the magnetic energy density,
$Y=u_\mathrm{rad}/\left(\left.B'\right.^{2}/8\pi\right)$. 
This quantity is time-dependent. However, when not stated
otherwise, $Y$ stands in this paper for the final value of the Compton parameter at $t'_\mathrm{ex}$.
As long as $Y<1$,
synchrotron losses dominate, the seed photons for inverse Compton
scatterings have the spectrum which is given above by
Eqs.~(\ref{eq:synfast}) and ~(\ref{eq:synslow}), and the
distribution of the electrons responsible for the scatterings is close
to the broken-power law distribution $\bar{n}(\gamma)$ described by
\citet{sari:98}. The corresponding spectral shape of the inverse Compton component
has been derived by \citet{sari:01} and is given in their appendix
A. It is based on the integration of the approximate relation 
\begin{equation}
\left.n_{\nu'}\right|_\mathrm{ic} \simeq  t'_\mathrm{ex}\int d\gamma\ \bar{n}(\gamma) \frac{\bar{P}^\mathrm{\,ic}_{\nu'}(\gamma)}{h\nu'}\ ,
\label{eq:ic}
\end{equation}
where $\bar{n}(\gamma)$ is the time-averaged electron distribution
predicted by \refeq{eq:nbarge} and $\bar{P}^\mathrm{\,ic}_{\nu'}(\gamma)$ is the
inverse Compton power radiated at frequency $\nu'$ by an electron with Lorentz
factor $\gamma$, computed assuming a seed photon distribution equal to the
standard, time-averaged, synchrotron spectrum given by \citet{sari:98}. 
Instead of the complete expression of $P^\mathrm{ic}_{\nu'}(\gamma)$
(see appendix~\ref{sec:radproc}), the authors use a simplified kernel
(which is equivalent to assume Thomson regime everywhere) so that the
integration can be made analytically. In the present study, both the
electron and photon distributions are time-dependant, which leads to
significant differences in the high-energy spectrum compared to the
time-averaged approach. This will be further discussed later on.\\

The intensity of the inverse Compton component is 
\begin{equation}
\left.\epsilon^{2}n_{\epsilon}\right|_\mathrm{ic} \simeq Y \left.\epsilon^{2}n_{\epsilon}\right|_\mathrm{syn}\ .
\end{equation}
The Compton parameter in this case equals\footnote{These expressions
assume that the maximum electron Lorentz factor 
$\Gamma_\mathrm{M}$ is greater than $\Gamma_\mathrm{c}$, which is always the case in the
fast cooling regime. On the other hand, a ``very slow'' cooling regime
is possible when
$\Gamma_\mathrm{m}<\Gamma_\mathrm{M}<\Gamma_\mathrm{c}$. In this case
the break at $\nu'_\mathrm{c}$ in the synchrotron spectrum is suppressed
as it is above the cutoff at $\nu'_\mathrm{M,syn}$. The Compton
parameter in this case equals 
$Y\simeq \frac{\epsilon_\mathrm{e}}{\epsilon_\mathrm{B}}\left(\frac{\Gamma_\mathrm{m}}{\Gamma_\mathrm{c}}\right)\left(\frac{\Gamma_\mathrm{m}}{\Gamma_\mathrm{M}}\right)^{-(3-p)}$
if $2<p<3$ and $Y\simeq
\frac{\epsilon_\mathrm{e}}{\epsilon_\mathrm{B}}\left(\frac{\Gamma_\mathrm{m}}{\Gamma_\mathrm{c}}\right)$
if $p>3$.
}
\begin{eqnarray}
Y & \simeq & 
\frac{4}{3}\frac{p-1}{p-2}\tau_\mathrm{T}^\mathrm{acc}\Gamma_\mathrm{m}\Gamma_\mathrm{c}
\times
\left\lbrace\begin{array}{cl}
1 & \mathrm{if}\ \Gamma_\mathrm{m} > \Gamma_\mathrm{c}\ ,\\
\left(\frac{\Gamma_\mathrm{m}}{\Gamma_\mathrm{c}}\right)^{p-2} & \mathrm{if}\ \Gamma_\mathrm{m} < \Gamma_\mathrm{c}\ \mathrm{and}\ 2<p<3\ , \\
{\Gamma_\mathrm{m}}/{\Gamma_\mathrm{c}} & \mathrm{if}\ \Gamma_\mathrm{m} < \Gamma_\mathrm{c}\ \mathrm{and}\ p>3\ . \\
\end{array}\right.\nonumber\\
\label{eq:Y}
\end{eqnarray}
Note that the term 
$\frac{4}{3}\frac{p-1}{p-2}\tau_\mathrm{T}^\mathrm{acc}\Gamma_\mathrm{m}\Gamma_\mathrm{c}$
is simply equal to $u_\mathrm{e}^\mathrm{acc}/u_\mathrm{B}$,
i.e. $\epsilon_\mathrm{e}/\epsilon_\mathrm{B}$ using the standard
parameterization of the microphysics.
In synchrotron fast cooling regime, the value of $Y$ can be understood
as follows: 
due to a short synchrotron
timescale, the region populated with
relativistic electrons has a typical size $\sim
c\,t'_\mathrm{syn}\left(\Gamma_\mathrm{m}\right)$. 
The effective Thomson
optical depth for relativistic electrons is $\sim \sigma_\mathrm{T}
n_\mathrm{e}^\mathrm{acc} c
t'_\mathrm{syn}\left(\Gamma_\mathrm{m}\right)$. In Thomson regime,
the frequency of a photon scattered by an electron with Lorentz factor
$\gamma$ is multiplied by $\gamma^{2}$. Therefore, the Compton parameter $Y$
is approximatively given by 
\begin{equation}
Y \simeq \frac{4}{3} \frac{p-1}{p-2}\left(\sigma_\mathrm{T} n_\mathrm{e}^\mathrm{acc} c t'_\mathrm{syn}\left(\Gamma_\mathrm{m}\right)\right) \Gamma_\mathrm{m}^{2}\ 
\end{equation}
which leads to the expression given above.
On the other hand, in synchrotron slow cooling regime,
the size of the region populated by relativistic electrons is
 now given by $\sim c t'_\mathrm{ex}$ and therefore the Thomson optical
 depth for relativistic electron becomes much larger $\sim 
 \sigma_\mathrm{T} n_\mathrm{e}^\mathrm{acc} c t'_\mathrm{ex}$, which
 leads to \refeq{eq:Y} for $p>3$. In the case where $2<p<3$,
 this equation is corrected by a factor
 $\left(\Gamma_\mathrm{c}/\Gamma_\mathrm{m}\right)^{3-p}>1$ as the
 typical Lorentz factor of scattering electrons is at an intermediate
 value between $\Gamma_\mathrm{m}$ and $\Gamma_\mathrm{c}$.\\

As shown in case (a) of \reffig{fig:testic}, except at very high
energies (above $h\nu'_\mathrm{KN}$),
our numerical calculations are in very good agreement with the
approximate spectrum given by \citet{sari:01}, as long as
$Y<1$. Above $\nu'_\mathrm{KN}$, a better estimate is obtained when integrating numerically
\refeq{eq:ic} with the same assumptions used by \citet{sari:01}
(time-averaged electron and seed photon distributions) but using a
complete kernel that includes Klein-Nishina corrections in the inverse
Compton power (thin black line in \reffig{fig:testic}). However, even
using a more accurate cross section, \refeq{eq:ic} always overpredicts
the inverse Compton emission at high energy. This systematic difference
appears because
the high energy photons in the inverse Compton
component are due to the scatterings of photons at $\nu'\sim
\nu'_\mathrm{p,syn}$ with high Lorentz factor electrons at
$\gamma>\max{\left\lbrace\Gamma_\mathrm{m}; \Gamma_\mathrm{c}\right\rbrace}$. In fast cooling regime, these two species
are not present at the same time in the shocked region, as the duration
necessary to form the synchrotron spectrum at $\nu'_\mathrm{p,syn}$ is
also the duration necessary to cool electrons above $\Gamma_\mathrm{m}$,
i.e. the synchrotron timescale
$t'_\mathrm{syn}\left(\Gamma_\mathrm{m}\right)$. As
\refeq{eq:ic} is based on a time-averaged approach, it cannot take into
account such effects, related to the way the radiation field is
built, and therefore it overestimates the spectrum above $\nu'_\mathrm{p,ic}$.
We checked that in synchrotron slow cooling regime the agreement is better
above $\nu'_\mathrm{p,ic}$ than what is observed in
\reffig{fig:testic}, but \refeq{eq:ic} is still overestimating the high-energy
component when evaluating the scatterings by fast cooling electrons
(i.e. electrons with $\gamma > \Gamma_\mathrm{c}$).\\

When $Y>1$, inverse Compton losses become dominant. Then, the effective
radiative timescale is shorter than the synchrotron timescale (by a
factor $\sim (1+Y)$), the effective critical Lorentz factor is reduced
($\tilde{\Gamma}_\mathrm{c}\sim \Gamma_\mathrm{c}/(1+Y)$) and the
corresponding frequency $\epsilon_\mathrm{c}$ in the synchrotron
spectrum (Eqs.~(\ref{eq:synfast}) and~(\ref{eq:synslow})) as
well. The intensity of the synchrotron component is reduced by a factor
$1/(1+Y)$. These corrections are however very approximate and valid only
in Thomson regime. Our tests
show that when inverse Compton scatterings become dominant, the modified cooling rate of
electrons affects the time-averaged distribution $\bar{n}(\gamma)$ (which
differs from the standard broken-law distribution given by
\refeq{eq:nbarge}), and therefore the distribution of seed synchrotron
photons becomes different from the
standard synchrotron spectrum given by Eqs.~(\ref{eq:synfast}) and~(\ref{eq:synslow}).
This is well seen in cases (b) and (c) in \reffig{fig:testic}. In
fact, in this case, the approach used by \citet{sari:01} is not appropriate
because the spectrum of the seed photons cannot be predicted by an \textit{a priori}
 calculation including the synchrotron process only: the resulting
spectrum has not enough time to be built when inverse Compton losses are
included.
This effect becomes stronger when Klein-Nishina corrections are
important, as the ratio of the inverse Compton to the synchrotron power
becomes highly dependant on the electron Lorentz factor.
As seen in \reffig{fig:testic}, the low-energy slope of the
synchrotron spectrum is steeper in that case. Such a behavior is in
agreement with the theoretical predictions made by \cite{derishev:01}. We plan to investigate in
a forthcoming paper if this could reconcile the synchrotron radiation
with the observed distribution of the low-energy photon index in {BATSE}
bursts \citep{preece:00}, which differs from the simplest prediction of
the fast cooling synchrotron spectrum \citep{ghisellini:00} as its mean value is close to  $\alpha\sim
-1$.

\paragraph{Formation of the radiation field.} 
\label{sec:Yex}
These results show that the high energy component of the photon spectrum cannot be
estimated accurately without understanding  
how the radiation
field (seed photons for inverse Compton scatterings) is formed. 
Initially, no photons are present and synchrotron radiation is always
dominant ($Y(t'=0)=0$). The Compton parameter is an increasing
function of time, due to the progressive building of the radiation field
(see \reffig{fig:testY}).
When synchrotron radiation is the
dominant process, the radiation field increases up to
$t'\sim t'_\mathrm{syn}\left(\Gamma_\mathrm{m}\right)= t'_\mathrm{ex} {\Gamma_\mathrm{c}}/{\Gamma_\mathrm{m}}$ in fast
cooling regime, and then it saturates. In slow cooling regime, it
increases up to $t'\simeq t'_\mathrm{ex}$. In both cases, 
the time evolution of the Compton parameter,
$Y(t')$, can be evaluated analytically (as shown in
appendix~\ref{sec:Y}) and its asymptotic value $Y_\mathrm{ex}=Y\left(t'_\mathrm{ex}\right)$
is given by \refeq{eq:Y}, as long as most scatterings occur in
Thomson regime. In fast cooling regime, a necessary condition to
have a dominant inverse Compton component in the final spectrum is
therefore
$u_\mathrm{e}^\mathrm{acc}/u_\mathrm{B}=\epsilon_\mathrm{e}/\epsilon_\mathrm{B}>1$.
It is however not a sufficient condition, as the intensity of the
inverse Compton component can be attenuated by Klein-Nishina effects,
and also by $\gamma\gamma$ annihilation.\\

When $Y_\mathrm{ex}>1$, the impact of inverse Compton scatterings on the
electron distribution will depend on the time $t'_\mathrm{ic}$ where
$Y(t'_\mathrm{ic})\sim 1$, i.e. the time when inverse Compton
scatterings become the dominant process of cooling. Indeed, only the
distribution of electrons below $\Gamma_\mathrm{c}(t'_\mathrm{ic})$ can
be affected by the new dominant cooling process, as electrons at higher
Lorentz factor have already cooled by synchrotron radiation. Here, the
Lorentz factor $\Gamma_\mathrm{c}(t')$ is defined as the Lorentz factor
giving a synchrotron timescale of the order of $t'$,
i.e. $t'_\mathrm{syn}\left(\Gamma_\mathrm{c}(t')\right)\simeq t'$. With
this definition
$\Gamma_\mathrm{c}=\Gamma_\mathrm{c}(t'_\mathrm{ex})$. In the
synchrotron fast cooling case, the synchrotron spectrum around the peak
at $\nu'_\mathrm{m}$ will be
affected by inverse Compton scatterings if this process becomes dominant
at very early times, i.e if $\Gamma_\mathrm{c}(t'_\mathrm{ic}) >
\Gamma_\mathrm{m}$, which is equivalent to $t'_\mathrm{ic} <
t'_\mathrm{syn}\left(\Gamma_\mathrm{m}\right)=t'_\mathrm{ex} \Gamma_\mathrm{c}/\Gamma_\mathrm{m}$.\\

When inverse Compton scatterings are extremely efficient, they can
represent
 the dominant electron cooling process, even at early times. When this happens,
 the maximum Lorentz factor of accelerated electrons 
$\Gamma_\mathrm{M}$ is overestimated in \refeq{eq:gmax}. From the
evolution of $Y(t')$ discussed in appendix~\ref{sec:Y}, one can deduce
the value reached by the Compton parameter when electrons at
$\Gamma_\mathrm{M}$ have cooled, i.e.
\begin{eqnarray}
Y_\mathrm{M} & = & Y\left(t'_\mathrm{syn}\left(\Gamma_\mathrm{M}\right)\right) \simeq 
\frac{4}{3}\frac{p-1}{3-p} \tau_\mathrm{T}^\mathrm{acc} \Gamma_\mathrm{c}\Gamma_\mathrm{m}\left(\frac{\Gamma_\mathrm{m}}{\Gamma_\mathrm{M}}\right)^{p-2}\left[1-\left(\frac{\Gamma_\mathrm{m}}{\Gamma_\mathrm{M}}\right)^{3-p}\right]
\ .\nonumber\\
\end{eqnarray}
Here it is assumed  that
$t'_\mathrm{syn}\left(\Gamma_\mathrm{M}\right) \ll t'_\mathrm{ex}$ in \refeq{eq:Yt},
which is always the case when the radiative cooling is the dominant
limiting process for electron acceleration.
If $Y_\mathrm{M} < 1$, the estimate of the electron maximum Lorentz
factor $\Gamma_\mathrm{M}$ given by \refeq{eq:gmax} is valid as the radiative
timescale of electrons at $\Gamma_\mathrm{M}$ is accurately given by
their synchrotron timescale. On the other hand, if $Y_\mathrm{M} > 1$,
it is possible that the value of
$\Gamma_\mathrm{M}$ given by \refeq{eq:gmax} is overestimated. It is not
always the case as the value of $Y_\mathrm{M}$ is computed assuming
inverse Compton scatterings in Thomson regime. The true value of
$Y_\mathrm{M}$  can therefore be reduced by Klein-Nishina
corrections. In practice, we checked that our assumptions regarding
the maximum Lorentz factor of electrons are consistent in all the cases
presented in this paper.

\paragraph{Photon--photon annihilation.}
The timescale associated with $\gamma\gamma$ annihilation is given by
\begin{equation}
t'_\mathrm{\gamma\gamma}(\nu')\simeq \left(\int d\tilde{\nu}'\ n_{\tilde{\nu}'}c\sigma_{\gamma\gamma}(\nu',\tilde{\nu}')\right)^{-1}
\!\!\! \simeq \left.
\frac{5h}{c\sigma_\mathrm{T}}
\frac{1}{\left(h\tilde{\nu}'n_{\tilde{\nu}'}\right)}
\right|_{\tilde{\nu}'=\frac{\left(m_\mathrm{e}c^{2}\right)^{2}}{h^{2}\nu'}}
\ ,
\label{eq:tgg}
\end{equation}
where  
a Dirac approximation  has been used for the cross section \citep{gould:67}. 
The cutoff will occur at high energy and the corresponding photons will
annihilate with low-energy photons whose distribution is approximatively
given by the synchrotron spectrum described in
Eqs.~(\ref{eq:synfast}) and~(\ref{eq:synslow}). An approximate
shape of the absorbed spectrum can then be computed by attenuating the
emitted spectrum by a factor
\begin{equation}
e^{-\tau_{\gamma\gamma}(\nu')} = \frac{t'_\mathrm{\gamma\gamma}(\nu')}{t'_\mathrm{ex}}\left(1-e^{-t'_\mathrm{ex}/t'_\mathrm{\gamma\gamma}(\nu')}\right)\ ,
\end{equation}
where $\tau_{\gamma\gamma}(\nu')$ is the $\gamma\gamma$ annihilation optical depth at frequency $\nu'$.
A comparison with the results of the detailed radiative code shows that
this approximate treatment is again accurate as long as inverse Compton
losses are not the dominant cooling process for electrons, i.e. as long
as the low-energy photon distribution is well described by the standard
synchrotron spectrum. In \reffig{fig:testic}, the
attenuation of the spectrum at high energy due to
$\gamma\gamma$ annihilation is shown in different cases with an increasing
importance of inverse Compton scatterings.

\subsection{The shape of the radiated spectrum}
\begin{figure*}[!t]
\begin{center}
\begin{tabular}{cc}
Minimum electron Lorentz factor & Magnetic field\\
\includegraphics[width=0.49\textwidth]{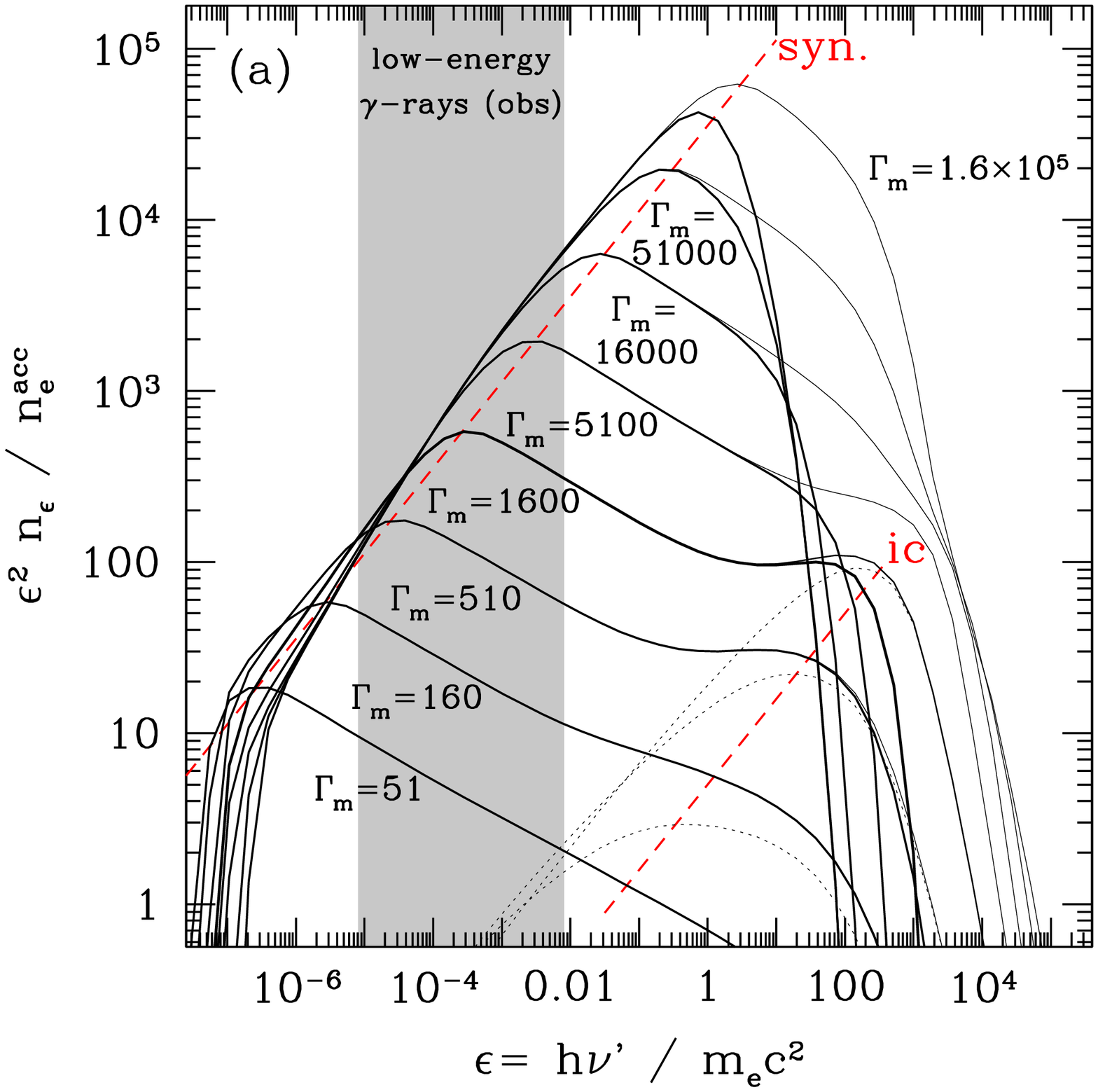} & 
\includegraphics[width=0.49\textwidth]{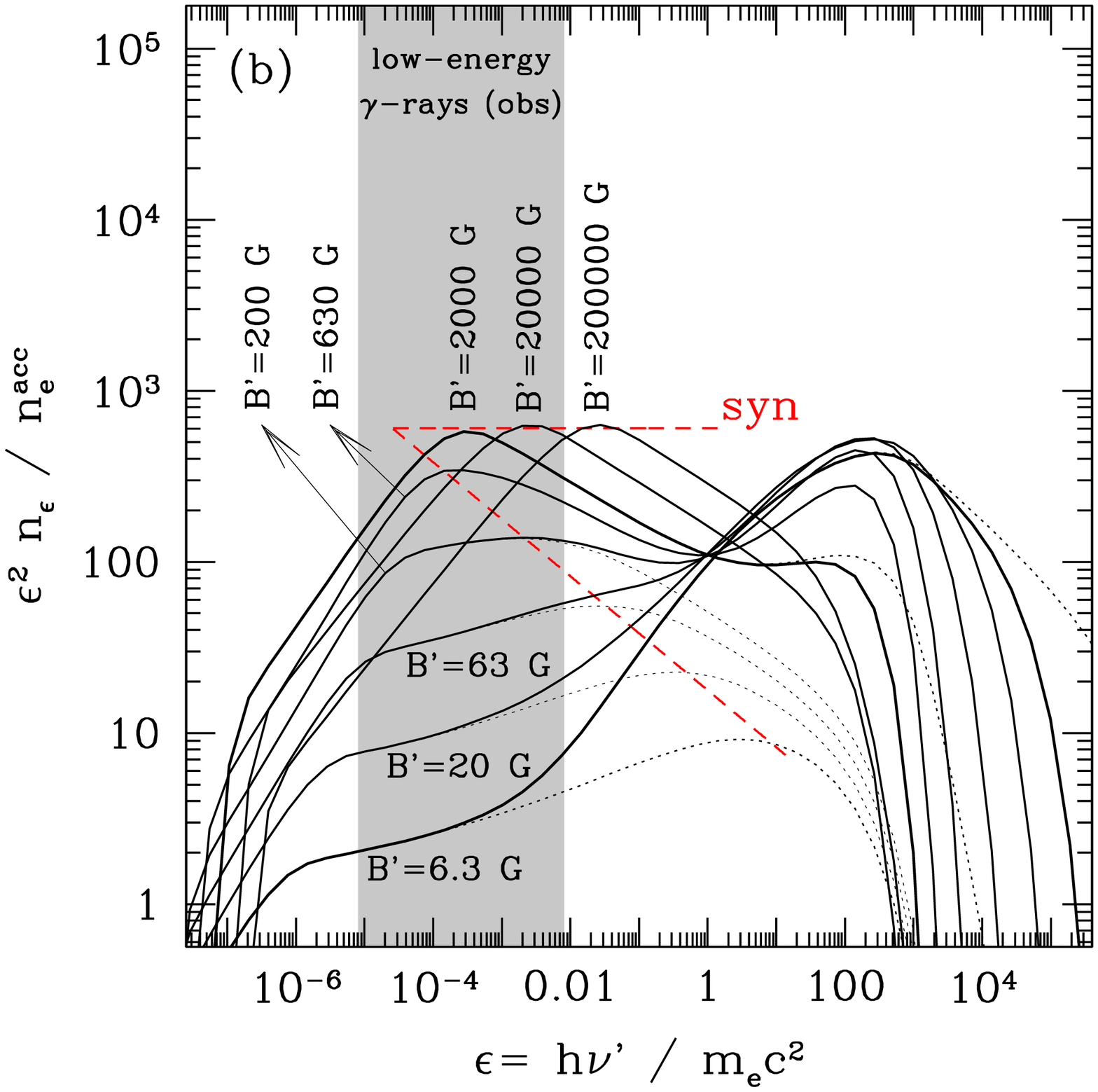} \\
Adiabatic cooling timescale & Density of accelerated electrons\\
\includegraphics[width=0.49\textwidth]{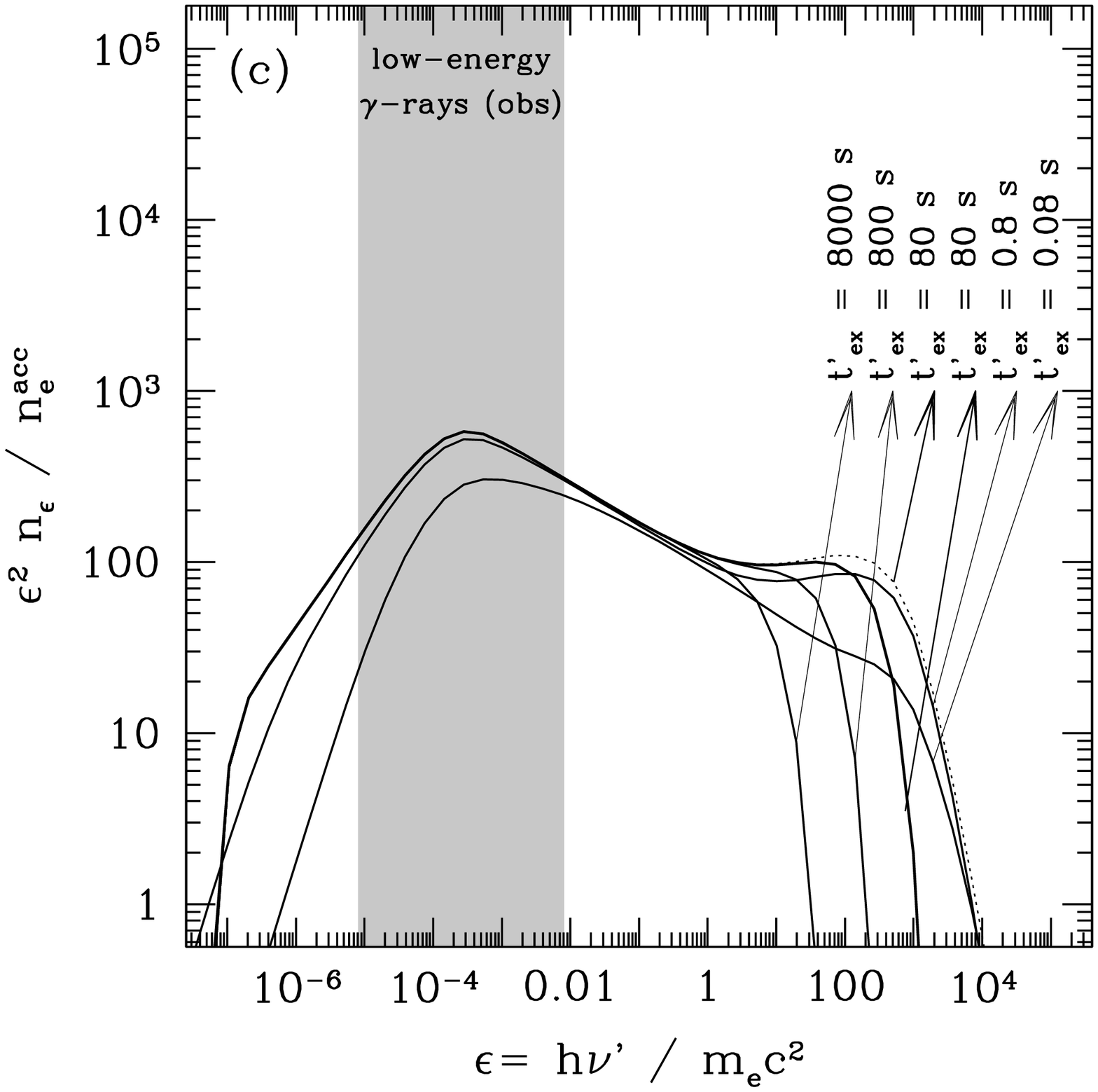} & 
\includegraphics[width=0.49\textwidth]{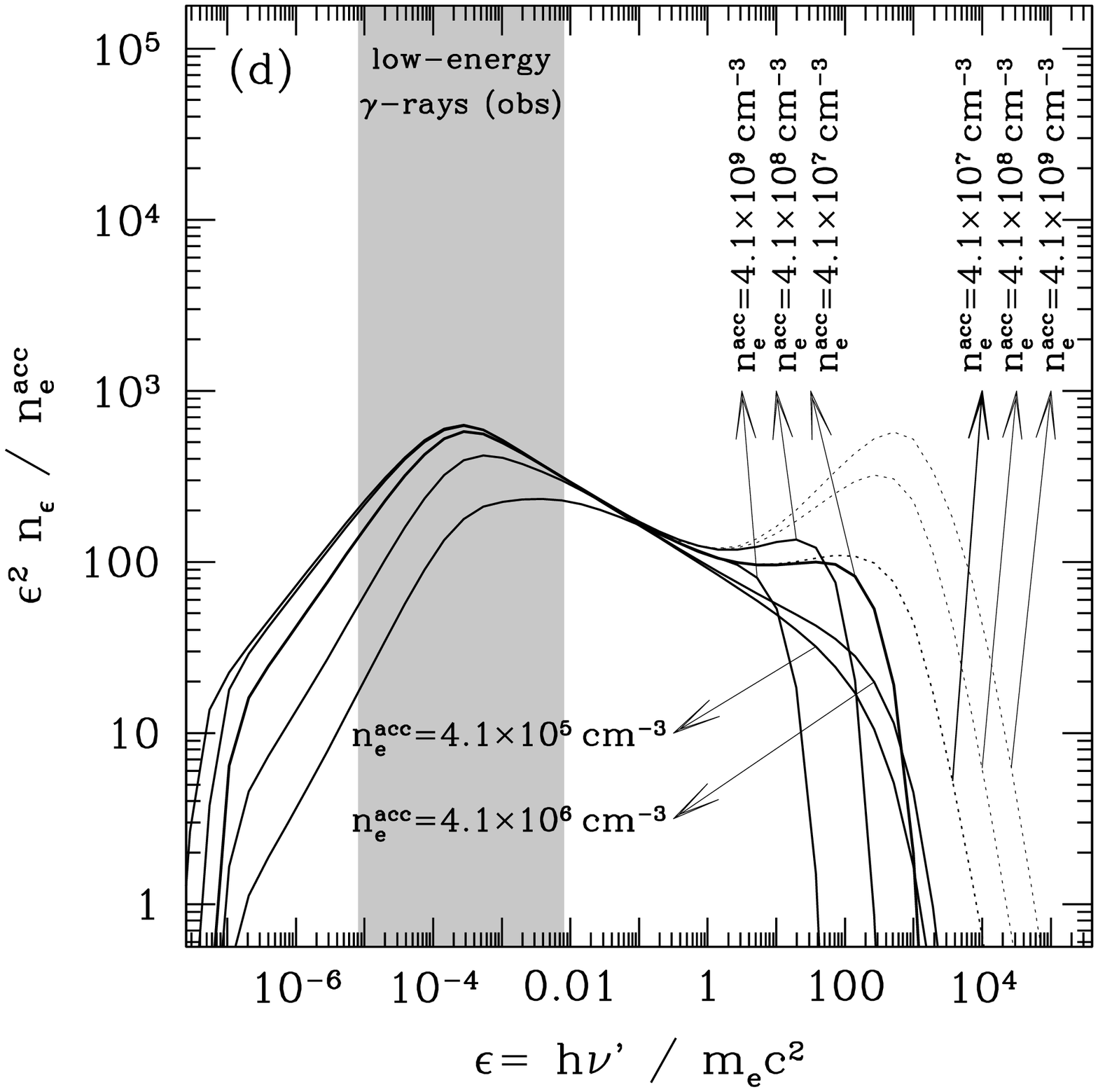} \\
\end{tabular}
\end{center}
\caption{\textbf{Emitted spectrum in the comoving frame.} We consider
 a ``reference case'' defined by $\Gamma_\mathrm{m}=1600$, $B'=2000\ \mathrm{G}$, $t'_\mathrm{ex}=80\ \mathrm{s}$ and $n_\mathrm{e}^\mathrm{acc}=4.1\times 10^{7}\ \mathrm{cm^{-3}}$ corresponding to the physical conditions in the shocked material at $t=1.9\times 10^{4}\ \mathrm{s}$ in the example shown in \reffig{fig:exampledyn}, for $\epsilon_\mathrm{e}=\epsilon_\mathrm{B}=1/3$, $p=2.5$ and $\zeta=10^{-2}$ (see text). Each panel shows the evolution of the spectrum when one parameter is varied, while all other parameters are maintained constant: (a) effect of the initial minimum Lorentz factor of the electrons $\Gamma_\mathrm{m}$. For low $\Gamma_\mathrm{m}$, the weak inverse Compton component is shown with a dotted line. The high-energy component that would be obtained without $\gamma\gamma$ annihilation is plotted with a thin solid line. Two dashed lines indicate the predicted position of the synchrotron peak in fast cooling regime, and of the inverse Compton peak in Thomson regime; (b) effect of the magnetic field $B'$. For low $B'$, the weak synchrotron component is plotted with a dotted line. The high-energy component that would be obtained without $\gamma\gamma$ annihilation is plotted with a dotted line for $B'=6.3\ \mathrm{G}$ and $B'=2000\ \mathrm{G}$. The dashed line indicates the predicted position of the synchrotron peak in slow and fast cooling regime. ; (c) effect of the adiabatic cooling timescale $t'_\mathrm{ex}$. For $t'_\mathrm{ex}=80\ \mathrm{s}$, the spectrum that would be obtained without $\gamma\gamma$ annihilation is plotted with a dotted line; (d) effect of the initial density of relativistic electrons $n_\mathrm{e}^\mathrm{acc}$. For high $n_\mathrm{e}^\mathrm{acc}$, the spectrum that would be obtained without $\gamma\gamma$ annihilation is plotted with a dotted line.
In all panels, the approximate range of low-energy gamma-rays (1 keV-1
 MeV) is indicated by the gray area, assuming $z=1$ and
 $\Gamma_{*}=240$.}
\label{fig:spec_com}
\end{figure*}

We define a ``reference case'' corresponding to the physical conditions
in the shocked material of the example in \reffig{fig:exampledyn} at
$t=1.9\times 10^{4}\ \mathrm{s}$, i.e. $\Gamma_{*}\simeq 240$,
$\rho_{*}\simeq 6.8\times 10^{-15}\ \mathrm{g~cm^{-3}}$,
$\epsilon_{*}/c^{2}\simeq 8.0\times 10^{-2}$ and $t'_\mathrm{ex}=80\
\mathrm{s}$. 
For $\epsilon_\mathrm{e}=\epsilon_\mathrm{B}=1/3$, $p=2.5$ and
$\zeta=10^{-2}$ (only 1 \%
of the electrons are accelerated),
this leads to $B'\simeq 2000\ \mathrm{G}$,
$n_\mathrm{e}^\mathrm{acc}\simeq 4.1\times 10^{7}\ \mathrm{cm^{-3}}$ and
$\Gamma_\mathrm{m}=1.6\times 10^{3}$. This choice of parameters is
motivated by the study presented in \citet{daigne:98} which favors the
case where the magnetic field is high and where only a small fraction of
electrons is accelerated, as these two conditions are required for
the synchrotron peak to be in the {BATSE} range. 
Starting from this ``reference case'',
 one of the parameters $\left(\Gamma_\mathrm{m}, B', t'_\mathrm{ex},
 n^\mathrm{acc}_\mathrm{e}\right)$ is varied, while all other
  parameters are maintained constant. The
resulting evolution of the spectrum is plotted in
\reffig{fig:spec_com}.
\paragraph{Effect of the initial minimum Lorentz factor of relativistic
    electrons.} Panel (a) shows the effect of  $\Gamma_\mathrm{m}$. The
  spectrum is the combination of a low-energy component due to the
  synchrotron emission in fast cooling regime, and a high-energy
  component due to inverse Compton emission, partially suppressed due
  to $\gamma\gamma$ annihilation. (i) \textit{Synchrotron emission.} The evolution of the low-energy peak
  (synchrotron emission) with $\Gamma_\mathrm{m}$ follows exactly the
  prediction of the analytical estimate in fast cooling regime: the
  spectrum peaks at the frequency $\nu'_\mathrm{m}$ (synchrotron
  frequency of electrons with $\gamma=\Gamma_\mathrm{m}$)
that scales as $\nu'_\mathrm{m}\propto \Gamma_\mathrm{m}^{2}$ and the 
 corresponding
peak intensity follows $\left(\left.\nu'\right.^{2}
n_{\nu'}\right)_{\nu'=\nu'_\mathrm{m}}\propto
\Gamma_\mathrm{m}$ \citep{sari:98}. A dashed line of slope $1/2$ in the
$\log{\nu'}$--$\log{\left(\left.\nu'\right.^{2}n_{\nu'}\right)}$ dia\-gram indicates
the predicted position of the synchrotron peak. The agreement with the
numerical calculation is excellent; (ii) \textit{inverse Compton
scattering.} For low $\Gamma_\mathrm{m}$, as
synchrotron photons peak at low energy, inverse Compton
scatterings occur in Thomson regime. Then, from \refeq{eq:Y}, the Compton
parameter $Y$ scales as $Y\propto \Gamma_\mathrm{m}$. This is indeed shown in
panel (a), where the intensity of the inverse Compton component
increases when $\Gamma_\mathrm{m}$ increases. The high-energy peak due
to inverse Compton emission follows exactly the predicted line of
slope $1/2$ in the $\log{\nu'}$--$\log{\left(\left.\nu'\right.^{2}n_{\nu'}\right)}$ diagram (the
peak energy scales as $\Gamma_\mathrm{m}^{2}\nu_\mathrm{m}\propto \Gamma_\mathrm{m}^{4}$ 
and the intensity scales as $Y
\left(\left.\nu'\right.^{2}n_{\nu'}\right)_\mathrm{m}\propto \Gamma_\mathrm{m}^{2}$).
As 
$\Gamma_\mathrm{m}$ increases, the synchrotron emission peaks at higher energy and
more and more photons have energies comparable with $m_\mathrm{e}c^{2}$
in the frame of electrons at $\gamma=\Gamma_\mathrm{m}$. Therefore the
efficiency of inverse Compton emission is strongly reduced by
Klein-Nishina corrections; (iii) \textit{$\gamma\gamma$ annihilation.}
this process becomes more efficient for high $\Gamma_\mathrm{m}$ as
the synchrotron emission peaks at higher energy and more photons are
above the threshold. In conclusion, these two  effects combine so that the high-energy component is
most intense for intermediate values of $\Gamma_\mathrm{m}$.
\paragraph{Effect of the magnetic field.} 
Panel (b) shows the effect of
$B'$. 
This effect is more complicated than for $\Gamma_\mathrm{m}$,
especially at low energy. 
\textit{(i) Synchrotron emission.} 
For very weak magnetic fields, inverse
Compton scatterings are so efficient that the effective electron radiative timescale is short
compared to the adiabatic cooling time and most of their energy is
radiated. This high radiative efficiency is reached despite the fact that
synchrotron radiation occurs in slow cooling regime ($t'_\mathrm{syn}\left(\Gamma_\mathrm{m}\right)>t'_\mathrm{ex}$).
In other terms, this situation corresponds to the following ordering of
characteristic electron Lorentz factors: $\tilde{\Gamma}_\mathrm{c} < \Gamma_\mathrm{m} < \Gamma_\mathrm{c}$.
For more intense magnetic fields, the electron synchrotron timescale is
reduced and synchrotron radiation operates in fast cooling regime
($\tilde{\Gamma}_\mathrm{c} < \Gamma_\mathrm{c} < \Gamma_\mathrm{m}$).
The complex evolution of the synchrotron peak with $B'$ observed in
panel (b) is then
due to the transition between synchrotron slow (low $B'$) and fast (high
$B'$) cooling regime. 
For low $B'$,
the synchrotron peak is given by the synchrotron
frequency $\nu'_\mathrm{c}$ of electrons with $\gamma=\Gamma_\mathrm{c}$
and decreases with $B'$ as $\nu'_\mathrm{c}\propto \left.B'\right.^{-3}$.
The corresponding peak intensity follows
$\left(\left.\nu'\right.^{2}n_{\nu'}\right)_{\nu'=\nu'_\mathrm{c}}\propto
 B^{2\left(p-2\right)}$ \citep{sari:98}, so that the low-energy peak
 follows a line of slope $-2(p-2)/3$ in the
 $\log{\nu'}$--$\log{\left(\left.\nu'\right.^{2}n_{\nu'}\right)}$ diagram. 
For larger values of $B'$,
 synchrotron radiation is in fast cooling regime  and the peak is given by
 $\nu'_\mathrm{m}$, which scales as $\nu'_\mathrm{m}\propto B'$. The
 corresponding peak intensity does not vary with $B'$ and therefore the
 low-energy peak falls into an horizontal line for high magnetic fields
 in the same diagram; (ii) \textit{inverse Compton scattering.} For low
 magnetic fields, when the synchrotron radiation is in slow cooling
 regime, the typical size of the region populated by relativistic
 electrons is $\sim c t'_\mathrm{ex}$, that is much larger than in the
 synchrotron fast cooling regime. Therefore, even in
 Klein-Nishina regime,
 inverse Compton scatterings are extremely efficient due to the
 increased optical depth. As
 $B'$ increases, the synchrotron peak energy decreases and inverse
 Compton scatterings can occur in the Thomson regime. However, as soon
as the magnetic field is high enough so that the synchrotron radiation
enters the fast cooling regime, the Thomson optical depth for
relativistic electrons is reduced as the size of the region populated
with relativistic electrons is $\sim c
t'_\mathrm{syn}\left(\Gamma_\mathrm{m}\right) \ll c t'_\mathrm{ex}$. In addition, the synchrotron peak energy increases,
so that Klein-Nishina corrections are present. These two effects result in a strong
decrease of the inverse Compton efficiency; (iii) \textit{$\gamma\gamma$ annihilation.}
This process becomes important for a low magnetic field $B'$, when the emission at high energy is most
efficient. In conclusion, weak
magnetic fields favor a strong high-energy emission, even when
$\gamma\gamma$ annihilation is non negligible.
\paragraph{Effect of the adiabatic cooling timescale.} The effect of $t'_\mathrm{ex}$
is shown in panel (c). For a magnetic field $B'=2000\ \mathrm{G}$ and an initial minimum electron
Lorentz factor $\Gamma_\mathrm{m}=1600$ as in our reference case, the
synchrotron timescale equals $t'_\mathrm{syn}\left(\Gamma_\mathrm{m}\right)\simeq
0.12\ \mathrm{s}$. 
(i) for high values, $t'_\mathrm{ex}\gg 0.1\ \mathrm{s}$, the spectrum at low energy does
not depend on $t'_\mathrm{ex}$ as it is produced by synchrotron radiation in
fast cooling regime. Inverse Compton scatterings are rare because
relativistic electrons are present only in a small fraction of the
volume of the
shocked region ($\sim t'_\mathrm{syn}/t'_\mathrm{ex}$, see above). On
the other hand the absorption due to $\gamma\gamma$ annihilation
increases with $t'_\mathrm{ex}$ because the effective size of the region
populated with photons is larger and the probability to have
photon-photon interactions is increased;
(ii) for low values, $t'_\mathrm{ex}\ll 0.1\ \mathrm{s}$, the
radiative efficiency decreases strongly as electrons are in slow cooling
regime; (iii) for intermediate values such as $t'_\mathrm{ex}\ga
t'_\mathrm{syn}\left(\Gamma_\mathrm{m}\right)$, relativistic electrons can radiate
efficiently and still populate most of the shocked region. This favors a
brighter high energy component as inverse Compton scatterings are more
frequent. In conclusion, an intense high energy component in the photon
spectrum is favored by intermediate values of the adiabatic cooling timescale.
\paragraph{Effect of the initial density of relativistic electrons.} The
effect of $n_\mathrm{e}^\mathrm{acc}$ is shown in panel (d). At low
densities, the optical depth for inverse Compton scatterings is low, so
the spectrum is simply a synchrotron spectrum in fast cooling regime. At
high densities the intensity of the inverse Compton component increases,
as well as the absorption due to $\gamma\gamma$ annihilation, because the
number of available
high-energy photons scales with the density of emitting electrons. Therefore the most intense high-energy
component is again obtained for intermediate values of the relativistic electron density, when inverse Compton
scatterings are efficient and the attenuation due to $\gamma\gamma$
annihilation is still not too
strong.\\

\begin{figure}[t]
\begin{center}
\includegraphics[width=\linewidth,viewport=1cm 1.25cm 20.25cm 10.5cm]{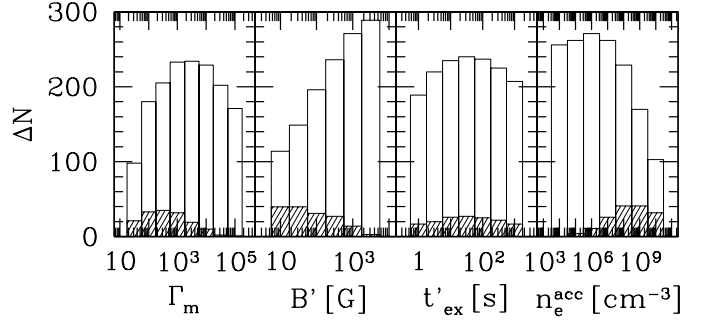}
\end{center}
\caption{\textbf{Physical conditions in the shocked region (comoving frame) that favor
 an intense high energy component.} From our exploration of the
 parameter study of the internal shock model (see
 \refsec{sec:spec_com}) the histograms of the electron
 minimum Lorentz factor $\Gamma_\mathrm{m}$, the magnetic field $B'$,
 the adiabatic cooling timescale $t'_\mathrm{ex}$ and the initial
 density of accelerated electrons $n_\mathrm{e}^\mathrm{acc}$ are plotted
 for all
 models that satisfy the radiative efficiency and transparency
 conditions (see text). The shaded histograms show the models that
 fulfill the same conditions and, in addition, have 
 an efficient inverse Compton emission (most of 50\% of
 the electron energy which is radiated is due to inverse Compton
 scatterings) and a negligible $\gamma\gamma$ annihilation (defined as
 cases where the peak in
 $\nu F_{\nu}$ of the high-energy component is shifted by less than 10
 \% when including $\gamma\gamma$ annihilation).}
\label{fig:distrib_com}
\end{figure}

{This parameter study
aims at identifying the physical conditions in the comoving frame
leading to an intense high energy emission.
\reffig{fig:distrib_com} shows the distributions of the parameters
corresponding to cases with an efficient inverse Compton emission (more
than 50 \% of the radiated energy is due to inverse Compton scatterings) and
a limited $\gamma\gamma$ annihilation (negligible attenuation at the
peak of the inverse Compton component).
The most intense high-energy components are  obtained for
low values of the magnetic field
($B' \la 100\ \mathrm{G}$) and for intermediate values of the electron minimum Lorentz factor
($\Gamma_\mathrm{m}\simeq 100-1000$), 
of the adiabatic time scale ($t'_\mathrm{ex}\simeq
3-30\ \mathrm{s}$) and 
of the density $n_\mathrm{e}^\mathrm{acc}$
($n_\mathrm{e}^\mathrm{acc}\simeq 10^{8}-10^{9}\ \mathrm{cm^{-3}}$).}

\section{Probing the parameter space of internal shocks}
\label{sec:parameterspace}
In the internal shock model, 
the four quantities studied in \refsec{sec:spec_com}
are not independent.
As described in \refsec{sec:method}, they are determined from 
two sets of parameters: the first set defines the
dynamical evolution. In the simple two shell version of the model, these
parameters are $\bar{\Gamma}$, $\kappa$, $\dot{E}$ and $\tau$. The
second set is related to the unknown microphysics in the shocked
region\,: 
$\epsilon_\mathrm{B}$, $\epsilon_\mathrm{e}$, $\zeta$ and $p$. 
Therefore, we have computed 
7200 spectra corresponding
to\,: (i) 4 values for the mean Lorentz factor in the outflow,
$\log{\bar{\Gamma}}=1.5$, $2$, $2.5$ and $3$; (ii) 4 values for the contrast
which
characterizes the amplitude of the variations in the initial distribution
of the Lorentz factor in the outflow,
$\kappa=2.5$, $5$, $7.5$ and $10$; (iii) 6 values of the injected kinetic power
during the relativistic ejection, $\log{\left(\dot{E}/1\ \mathrm{erg~s^{-1}}\right)}=50$, 
$51$, $52$, $53$, $54$ and $55$; (iv)  5 values for the
variability timescale, $\log{\left(\tau/1\ \mathrm{s}\right)}=-2$, $-1$,
$0$, $1$ and $2$; (v) 3 values for the fraction of the
dissipated energy which is injected in the magnetic field,
$\log{\epsilon_\mathrm{B}}=-3.5$, $-2$ and $-0.5$; (vi) 5 values for the fraction of electrons
that are accelerated, $\log{\zeta}=-4$, $-3$, $-2$, $-1$ and $0$.\\

The moderate efficiency of the conversion of kinetic energy into
internal energy by internal shocks imposes that a large fraction $\epsilon_\mathrm{e}$ of this
dissipated energy is injected in relativistic electrons to maintain a
reasonable total efficiency. Therefore we fix $\epsilon_\mathrm{e}=1/3$.  In
the example presented in \reffig{fig:exampledyn}, about $7\ \%$
of the kinetic energy is converted in internal energy by shock
waves. If electrons are radiating efficiently, about $2\ \%$
of the initial kinetic energy will be radiated.\\

We also assume a
slope $p=2.5$ for the electron distribution, except where mentioned
otherwise. This new set of spectra will allow us
to identify which properties of the outflow determine the shape of the
high energy spectrum, and therefore help to identify physical
diagnostics for future \textit{Fermi} data.
\begin{figure*}[!t]
\begin{center}
\begin{tabular}{cc} 
\hspace*{1cm} Bulk Lorentz factor & Contrast\\
\includegraphics*[height=6.25cm,viewport=0cm   0cm 20.2cm 20.2cm]{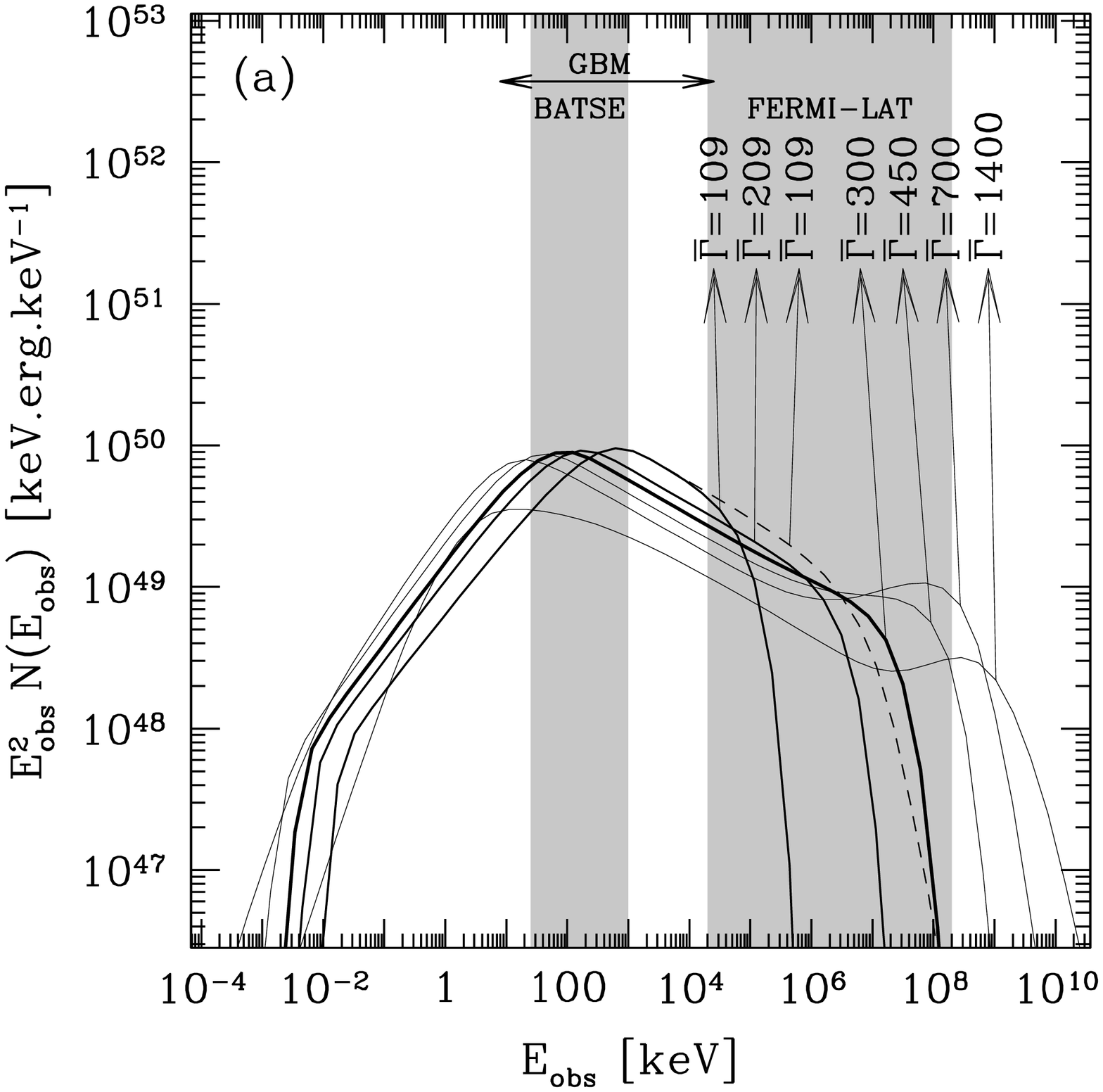} &
\includegraphics*[height=6.25cm,viewport=3.5cm 0cm 20.2cm 20.2cm]{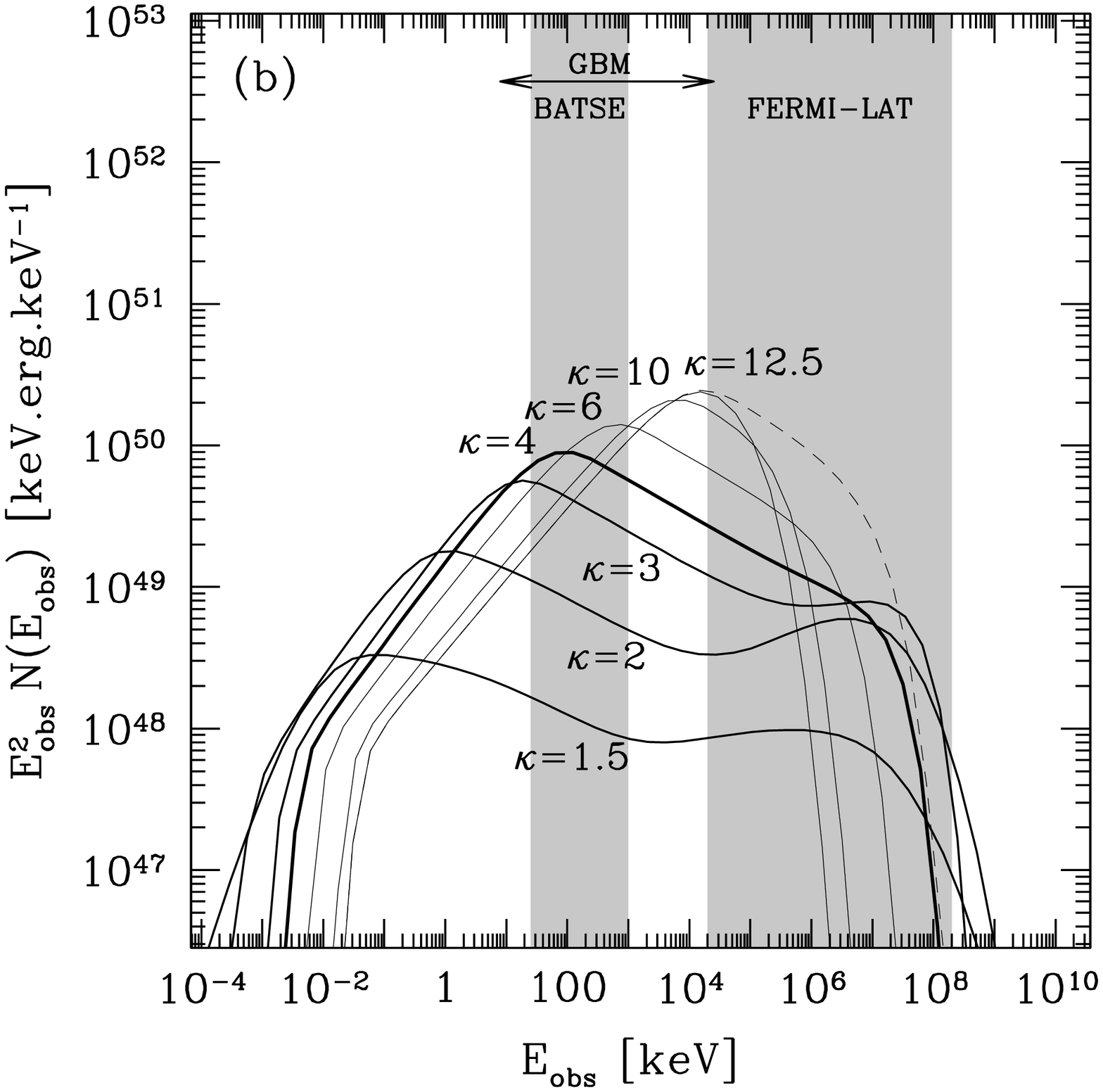} \\
\hspace*{1cm} Injected kinetic power & Variability timescale\\
\includegraphics*[height=6.25cm,viewport=0cm   0cm 20.2cm 20.2cm]{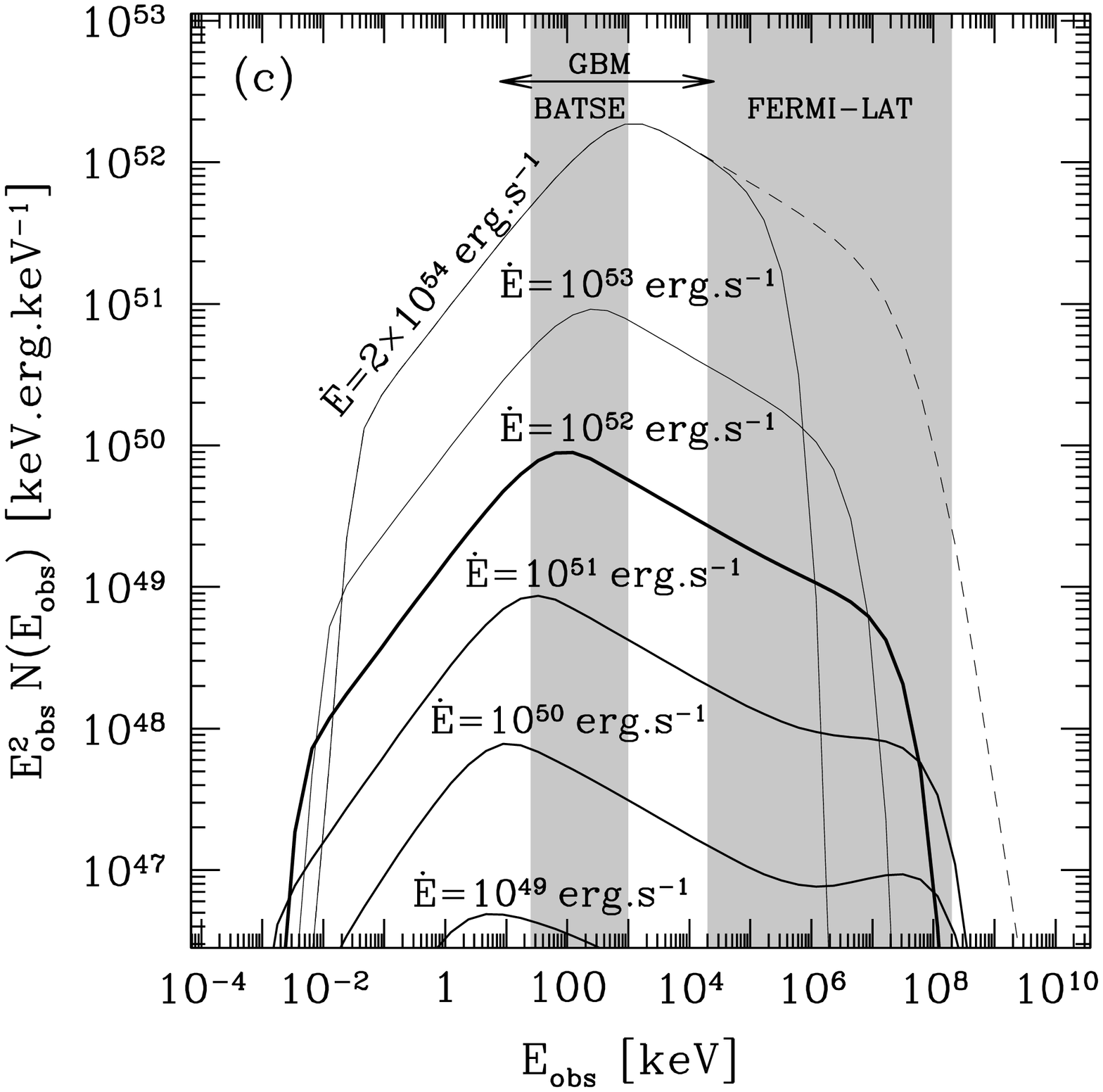} &
\includegraphics*[height=6.25cm,viewport=3.5cm 0cm 20.2cm 20.2cm]{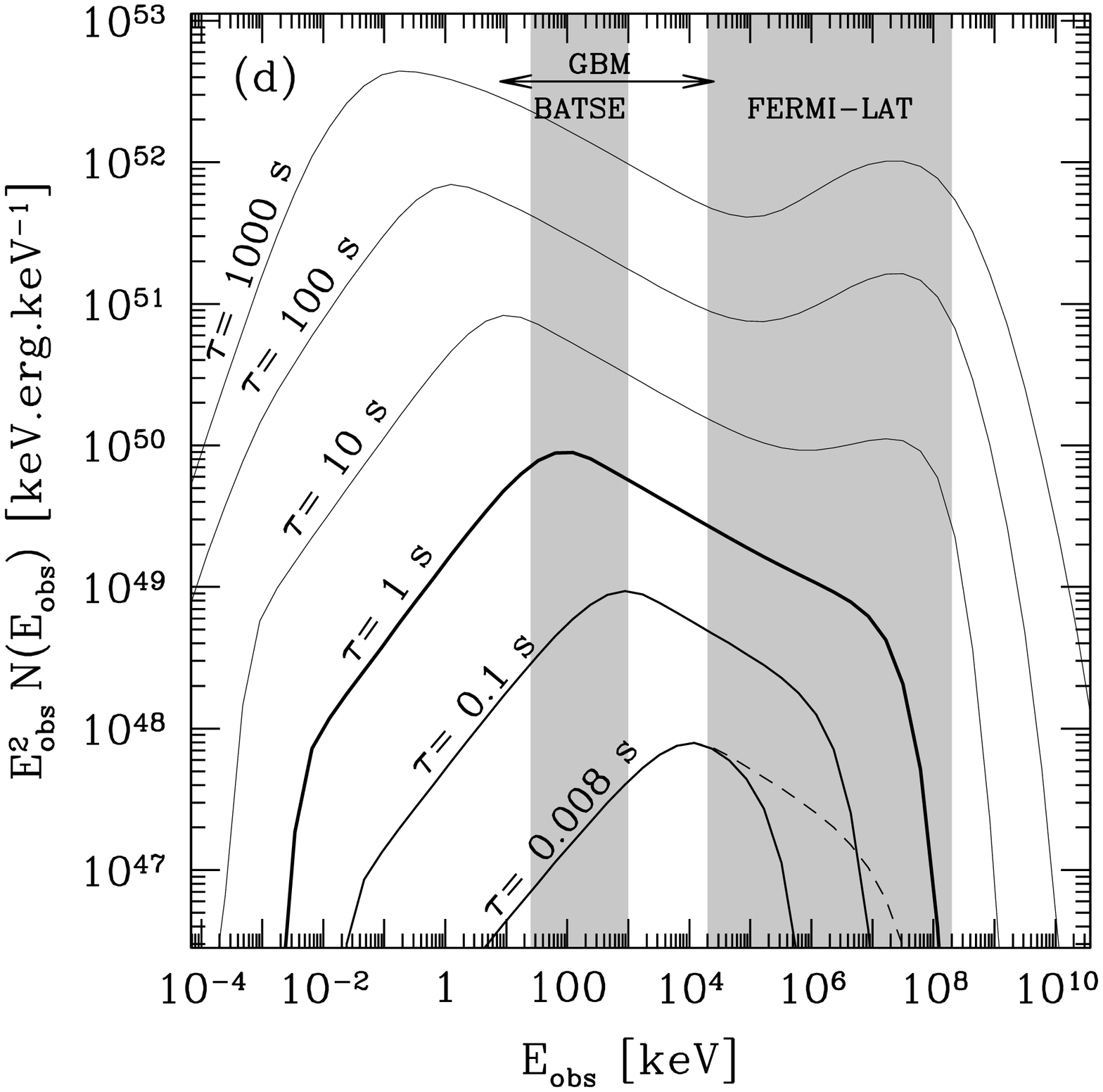} \\
\hspace*{1cm} Fraction of energy injected in the magn. field & Fraction of
     accelerated electrons \\
\includegraphics*[height=6.25cm,viewport=0cm   0cm 20.2cm 20.2cm]{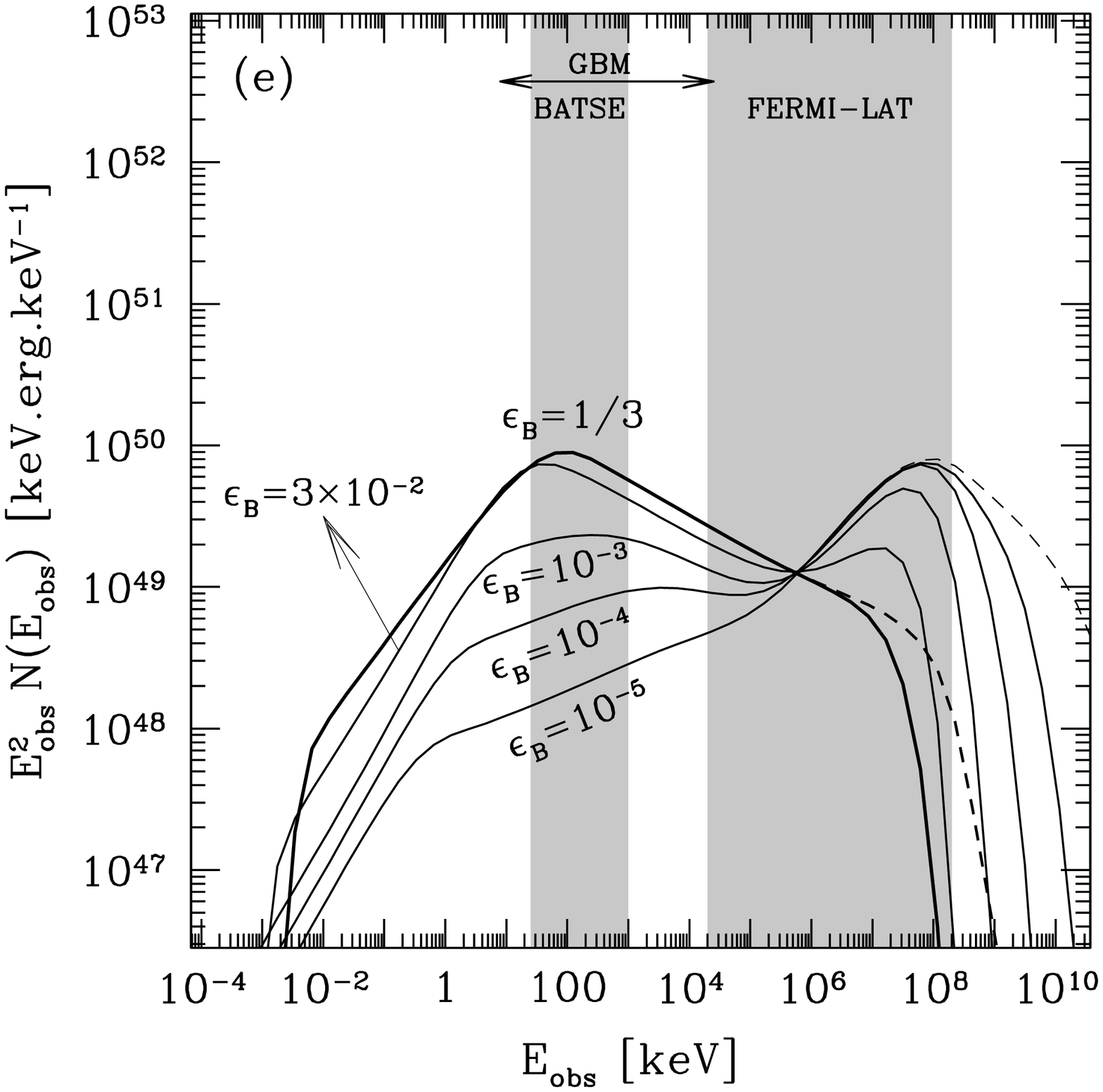} &
\includegraphics*[height=6.25cm,viewport=3.5cm 0cm 20.2cm 20.2cm]{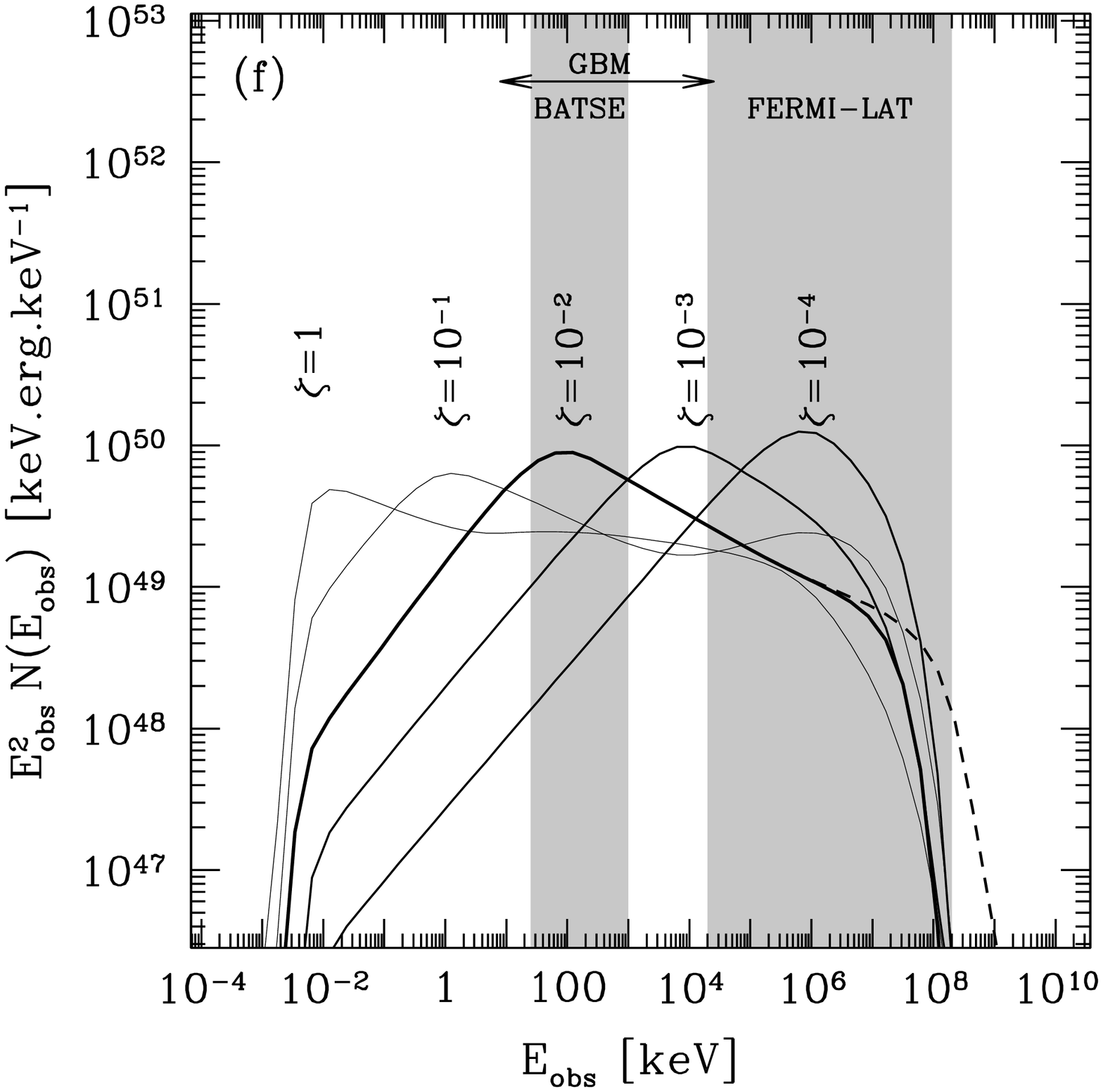} \\
\end{tabular}
\end{center}
\caption{\textbf{The effect of internal shock parameters on the emitted
 spectrum (``synchrotron case'').} We use the simple two shell version of the internal shock
 model (see text) and define a ``reference case'' by $\bar{\Gamma}=300$,
 $\kappa=4$, $\dot{E}=10^{52}\ \mathrm{erg~s^{-1}}$, $\tau=1\ \mathrm{s}$, $\epsilon_\mathrm{e}=\epsilon_\mathrm{B}=1/3$, $\zeta=10^{-2}$ and $p=2.5$. Each panel shows the evolution of the observed spectrum (assuming $z=1$) when one parameter is varied, while all other parameters are maintained constant. Two effects can limit the parameter range: electrons become radiatively inefficient (``efficiency limit'') or the medium becomes optically thick due to the intense production of pairs (``transparency limit''). For each limiting case corresponding to the transparency limit (panels a,b,c,d), the spectrum that would be observed without
 $\gamma\gamma$ annihilation is plotted with a dashed line.
 (a) effect of
 $\bar{\Gamma}$. The transparency limit is reached for  $\bar{\Gamma}< 109$ and the efficiency limit for $\bar{\Gamma} > 1400$ ; 
 (b) effect of $\kappa$. The efficiency limit is reached for
 $\kappa < 1.5$ and the transparency limit for $\kappa > 12.5$; 
(c) effect of $\dot{E}$. The transparency limit is reached for $\dot{E} > 2\times 10^{54}\ \mathrm{erg~s^{-1}}$; 
(d) effect of $\tau$. The transparency limit is reached for $\tau < 0.008\ \mathrm{s}$; 
(e) effect of $\epsilon_\mathrm{B}$. The transparency limit is never
 reached. The spectrum that would be observed without $\gamma\gamma$
 annihilation is plotted with a dashed line for $\epsilon_\mathrm{B}=10^{-5}$ and $\epsilon_\mathrm{B}=1/3$; (f) effect of $\zeta$. The transparency limit is never reached. The
 spectrum that would be observed without $\gamma\gamma$ annihilation is
 plotted with a dashed line for $\zeta=10^{-2}$.}
\label{fig:spec_is} 
\end{figure*} 

\subsection{The spectral shape of internal shock emission}

The effect
of the six parameters $\left(\bar{\Gamma}, \kappa, \dot{E}, \tau,
\epsilon_\mathrm{B}, \zeta\right)$ on the emitted spectrum is now studied. We define again
a ``reference case'' by $\bar{\Gamma}=300$, $\kappa=4$,
$\dot{E}=10^{52}\ \mathrm{erg~s^{-1}}$,
$\tau=1\ \mathrm{s}$, $\epsilon_\mathrm{B}=1/3$ and
$\zeta=10^{-2}$. Such a set of parameters corresponds to a ``typical''
GRB pulse with a peak energy $E_\mathrm{p}\simeq 200\ \mathrm{keV}$
(source frame) due to the synchrotron radiation.
\reffig{fig:spec_is} shows the evolution of the
observed spectrum when one of the parameters is varied, while all other parameters
are maintained
constant (assuming a redshift $z=1$). 
\paragraph{Maximum radius to maintain a high radiative efficiency.}
At very large distances from the source, the density becomes
 very low as well as the magnetic field. This increases the
synchrotron timescale. In an equivalent way $\Gamma_\mathrm{c}$ is
increasing and, at some maximum radius, can become of the order of
$\Gamma_\mathrm{m}$ which strongly reduces the radiative efficiency. For
such high radii, inverse Compton scatterings are rare due to a low
density. This limit can then be evaluated by taking into
account synchrotron radiation only. From
Eqs.~(\ref{eq:isradius}),~(\ref{eq:iscom}),~(\ref{eq:gm}) and~(\ref{eq:gc}) the condition
$\Gamma_\mathrm{m}>\Gamma_\mathrm{c}$ leads to
\begin{eqnarray}
\left(\frac{(p-2)/(p-1)}{1/3}\right)\left(\frac{\epsilon_\mathrm{B}}{1/3}\right)\left(\frac{\epsilon_\mathrm{e}}{1/3}\right)\left(\frac{\zeta}{10^{-2}}\right)^{-1} & & \nonumber\\
\times\left(\frac{f(\kappa)}{f(4)}\right)\left(\frac{\dot{E}}{10^{52}\ \mathrm{erg/s}}\right)\left(\frac{\bar{\Gamma}}{300}\right)^{-5}\left(\frac{\tau}{1\ \mathrm{s}}\right)^{-1} & \ga & 1.0\times 10^{-3}\ ,
\end{eqnarray}
with 
$f(\kappa)=(\kappa+1)^{6}(\kappa-1)(\sqrt{\kappa}-1)^{2}/\kappa^{9/2}$.
For our reference case, when varying one
parameter only, this condition leads to a maximum bulk Lorentz factor
$\bar{\Gamma} \la 1200$, 
a minimum contrast $\kappa\ga 1.2$, 
a minimum injected kinetic power $\dot{E}\ga 1.0\times 10^{49}\ \mathrm{erg/s}$, 
a maximum timescale $\tau \la 1000\ \mathrm{s}$ 
and a minimum fraction of the energy injected into the magnetic field 
$\epsilon_\mathrm{B}\ga 3\times 10^{-4}$. 
The numerical results shown in \reffig{fig:spec_is} agree
well with these analytical estimates.\\

Note that an additional condition should apply to limit the maximum
radius of internal shocks \citep[see e.g.][]{daigne:07}: most collisions should occur before the
deceleration radius, otherwise the propagation of the reverse shock in
the relativistic outflow will suppress the internal shock phase. For
reasonable estimates of the external density, the deceleration radius
is of the order of $10^{16}-10^{17}\ \mathrm{cm}$. From
\refeq{eq:isradius}, this leads to a new constraint
\begin{equation}
\left(\frac{g(\kappa)}{g(4)}\right)\left(\frac{\bar{\Gamma}}{300}\right)^{2}\left(\frac{\tau}{1\ \mathrm{s}}\right) \la 110 \left(\frac{R_\mathrm{dec}}{10^{17}\ \mathrm{cm}}\right)\ ,
\end{equation}
with $g(\kappa) = \kappa^{2}/\left((\kappa-1)(\kappa+1)^{3}\right)$. 

\paragraph{Maximum density to have an optically thin medium.}
On the other end, if internal shocks occur close to the central source, the
density will be high. The Thomson optical depth due to the ambient
electrons and the pairs produced by $\gamma\gamma$ annihilation can then
make
the outflow optically thick. For our reference model, when varying one
parameter only, $\tau_\mathrm{T}^\mathrm{tot}<1$ leads to a minimum value of
$\bar{\Gamma}\ga 110$, a maximum value of $\kappa\la 13$, a maximum
value of $\dot{E}\la 2\times 10^{54}\ \mathrm{erg~s^{-1}}$ and a
minimum value of $\tau\ga 0.008\ \mathrm{s}$. 
\paragraph{The synchrotron component at low energy.}
As the scaling given by
\refeq{eq:epsyn} for the synchrotron peak energy is quite accurate, it
is not surprising to find that 
in most cases, the position of
the synchrotron peak is simply given by \citep{barraud:05}
\begin{equation}
E_\mathrm{syn,obs} \propto \Gamma_{*} B' \Gamma_\mathrm{m}^{2}
\propto \Gamma_{*} \frac{\epsilon_\mathrm{B}^{1/2}\epsilon_\mathrm{e}^{2}}{\zeta^{2}} \rho_\mathrm{*}^{1/2} \epsilon_\mathrm{*}^{5/2}
\propto \frac{\epsilon_\mathrm{B}^{1/2}\epsilon_\mathrm{e}^{2}}{\zeta^{2}} \Phi\left(\kappa\right) \frac{\dot{E}^{1/2}}{\bar{\Gamma}^{2}\tau}
\end{equation}
with
\begin{equation}
\Phi(\kappa) = \frac{\left(\sqrt{\kappa}-1\right)^{5}\left(\kappa-1\right)\left(\kappa+1\right)^{3}}{\kappa^{7}}\ .
\end{equation}
As predicted, the observed photon energy of the synchrotron peak 
increases (see \reffig{fig:spec_is}) when (i) the Lorentz factor $\bar{\Gamma}$ decreases; (ii) the contrast $\kappa$ increases; (iii) the injected kinetic power $\dot{E}$ increases; (iv) the duration of the ejection
$\tau$ decreases; (v) the fraction $\epsilon_\mathrm{B}$ increases; (vi)
the fraction $\zeta$ decreases. This confirms that a low fraction of
accelerated electrons is necessary to have a synchrotron peak in the
gamma-ray range \citep{daigne:98} and that X-ray flashes and X-ray rich
gamma-ray bursts can be produced by internal shocks within ``clean
fireballs'', i.e. outflows having a
high Lorentz
factor $\bar{\Gamma}$ (small baryonic pollution) and a small contrast
$\kappa$ \citep{barraud:05}.\\

There are two situations when this scaling for the synchrotron peak is
not valid anymore:\\
\noindent--~if synchrotron radiation occurs in slow cooling
regime. This situation would normally be rejected due to its low
radiative efficiency. However, the synchrotron slow cooling regime can be
compensated by efficient inverse Compton scatterings. It has been shown in
the previous section (\S~\ref{sec:spec_com}) that the scaling given by \refeq{eq:epsyn} is not
accurate in this case. Even the shape of the synchrotron spectrum can be
modified. Such cases can be found for instance in panel (e) of
\reffig{fig:spec_is} for $\epsilon_\mathrm{B}\la 10^{-3}$;\\
\noindent--~if the medium is dense enough so that the synchrotron
self-absorption frequency is above the expected synchrotron peak. Such
highly self-absorbed cases require a large density of relativistic
electrons. As most of the spectra shown in \reffig{fig:spec_is} are computed with a low fraction
$\zeta$ of accelerated electrons, this is usually not the case. In the
full exploration of the parameter space of the internal shock model, we
find that highly absorbed synchrotron spectrum can be found for
$\zeta=1$. However, in this case the emission detected in the {BATSE}
range corresponds to the inverse Compton component. This will be
discussed below (\S~\ref{sec:iccase}).

\subsection{Spectral components in the \textit{Fermi}-{LAT} energy range}
\paragraph{Conditions for intense inverse Compton emission.}
From the study made in the previous section (\S~\ref{sec:spec_com}), it
is expected that the efficiency of
the inverse
Compton scatterings is increased by (i) a moderate electron minimum Lorentz
factor $\Gamma_\mathrm{m}$,
which corresponds to internal shocks with 
a moderate contrast $\kappa$ between the Lorentz factors of the
colliding shells, and/or 
a large fraction $\zeta$ of accelerated electrons; 
(ii) a low magnetic field $B'$, which corresponds to internal shocks
with a high bulk Lorentz factor $\bar{\Gamma}$,
a moderate contrast $\kappa$,
a moderate injected kinetic power $\dot{E}$,
a large variability timescale $\tau$,
and/or a low fraction $\epsilon_\mathrm{B}$ of the energy injected in the
magnetic field; 
(iii) a low $t'_\mathrm{ex}$, i.e. by internal shocks with a
 high bulk Lorentz factor $\bar{\Gamma}$,
a moderate contrast $\kappa$,
and/or a short variability timescale $\tau$. This is in good agreement
with the results shown in \reffig{fig:spec_is}. However, even when
the inverse Compton emission is efficient, the corresponding spectral
component is not necessarily intense, as it can be suppressed by
photon--photon annihilation.
\paragraph{Conditions for strong photon--photon annihilation.}
As shown in the previous section (\S~\ref{sec:spec_com}), 
$\gamma\gamma$ annihilation is important for large values of
the optical depth $\tau_\mathrm{T}^\mathrm{acc}$.  
The reason is that the $\gamma\gamma$ annihilation and inverse Compton (in Thomson regime)
cross sections
 are of the same order.
To investigate this effect, we have considered the transparency
condition $\tau_\mathrm{T}^\mathrm{tot}<0.1$ (see \S~\ref{sec:transparency}). It corresponds (from
Eqs.~(\ref{eq:tauacc}), (\ref{eq:tex}), (\ref{eq:iscom}) and
(\ref{eq:isradius})) to low bulk Lorentz factors $\bar{\Gamma}$, high
contrasts $\kappa$, large injected kinetic power $\dot{E}$, short
timescales $\tau$ and high fraction of accelerated electrons
$\zeta$. However, it is also favored by a high peak energy of the synchrotron
component, which is also obtained for low $\bar{\Gamma}$, high $\kappa$,
large $\dot{E}$ and short $\tau$, but low $\zeta$ and high
$\epsilon_\mathrm{B}$. These effects are well observed in
\reffig{fig:spec_is}, panels (a-d) for the dynamical parameters and
panels (e-f) for the microphysics, where it is seen in particular that
$\gamma\gamma$ annihilation is strongest for intermediate values of
$\zeta$.
\begin{figure}[t!]
\centerline{\includegraphics[width=\linewidth]{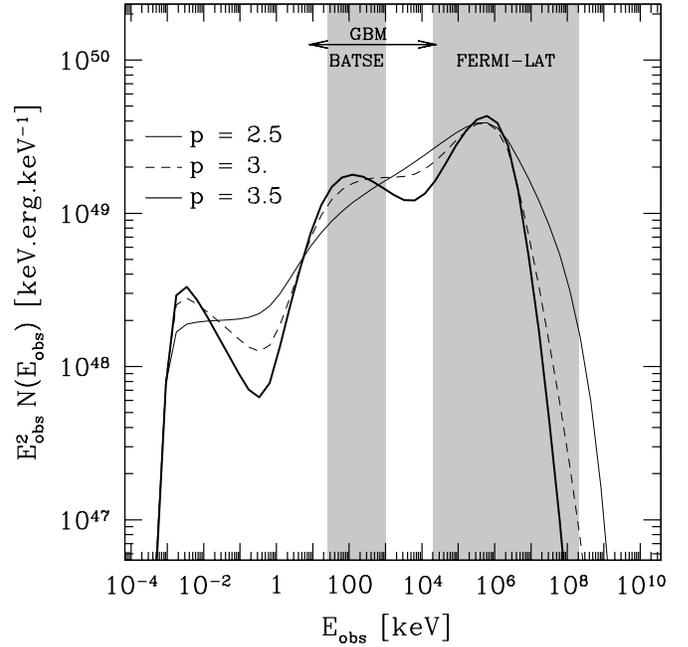}}
\caption{\textbf{Effect of the slope of the accelerated electron
 distribution in the ``inverse Compton case''.} The observed spectrum
 (assuming $z=1$) obtained including all radiative processes
 is plotted for $\bar{\Gamma}=300$, $\kappa=4$, $\dot{E}=10^{52}\ \mathrm{erg~s^{-1}}$, $\tau=1\ \mathrm{s}$, $\epsilon_\mathrm{B}=10^{-3}$, $\epsilon_\mathrm{e}=1/3$, $\zeta=1$ and three different values of $p$. The first inverse Compton peak is well defined for $p> 3$.}
\label{fig:ic_slope}
\end{figure}

\subsection{Dominant radiative process in the keV-MeV range and
    consequences at higher energy}

From our exploration of the parameter space of the internal shock model
we find, as expected from previous studies
\citep{papathanassiou:96,daigne:98,meszaros:00}, that there are two
classes of spectra, 
depending on the radiative process
responsible for the prompt emission in the keV-MeV range. This energy
range is
detected for instance by instruments such as {BATSE},
\textit{Beppo-SAX}, \textit{HETE-2}, \textit{Integral}, 
\textit{Swift} or \textit{Fermi}-{GBM}. These two cases have very different behavior in the
MeV-GeV range and therefore \textit{Fermi}-{GBM}+{LAT} observations will
allow us to distinguish between the two possibilities.

\paragraph{``Synchrotron case''.} The synchrotron component peaks in the {BATSE}
range (keV-MeV). This case is favored in internal shocks as it predicts
pulse shapes and spectral evolution in GRB lightcurves that are in
better agreement with observations \citep{daigne:98,daigne:03}.
The ``synchrotron case'' is found in most spectra plotted in
\reffig{fig:spec_is}. In this case, the inverse Compton component
peaks at higher energy (MeV-GeV range). These spectra are characterized
by a high magentic field and by a low fraction $\zeta$ of accelerated electrons\footnote{This
is why the ``synchrotron case'' in fast cooling regime, which is our preferred case, is disfavored by
\citet{kumar:08}. 
Their study does not consider the possibility to have $\zeta<1$.
Therefore, the authors conclude that the ``synchrotron case'' in
fast cooling regime is very unlikely, as it
would involve very high contrasts $\kappa=\Gamma_{2}/\Gamma_{1}$ in
internal shocks. With $\zeta=1$, the only possibility to reach high
electron Lorentz factor is indeed to dissipate more energy per particle
in shocks. The assumption $\zeta \ll 1$ made in the present study solves
this problem. Note that \citet{kumar:08} also disfavor the
``synchrotron case'' in slow cooling regime, as it implies a typical
radius for internal shocks which is too large (of the order of the
deceleration radius or larger). We do not discuss this case in the
present study as it reduces even more the efficiency of the conversion of
the kinetic energy of the outflow into radiation by internal shocks, which is already low
in the fast cooling regime.}, which allows high
values of the electron Lorentz factor $\Gamma_\mathrm{m}$. In the ``synchrotron case'' most inverse Compton scatterings occur in
Klein-Nishina regime. This leads to four types of spectra at high
energy ({LAT} range): 
\begin{enumerate}
\item a strong second peak with a large
$\gamma\gamma$ attenuation. This case is found for example for $\epsilon_\mathrm{B}\la 0.03$  in
\reffig{fig:spec_is}, panel (e);
\item a weak second peak with a negligible $\gamma\gamma$
attenuation. This case is found for example 
for $\bar{\Gamma}\sim 450-600$ in \reffig{fig:spec_is}, panel (a);
\item a weak second peak with a strong
$\gamma\gamma$ annihilation. This case is found for example for
$\bar{\Gamma}\sim 250-450$ in
\reffig{fig:spec_is}, panel (a);
\item no second peak, the high-energy emission is only the tail of the
synchrotron component, with a cutoff in the 100 MeV - 10 GeV range due
to $\gamma\gamma$ annihilation. This case is found for example for $\bar{\Gamma}\la 250$ in
\reffig{fig:spec_is}, panel (a). 
\end{enumerate}
From case 1. to case 4., the bulk Lorentz factor $\bar{\Gamma}$ is
decreasing, the contrast $\kappa$ is increasing, the injected kinetic power
 $\dot{E}$ is increasing, and/or the timescale $\tau$ is
decreasing. 
We discuss below the corresponding possible physical diagnostics using
\textit{Fermi} data (\S~\ref{sec:FermiDiag}).
\begin{figure*}[!t]
\begin{center}
\begin{tabular}{cc} 
\hspace*{1cm} Bulk Lorentz factor & Contrast\\
\includegraphics*[height=6.25cm,viewport=0cm   0cm 20.2cm 20.2cm]{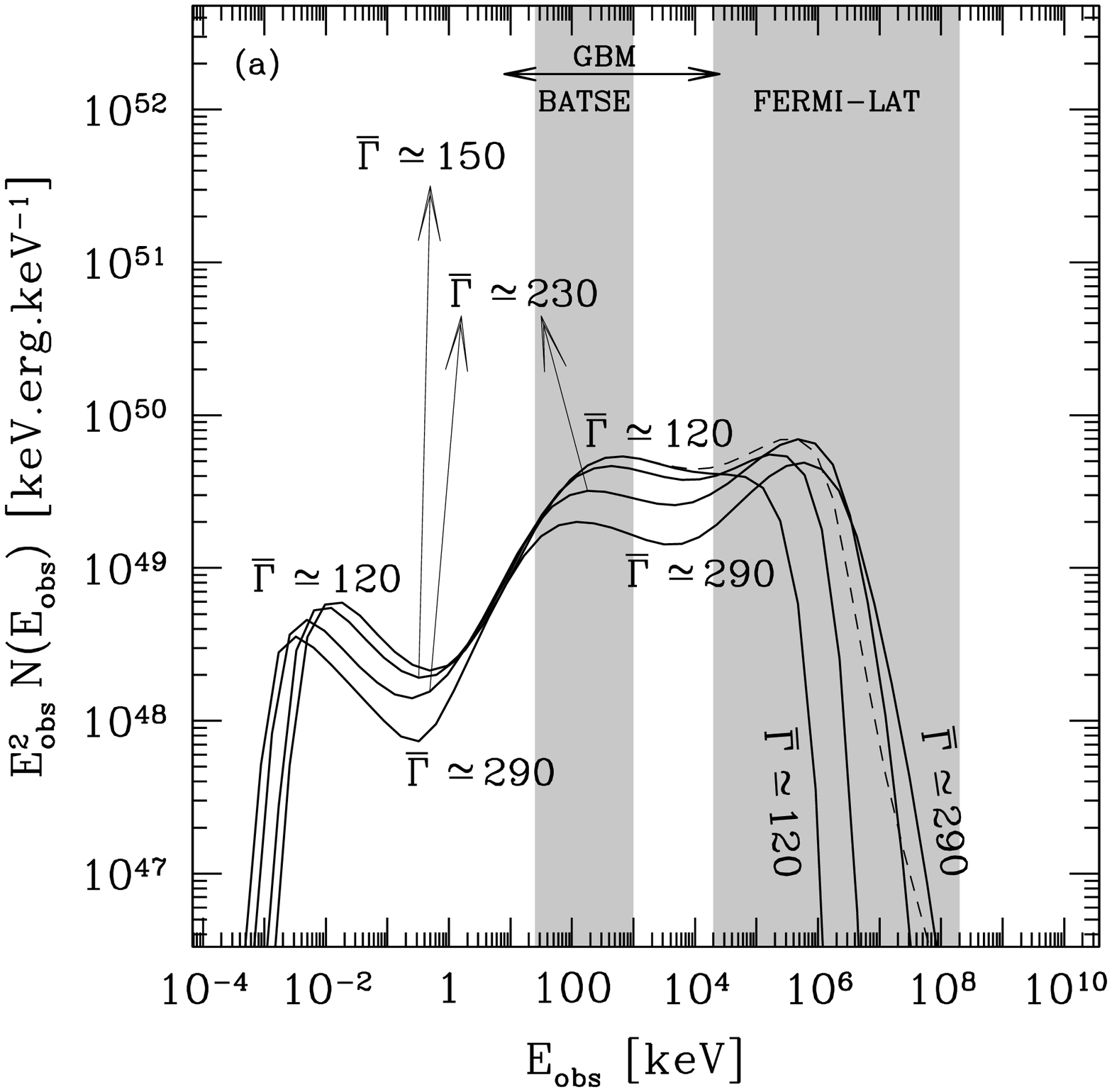} &
\includegraphics*[height=6.25cm,viewport=3.5cm 0cm 20.2cm 20.2cm]{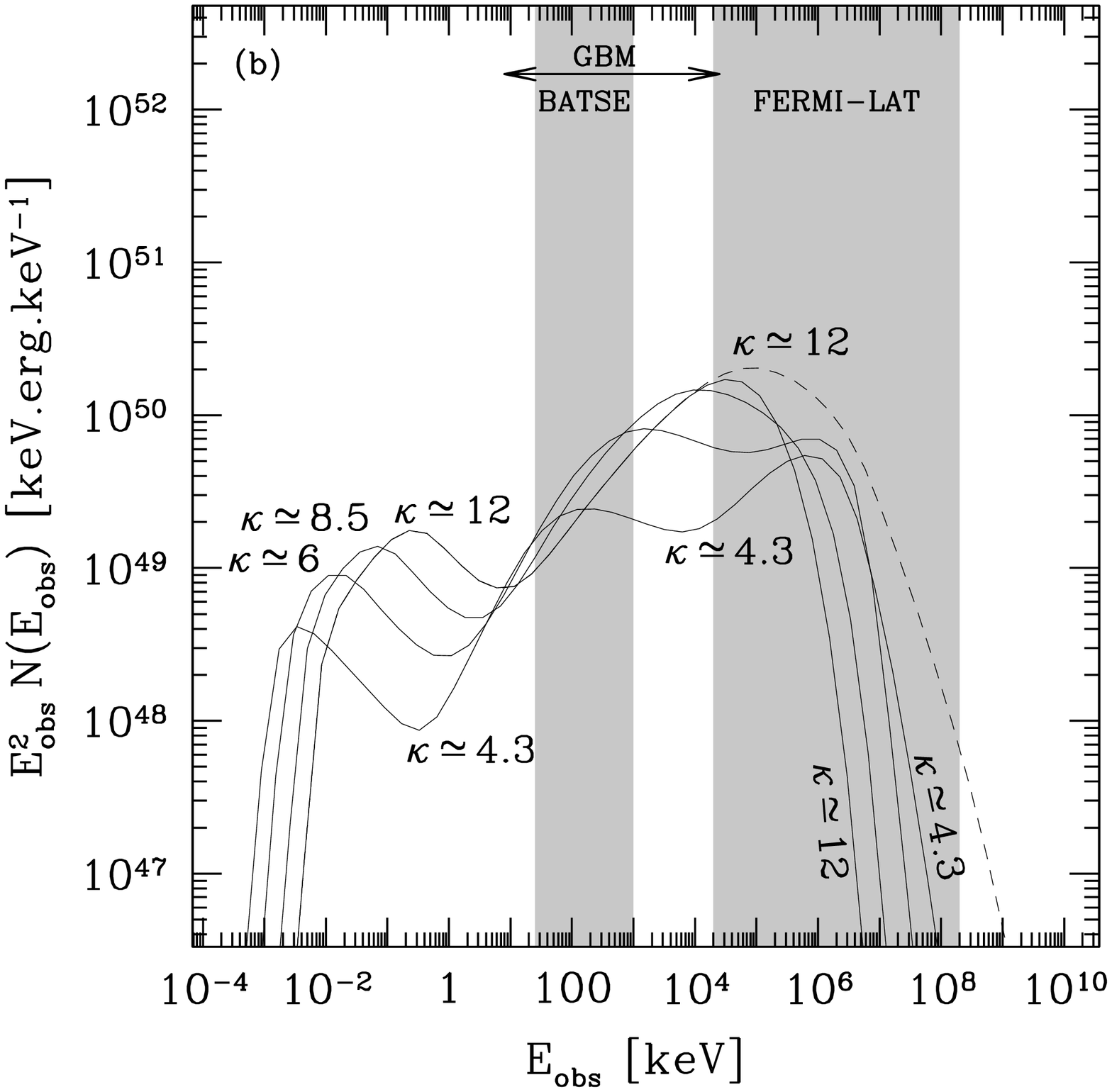} \\
\hspace*{1cm} Injected kinetic power & Variability timescale\\
\includegraphics*[height=6.25cm,viewport=0cm   0cm 20.2cm 20.2cm]{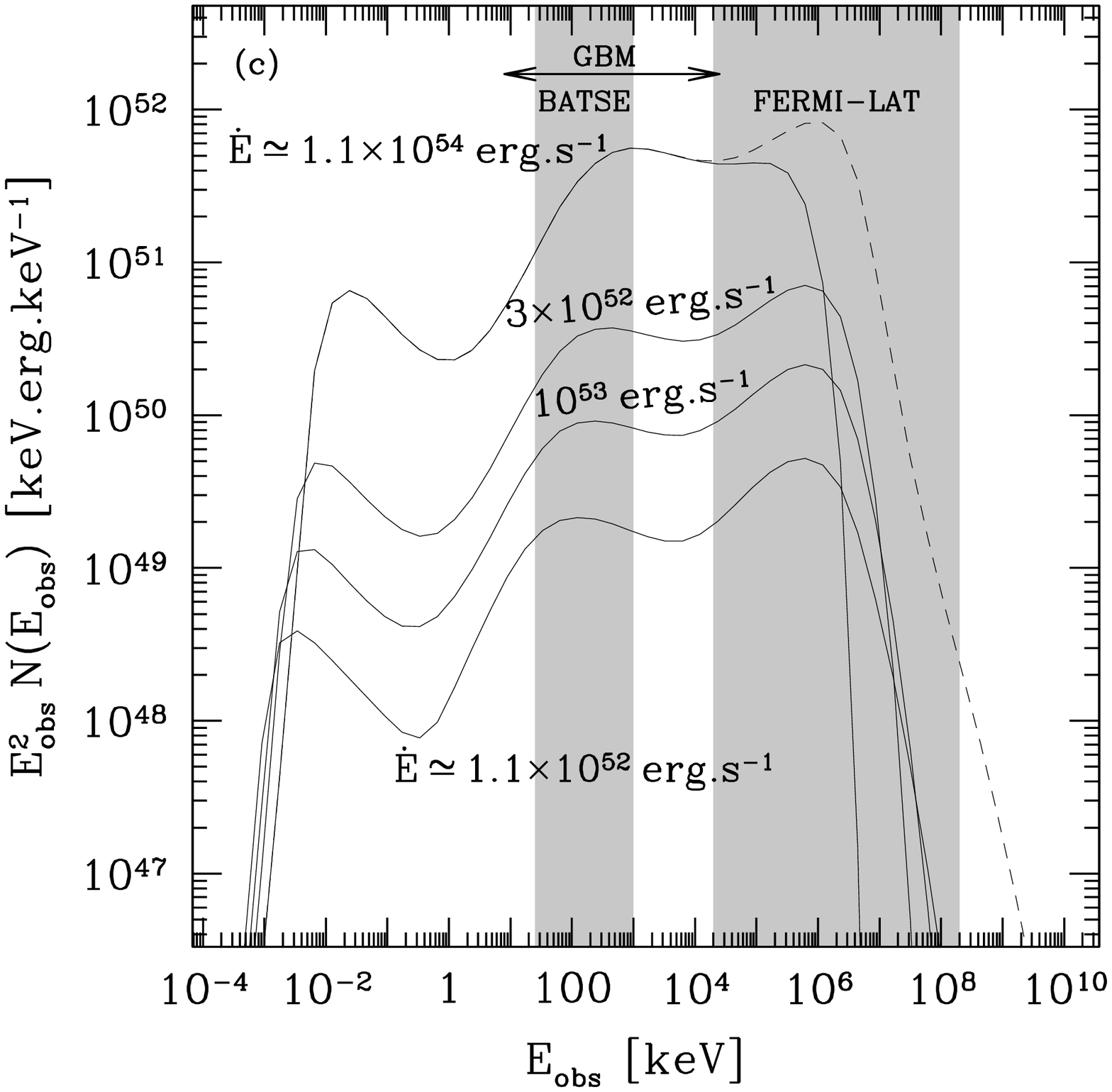} &
\includegraphics*[height=6.25cm,viewport=3.5cm 0cm 20.2cm 20.2cm]{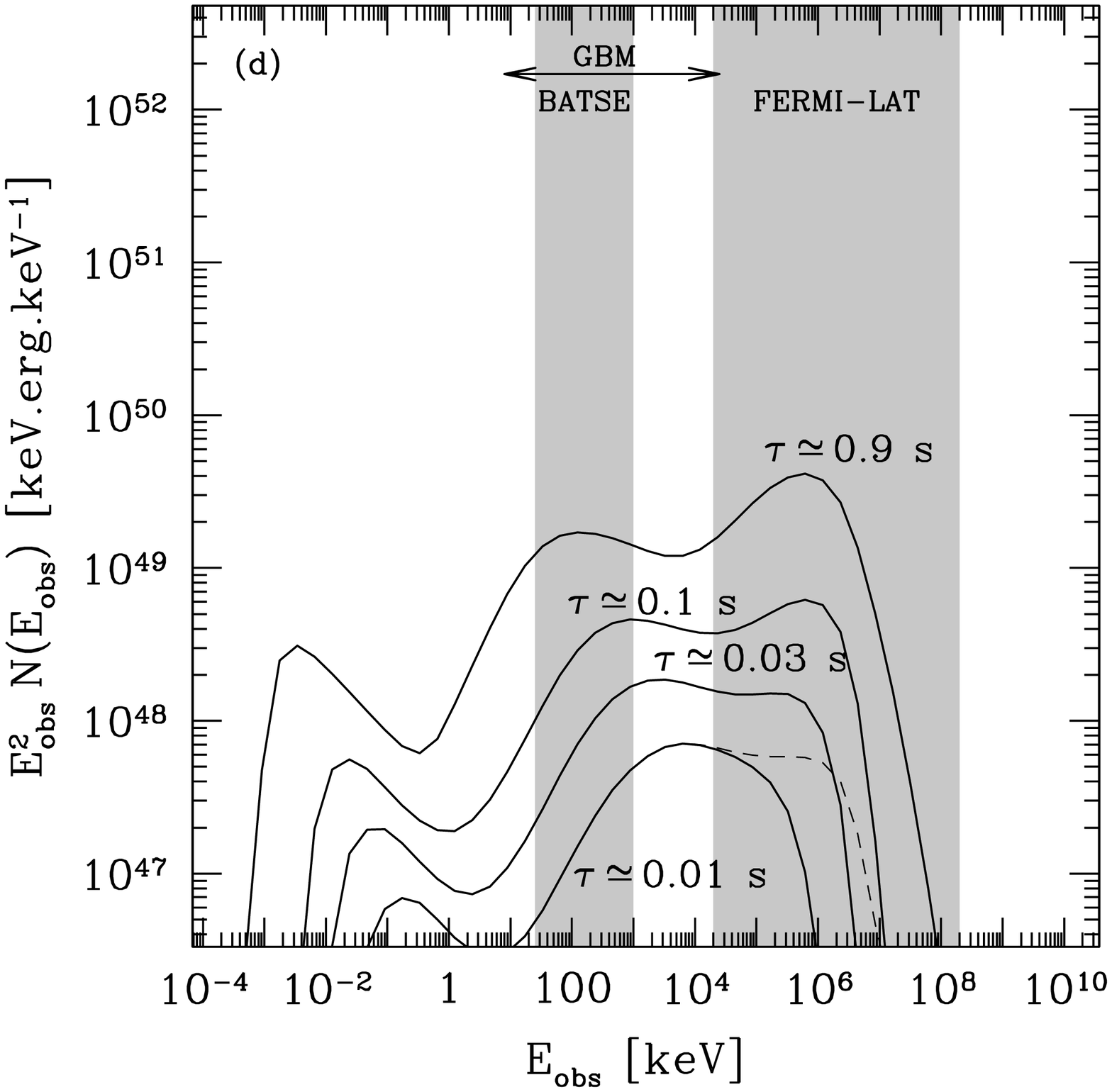} \\
\end{tabular}
\end{center}
\caption{\textbf{The effect of internal shock parameters on the emitted
 spectrum (``inverse Compton case'').} Same as in
 \reffig{fig:spec_is} with a new ``reference case'' defined by $\bar{\Gamma}=300$,
 $\kappa=4$, $\dot{E}=10^{52}\ \mathrm{erg~s^{-1}}$, $\tau=1\ \mathrm{s}$,  $\epsilon_\mathrm{B}=10^{-3}$, $\epsilon_\mathrm{e}=1/3$, $\zeta=1$ and $p=3.5$. 
 (a) effect of
 $\bar{\Gamma}$. The transparency limit is reached for
 $\bar{\Gamma}\simeq 120$ and the efficiency limit for $\bar{\Gamma}\simeq 290$ ; 
 (b) effect of $\kappa$. The efficiency limit is reached for
 $\kappa\simeq 4$ and the transparency limit for $\kappa\simeq 12$; 
(c) effect of $\dot{E}$. The efficiency limit is reached for
 $\dot{E}\simeq 1.1\times 10^{52}\ \mathrm{erg~s^{-1}}$ and the transparency limit is reached for
 $\dot{E}\simeq 1.1\times 10^{54}\ \mathrm{erg~s^{-1}}$; 
(d) effect of $\tau$. The transparency limit is reached for $\tau\simeq 0.01\ \mathrm{s}$ and the efficiency limit is reached for $\tau\simeq 0.9\ \mathrm{s}$.
In each panel, the observed spectrum obtained without including $\gamma\gamma$
 annihilation is plotted with a dashed line for the value of the parameter
 corresponding to the transparency limit.
}
\label{fig:spec_is_ic} 
\end{figure*} 
\begin{figure*}[!t]
\begin{center}
\begin{tabular}{cc}
\multicolumn{2}{c}{\textbf{``Synchrotron case''}}\\
\hspace*{1cm} High magnetic field & \hspace*{1cm} Low magnetic field\\
\includegraphics[width=0.45\textwidth]{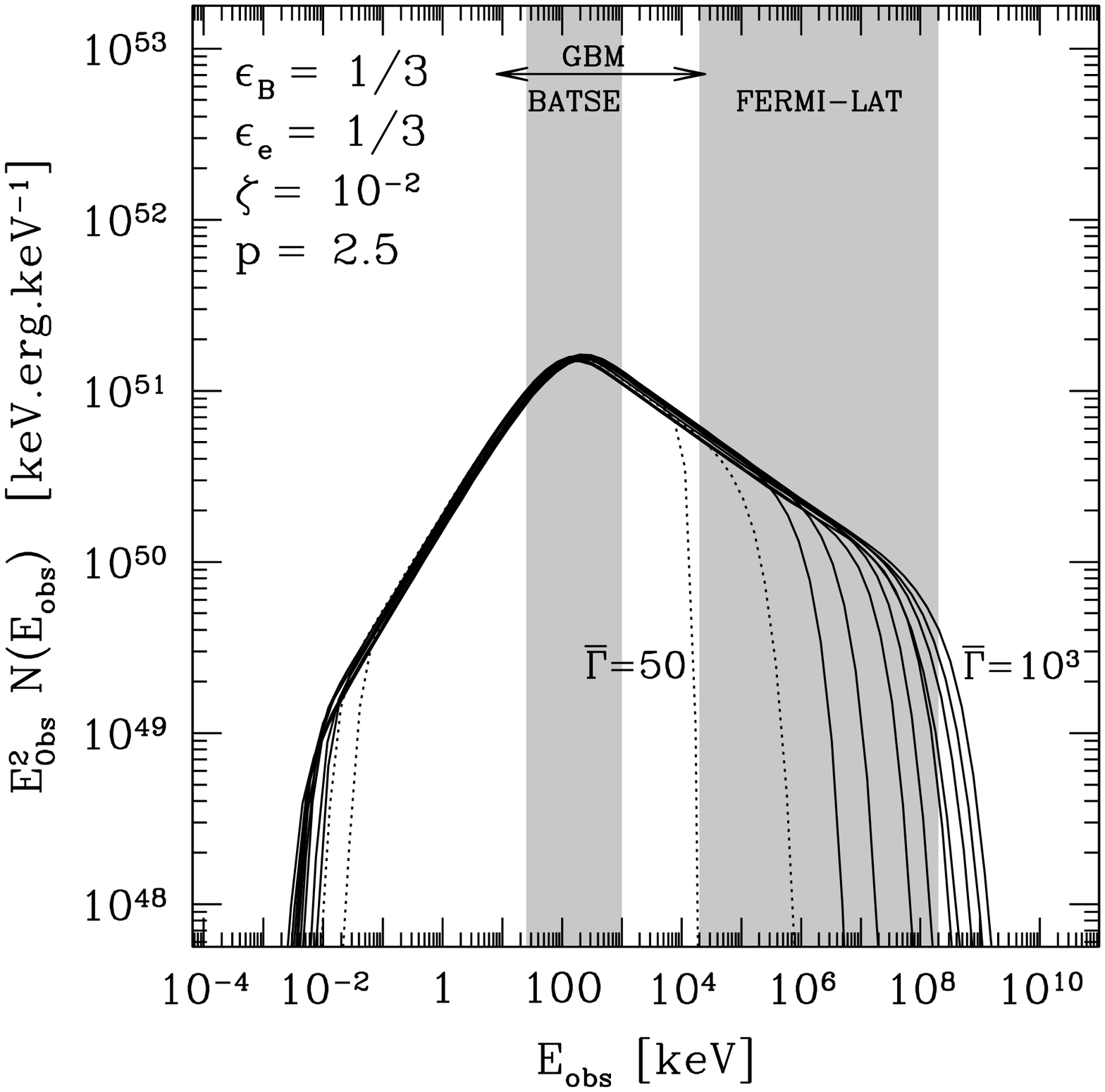} &
\includegraphics[width=0.45\textwidth]{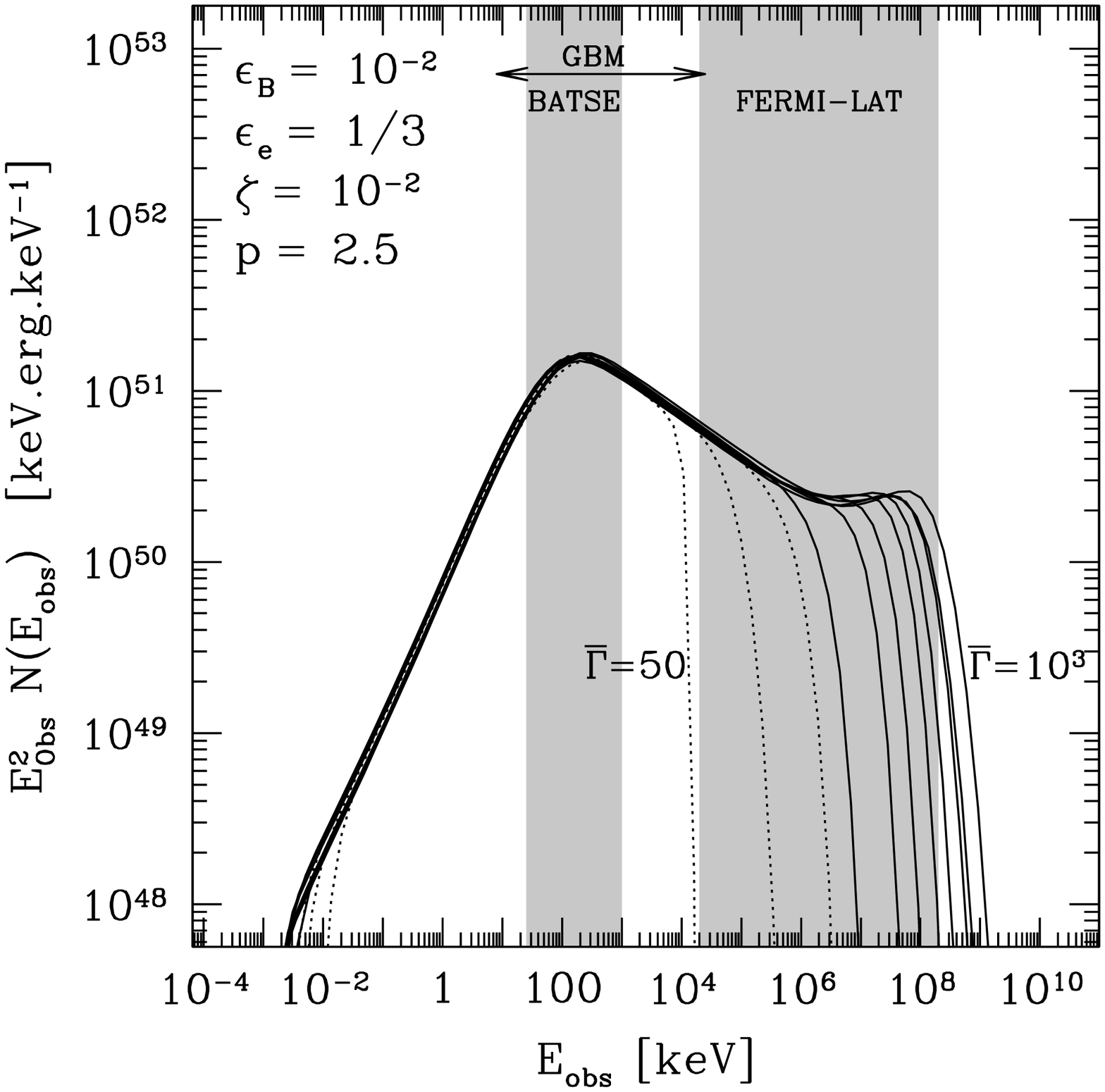} \\
\multicolumn{2}{c}{\textbf{``Inverse Compton case''}}\\
\hspace*{1cm} High magnetic field & \hspace*{1cm} Low magnetic field\\
\includegraphics[width=0.45\textwidth]{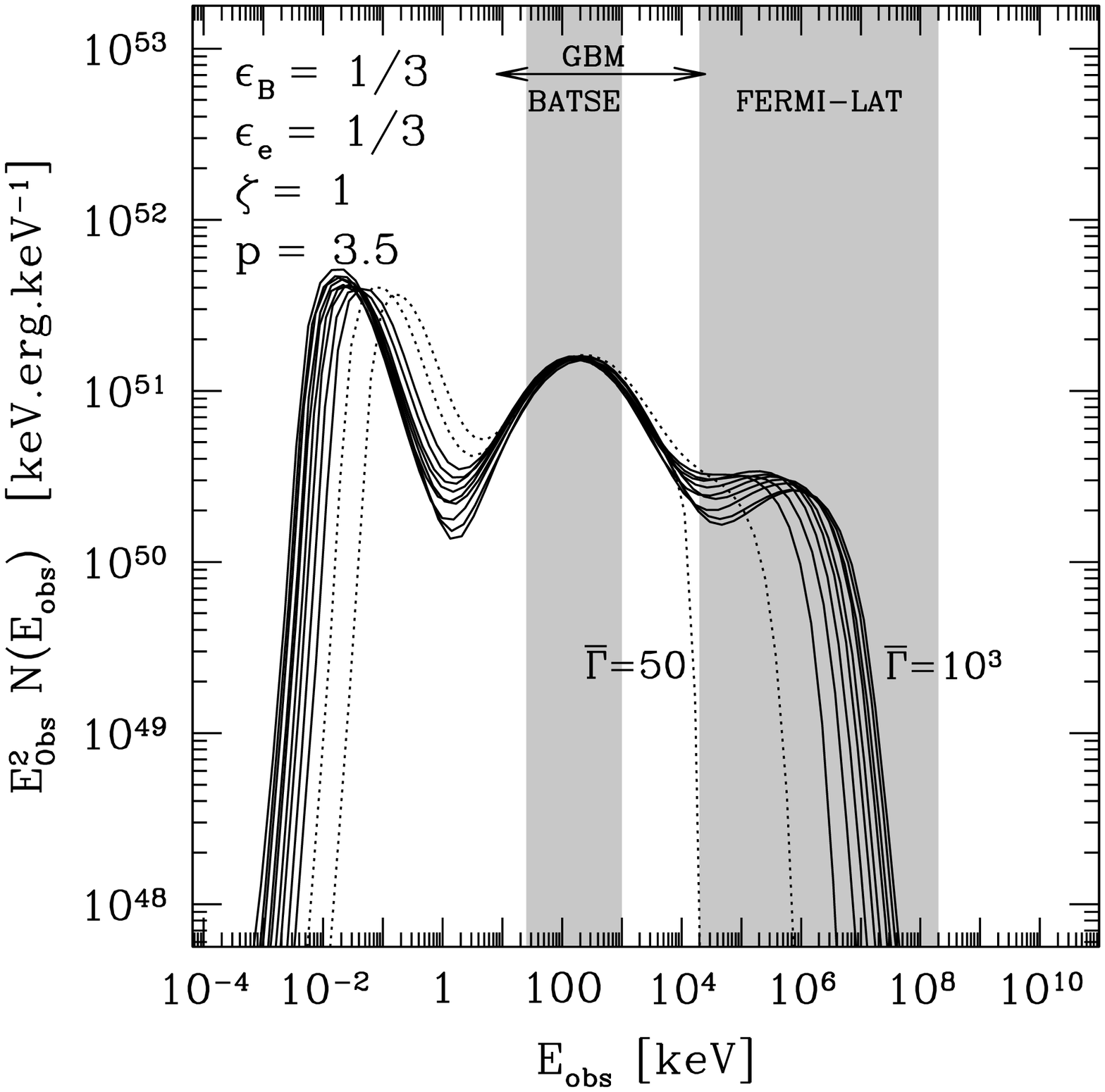} &
\includegraphics[width=0.45\textwidth]{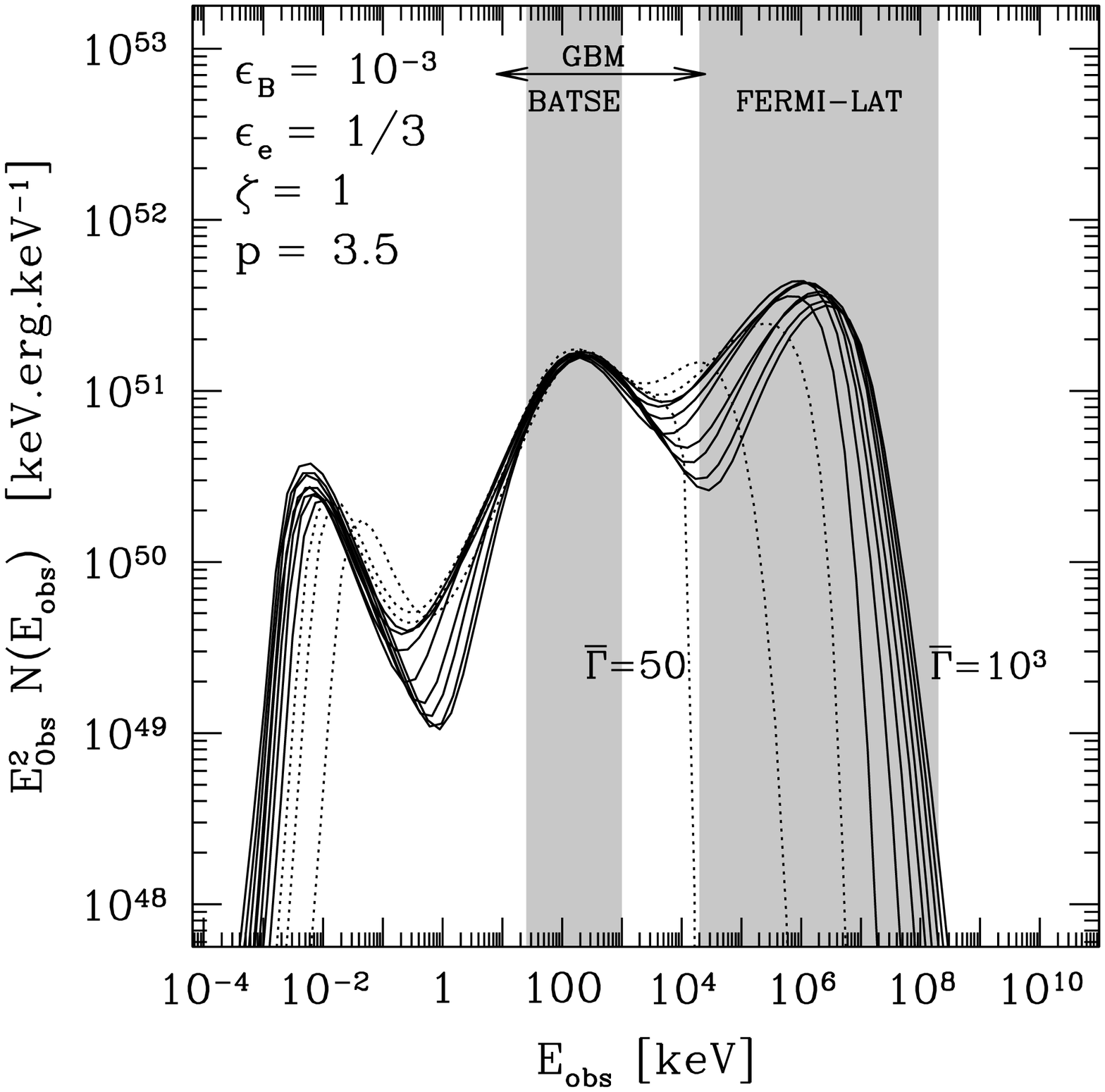} \\
\end{tabular}
\end{center}
\caption{\textbf{Physical diagnostics from \textit{Fermi} observations:
spectral shape.}
A GRB pulse at $z=1$ characterized by a peak energy $E_\mathrm{p,obs}=200\ \mathrm{keV}$,
an isotropic radiated energy in the {BATSE} range
 $E_\mathrm{25\,\mathrm{keV}-1\,\mathrm{MeV}}=5\times 10^{51}\ \mathrm{erg}$ and a duration $(1+z)\tau=2\ \mathrm{s}$ is modelled using the two-shells version of the internal shock model and different assumptions for the microphysics. \textit{Top-left:} ``synchrotron case'' with a high magnetic field ($\epsilon_\mathrm{B}=\epsilon_\mathrm{e}=1/3$, $\zeta=10^{-2}$ and $p=2.5$); \textit{Top-right:} ``synchrotron case'' with a low magnetic field ($\epsilon_\mathrm{B}=10^{-2}$, $\epsilon_\mathrm{e}=1/3$, $\zeta=10^{-2}$ and $p=2.5$); \textit{Bottom-left:} `` inverse Compton case'' with a high magnetic field ($\epsilon_\mathrm{B}=\epsilon_\mathrm{e}=1/3$, $\zeta=1$ and $p=3.5$); \textit{Bottom-right:} `` inverse Compton case'' with a low magnetic field ($\epsilon_\mathrm{B}=10^{-3}$, $\epsilon_\mathrm{e}=1/3$, $\zeta=1$ and $p=3.5$). 
In each case, the evolution of the observed spectrum is plotted for an
 increasing bulk Lorentz factor $\bar{\Gamma}$ (calculation including
 all radiative processes), while the other parameters ($\kappa$, $\dot{E}$,
 $\tau$) are adjusted to match the imposed observed quantities.  Spectra
 in dotted lines correspond to cases where pair creation should not be neglected ($\tau_\mathrm{T}^\mathrm{tot}>0.1$).}
\label{fig:diag}
\end{figure*}\begin{figure*}[!t]
\begin{center}
\begin{tabular}{cccc}
\multicolumn{2}{c}{\hspace*{0.5cm} \textbf{``Synchrotron case''}} &
\multicolumn{2}{c}{\hspace*{0.5cm} \textbf{``Inverse Compton case''}}\\
\hspace*{1cm} High magnetic field &
Low magnetic field  &
\hspace*{1cm} High magnetic field &
Low magnetic field  \\
\includegraphics*[width=0.263\textwidth,viewport=0cm 0cm 20.2cm 20.2cm]{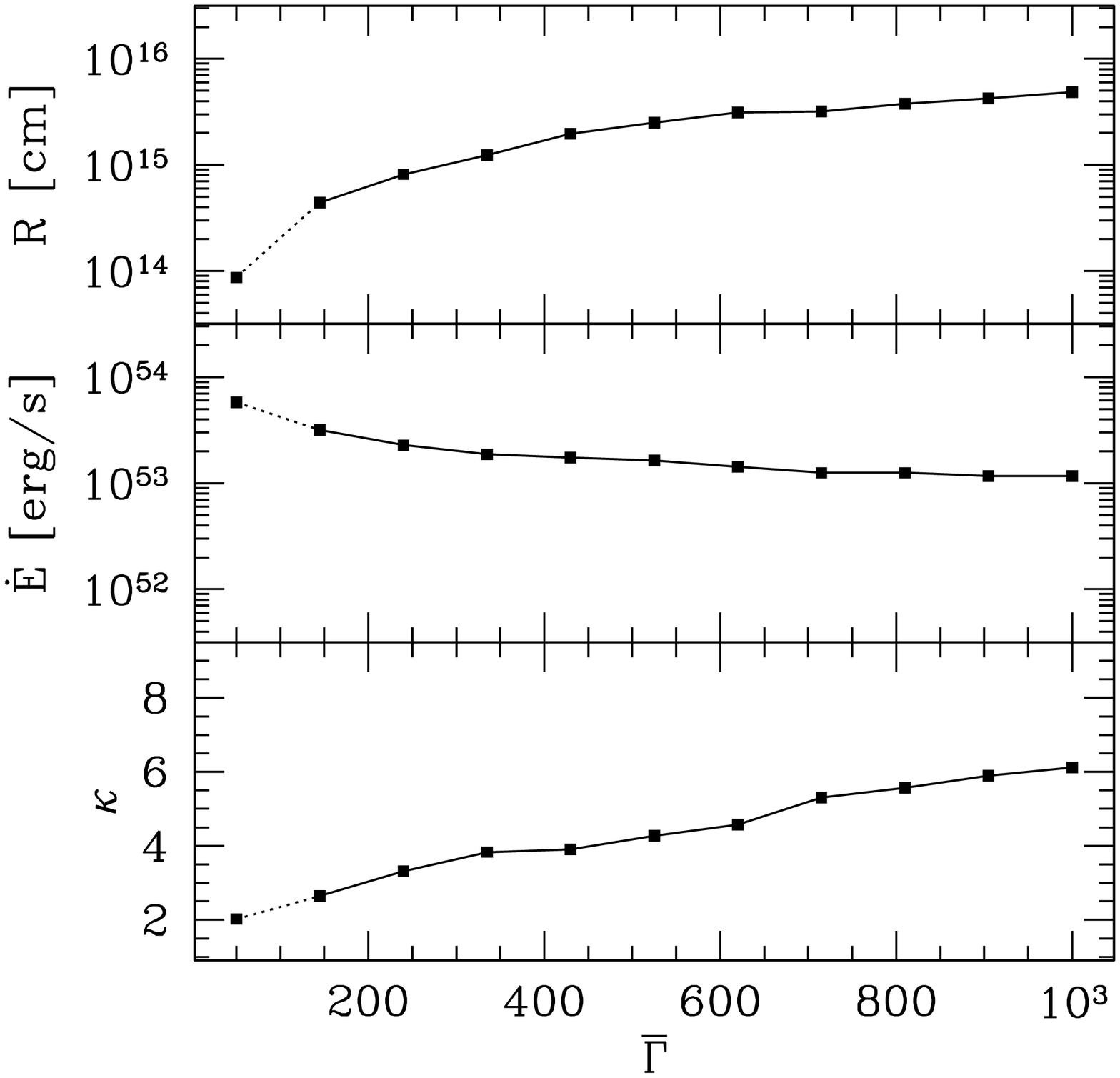} & 
\includegraphics*[width=0.217\textwidth,viewport=3.5cm 0cm 20.2cm 20.2cm]{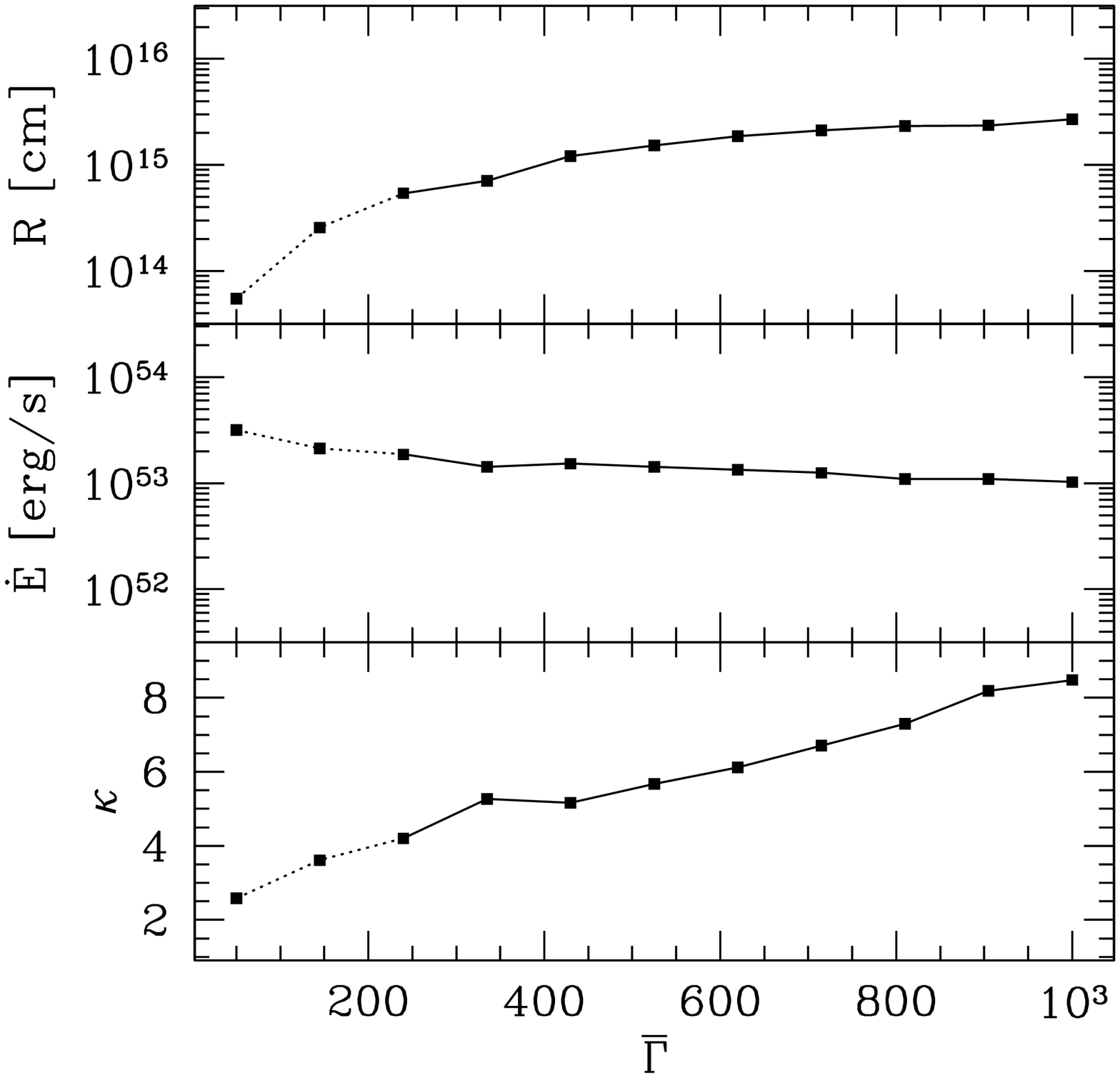} &
\includegraphics*[width=0.263\textwidth,viewport=0cm 0cm 20.2cm 20.2cm]{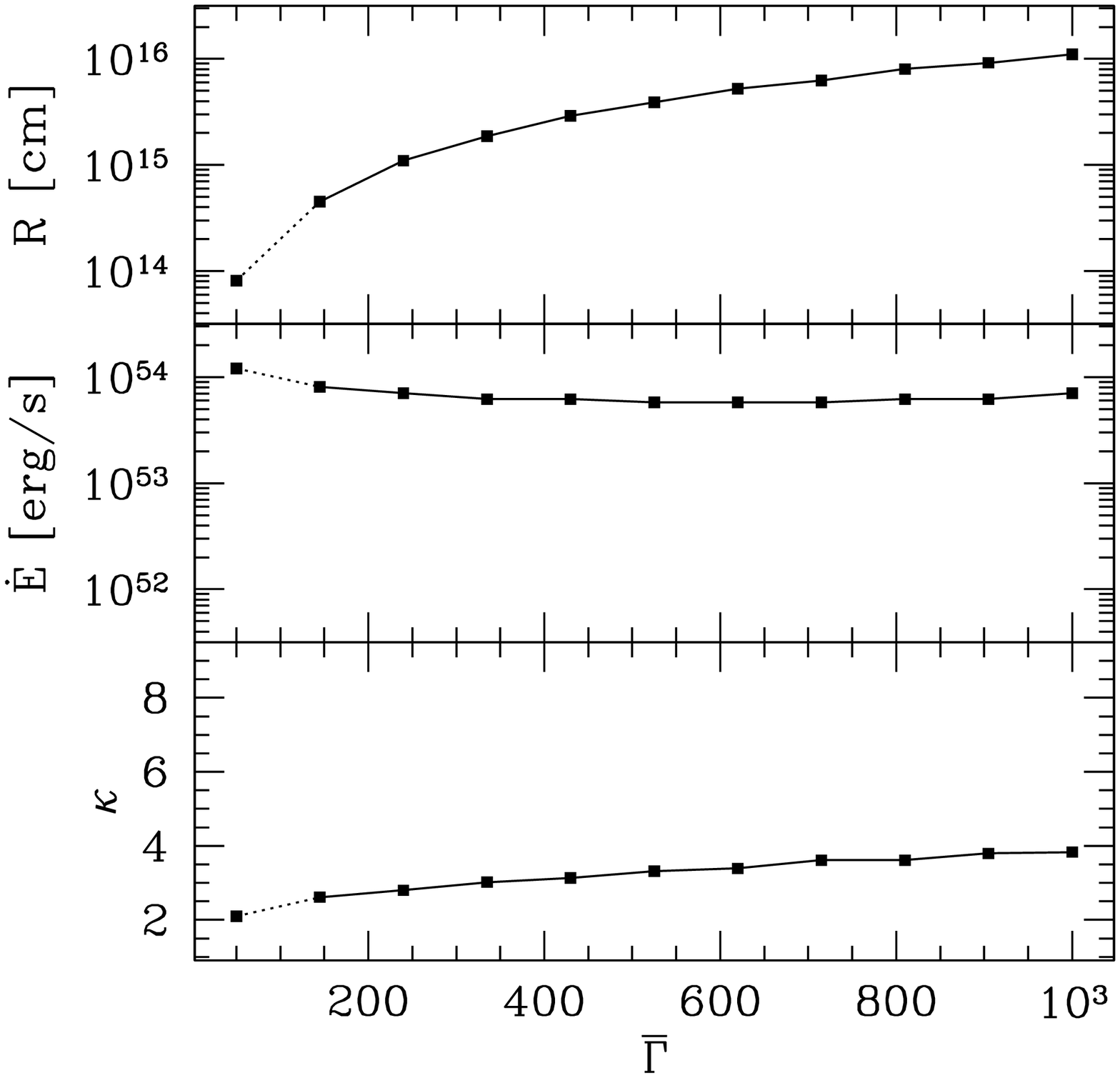} &
\includegraphics*[width=0.217\textwidth,viewport=3.5cm 0cm 20.2cm 20.2cm]{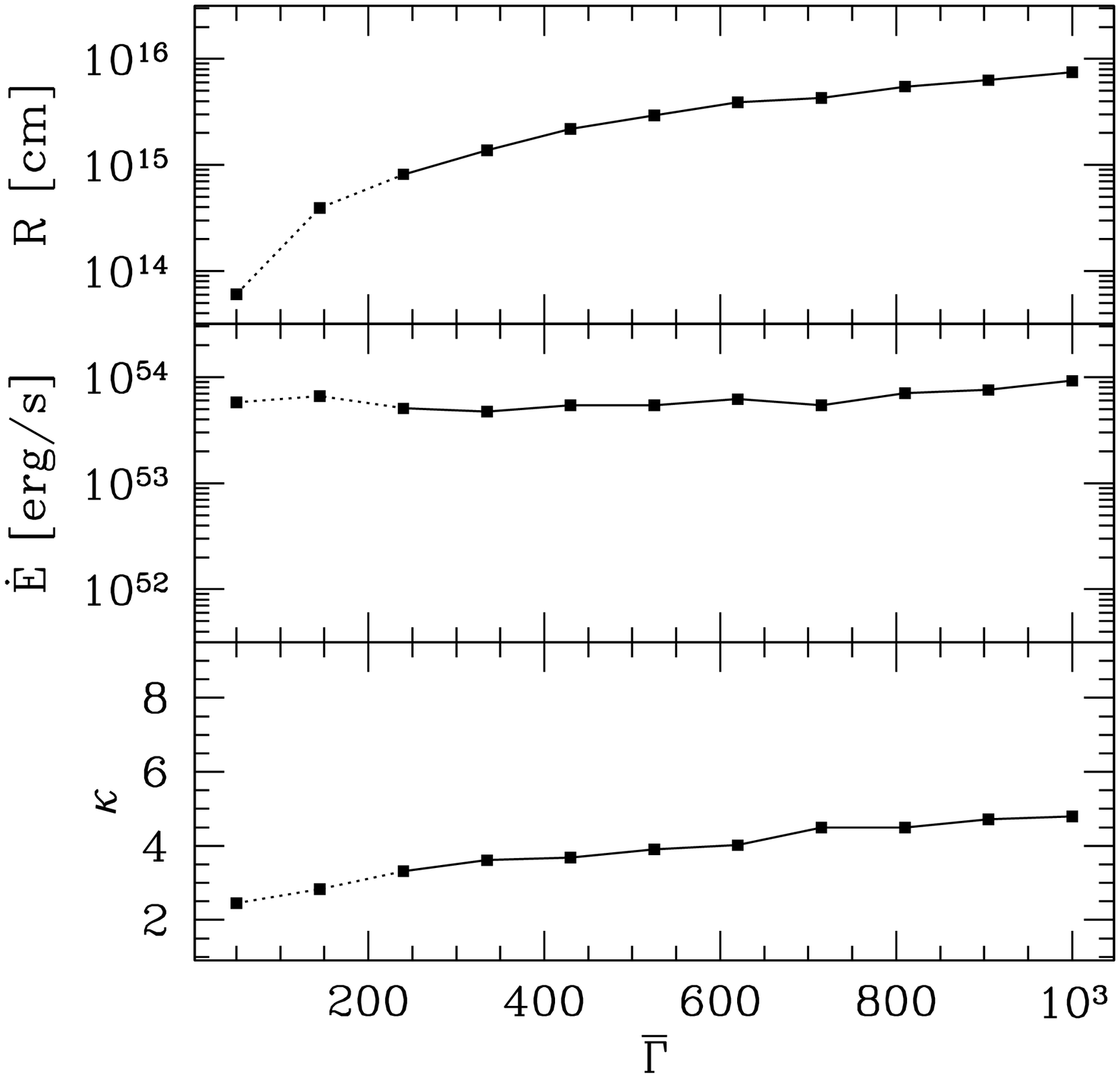} \\
\end{tabular}
\end{center}
\caption{\textbf{Physical diagnostics from \textit{Fermi} observations:
 properties of the relativistic outflow.}
For each case considered in \reffig{fig:diag},
we plotted
 the values of the injected kinetic power $\dot{E}$ and the contrast
 $\kappa$ that have been adjusted for each bulk Lorentz factor
 $\bar{\Gamma}$ to match the imposed observed quantities (redshift
 $z=1$, peak energy
 $E_\mathrm{p,obs}=200\ \mathrm{keV}$, isotropic
 radiated energy $E_\mathrm{25\ keV-1\ MeV}=5\times 10^{51}\ \mathrm{erg}$ and duration $(1+z)\tau=2\ \mathrm{s}$). This adjustement is done within 10\%
 uncertainty, which explains why the curves are not perfectly smooth.
 The corresponding typical internal shock radius $R_\mathrm{is}$ is also
 shown. The dotted lines correspond to cases where pair creation should not be neglected ($\tau_\mathrm{T}^\mathrm{tot}>0.1$).
}
\label{fig:diag_param}
\end{figure*}

\paragraph{``Inverse Compton case''.}\label{sec:iccase} The synchrotron component
peaks at low energy and the inverse Compton component peaks in the {BATSE}
range (keV-MeV). This case is usually called ``Synchrotron
      Self-Compton'' in the literature, and emerges naturally in the often considered
situation where 
all electrons are accelerated ($\zeta=1$). It has been shown by
\citet{panaitescu:00,stern:04} that it can reproduce the steep low-energy
spectral slopes observed in the {BATSE} range. Because of the bright
synchrotron component at low energy, possibly in the optical range, this case has recently been
proposed to explain the prompt emission of the ``naked eye burst'' GRB
080319b \citep{racusin:08,kumar:08b} and more generally of bursts with
a bright prompt optical emission \citep{panaitescu:08}. The
``inverse Compton'' case is characterized by a low magnetic field 
and a
high fraction of accelerated electrons. In addition, having a well defined
IC peak in the {BATSE} range requires a steep
slope for the electron distribution, $p>3$ \citep[see
also][]{peer:04}. While, the results obtained
in the ``synchrotron case'' are very similar for all slopes $p>2$,
we
find a large difference in the ``inverse Compton case'' between spectra
computed assuming a slope $2<p<3$ or assuming a slope $p>3$. It is only
for $p>3$ that the first inverse Compton peak is well defined (the
scatterings by electrons above $\Gamma_\mathrm{p}$ become negligible). This is
illustrated in \reffig{fig:ic_slope}.\\

 \reffig{fig:spec_is_ic}
shows the same parameter study as in \reffig{fig:spec_is}, but for
a reference set of parameters corresponding to the ``inverse Compton
case''. Note that the allowed range for each parameter $\bar{\Gamma}$, $\kappa$,
$\dot{E}$ and $\tau$ is usually more limited than in the ``synchrotron
case'', especially due to the requirement of a high radiative efficiency
($u_\mathrm{rad}/u_\mathrm{e}^\mathrm{acc}>0.5$), as a lower magnetic
field leads to longer synchrotron timescales.
In the ``inverse Compton case'', most scatterings occur in Thomson regime, due to a low magnetic
field and a low minimum electron Lorentz factor leading to
$\Gamma_\mathrm{p}\ll \Gamma_\mathrm{KN}$ (see \refeq{eq:gkn}). The
condition necessary to have the possibility of a second scattering in
Thomson regime is 
$\Gamma_\mathrm{p} \epsilon_\mathrm{p,ic} \ll 1$, i.e.
\begin{equation}
\Gamma_\mathrm{p} \ll \Gamma_\mathrm{KN}^\mathrm{(2^{nd})} \simeq 200 \left(\frac{B'}{100\ \mathrm{G}}\right)^{-1/5}\ .
\end{equation}  
For most parameters in the ``inverse Compton case'' this condition is
fulfilled and efficient second scatterings occur, leading to a 
second inverse Compton component at high energy (\textit{Fermi}
range). The first inverse Compton component is never affected by
$\gamma\gamma$ annihilation. Therefore the spectra differ again mainly
by their high-energy component, i.e. by the intensity of the second inverse Compton
component. This intensity depends on whether most second
scatterings occur in Thomson regime or are affected by Klein-Nishina
corrections, and also on the strength of the attenuation due
to $\gamma\gamma$ annihilation. 
As long as
Klein-Nishina corrections and $\gamma\gamma$ absorption are not too
strong at very high energy, the
synchrotron, first and second inverse Compton components have relative
intensities $1:Y:Y^{2}$, where $Y$ has to be large to avoid that most of
the energy is radiated in the synchrotron component in the sub-keV
range. Therefore, it is difficult to avoid that most of the
energy is radiated in the MeV-GeV range. The isotropic equivalent
 radiated energy 
 in the {BATSE} range is typically $E_\mathrm{rad,BATSE}\sim
10^{51}-10^{54}\ \mathrm{erg}$. If the Compton parameter is $Y\sim 10$ or
more, the resulting total radiated energy is greater than
$E_\mathrm{rad}\sim 10^{52}-10^{55}\ \mathrm{erg}$.
This can lead to a crisis for
the GRB energy budget and is another reason to disfavor the ``inverse
Compton case'' as pointed out recently by \citet{piran:08}. High
magnetic field can lead to smaller values of $Y$ but most of the energy
is radiated in the synchrotron component in this case. Having the
first inverse Compton peak dominant requires to fine-tune
$\epsilon_\mathrm{B}$. Moreover the peak energy of the first inverse
Compton component has a stronger dependence on the variations of the
physical conditions in the shocked regions (compare Eqs.~(\ref{eq:epsyn})
and~(\ref{eq:epic})). Thus the ``inverse Compton case'' also predicts a
faster spectral evolution during GRB pulses than in the ``synchrotron case'' and
is therefore disfavored by the  observed pulse shape
and spectral evolution in {BATSE} bursts \citep{daigne:98,daigne:03}.

\subsection{Physical diagnostics from \textit{Fermi} observations.}
\label{sec:FermiDiag}
As can be seen from this study, the high-energy emission component is
shaped by several physical parameters of the internal shock model. It is
therefore difficult to identify simple diagnostics that could be applied
to forthcoming \textit{Fermi} data. It is only a detailed spectral
fitting covering a broad spectral range that will allow us
to measure fundamental quantities which are still largely unknown for
GRBs (e.g. the radius and the Lorentz factor of the emitting
material, the typical Lorentz factor of radiating electrons or the
magnetic field in the shocked region).\\

\paragraph{Diagnosing the dominant radiative process and the physical
    conditions in the shocked region.}
As seen in \reffig{fig:diag}, one can distinguish between
the ``synchrotron case'' and the ''inverse Compton case'' from the
spectral shape and then identify the dominant
radiative process. This requires however a broad spectral range, like
the one  available with {GBM}+{LAT}. 
More precise informations about the
physical conditions in the shocked region can be obtained from such observations using the following
procedure: (a) assume microphysics parameters (the initial choice is
suggested by
 the general spectral shape, for instance a  
high $\epsilon_\mathrm{B}$ and a low fraction
$\zeta$ if the ``synchrotron case'' without bright IC component at high
energy is favored); (b) estimate $\tau$ from the observed lightcurve; (c) vary
$\bar{\Gamma}$ and for each $\bar{\Gamma}$ adjust $\kappa$ and $\dot{E}$
to match the correct observed peak energy and fluence in the
{BATSE} range; (d) determine $\bar{\Gamma}$ from the spectral
shape at high energy. This procedure is illustrated in
Figs.~\ref{fig:diag} and~\ref{fig:diag_param}. Good quality high energy
observations can help in reducing the 
uncertainty related to the assumptions made for $\epsilon_\mathrm{B}$ and
$\zeta$. \\
\begin{figure}[b!]
\centerline{\includegraphics[width=\linewidth]{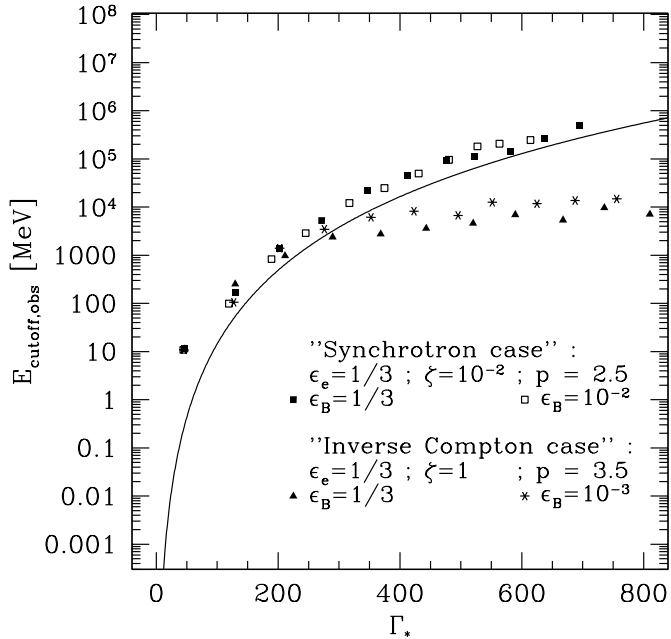}}
\caption{\textbf{Measuring the Lorentz factor of the emitting region
 from the observed cutoff energy.} The cutoff energy
 $E_\mathrm{cut,obs}$ due to $\gamma\gamma$ annihilation is plotted as a
 function of the Lorentz factor $\Gamma_{*}$ of the emitting region, as
 predicted from \refeq{eq:cutoff}, assuming $z=1$,
 $E_\mathrm{rad,iso}=5\times 10^{51}\ \mathrm{erg}$,
 $E_\mathrm{p,obs}=200\ \mathrm{keV}$ and $\beta=2.3$. The cutoff
 energies measured in all the spectra shown in \reffig{fig:diag}  are indicated
 with four different
 symbols corresponding to the four different assumptions regarding the
 microphysics parameters. 
}
\label{fig:cutoff}
\end{figure}

\paragraph{Measuring the Lorentz factor of the outflow.}
\label{sec:cutoff}
It has been proposed by several authors to measure the Lorentz factor of
the relativistic outflow from the position of the cutoff in the high
energy spectrum due to $\gamma\gamma$ annihilation \citep[see
e.g.][]{baring:97,lithwick:01,baring:06,gupta:08,murase:08}. It is assumed that three quantities have been measured: the isotropic equivalent radiated energy
$E_\mathrm{rad,iso}$, the peak energy $E_\mathrm{p,obs}$ and the typical
pulse duration $\tau_\mathrm{obs}=(1+z)\tau$. From \refsec{eq:isradius},
the radius where the emission takes place is related to the unknown
Lorentz factor of the emitting region $\Gamma_{*}$ by $R=
2\kappa^{2}/(\kappa^{2}-1)\Gamma_{*}^{2}c\tau \simeq 2 \Gamma_{*}^{2} c
\tau$ (the difference is less than 10\% for $\kappa>3$). The 
comoving photon density can then be estimated as a function of $\Gamma_{*}$\,:
\begin{equation}
n'\left(E'\right) \simeq \frac{\Gamma_{*} E_\mathrm{rad,iso}}{4\pi R^{2} \left(1+z\right)^{2} E_\mathrm{p,obs}^2\Delta '}\left(\frac{\Gamma_{*}E'}{(1+z)E_\mathrm{p,obs}}\right)^{-\beta}
\ ,
\label{eq:neprime}
\end{equation}
where $-\beta$ is the observed slope of the photon
spectrum in the {BATSE} range and
 $\Delta'$ is the comoving width of the emitting region. Using the
Dirac approximation for the $\gamma\gamma$ annihilation cross section
\citep{gould:67}, one gets the following expression for the optical depth at
observed energy $E_\mathrm{obs}$\,:
\begin{equation}
\tau_\mathrm{\gamma\gamma}\left(E_\mathrm{obs}\right)\simeq
\sigma_\mathrm{T}\Delta'\frac{\Gamma_{*}\left(m_\mathrm{e}c^{2}\right)^{2}}{(1+z)E_\mathrm{obs}} n'\left(E'=\frac{\Gamma_{*}\left(m_\mathrm{e}c^{2}\right)^{2}}{(1+z)E_\mathrm{obs}}\right)\ .
\end{equation}
The cutoff energy can be estimated from the condition
$\tau_\mathrm{\gamma\gamma}\left(E_\mathrm{cut,obs}\right)\sim 1$, which
gives
\begin{eqnarray}
E_\mathrm{cut,obs} & \simeq & \frac{m_\mathrm{e}c^{2}}{1+z} 
\left(\frac{(1+z)^{2}\sigma_\mathrm{T}E_\mathrm{rad,iso}}{16\pi m_\mathrm{e}c^{4}\tau_\mathrm{obs}^{2}}\right)^{-\frac{1}{\beta-1}}
\Gamma_{*}^{2\frac{\beta+1}{\beta-1}}
\left(\frac{(1+z)E_\mathrm{p,obs}}{m_\mathrm{e}c^{2}}\right)^{-\frac{\beta-2}{\beta-1}}\ .\nonumber\\
\label{eq:cutoff}
\end{eqnarray}
This estimate of the observed cutoff energy $E_\mathrm{cut,off}$ is
plotted as a function of the Lorentz factor $\Gamma_{*}$ of the emitting
material in \reffig{fig:cutoff} for the same observed quantities
as in \reffig{fig:diag}, i.e. $z=1$, $E_\mathrm{rad,iso}=5\times
10^{51}\ \mathrm{erg}$, $E_\mathrm{p,obs}=200\ \mathrm{keV}$ and
$\tau_\mathrm{obs}=2\ \mathrm{s}$, and assuming a slope $\beta=2.3$
(note that -- for most of the values of the cutoff energy $E_\mathrm{cut,obs}$ plotted in
\reffig{fig:cutoff} -- the typical energy $\sim
\Gamma_{*}^{2}\left(m_\mathrm{e}c^{2}\right)^{2}/(1+z)^{2}/E_\mathrm{cut,obs}$
of low-energy photons that annihilate preferentially with photons at
energy $E_\mathrm{cut,obs}$ is above $E_\mathrm{p,obs}$, which justifies
our choice of $\beta$). On
the same figure the measured values of $E_\mathrm{cut,obs}$ are plotted
for all spectra shown in \reffig{fig:diag}, i.e. for three
different cases regarding the microphysics. This cutoff energy is
measured as the energy where the slope of the photon spectrum falls
below -3. It appears that the precise calculation agrees well with the
approximate expression given by \refeq{eq:cutoff} in the ``synchrotron
case'' (except for a normalization factor of about $\sim 3$, which could
be improved with a more accurate description of the low energy spectrum
in equation~\ref{eq:neprime}) but that there is a
large discrepancy in the ``inverse Compton case'' where
\refeq{eq:cutoff} overestimates $E_\mathrm{cut,obs}$. The $\gamma\gamma$
annihilation in this case is indeed negligible and the cutoff observed
at high energy is due to the limitation of inverse Compton scatterings
by Klein-Nishina corrections.\\
\begin{figure*}[t!]
\begin{center}
\begin{tabular}{cc}
\includegraphics[width=0.45\textwidth]{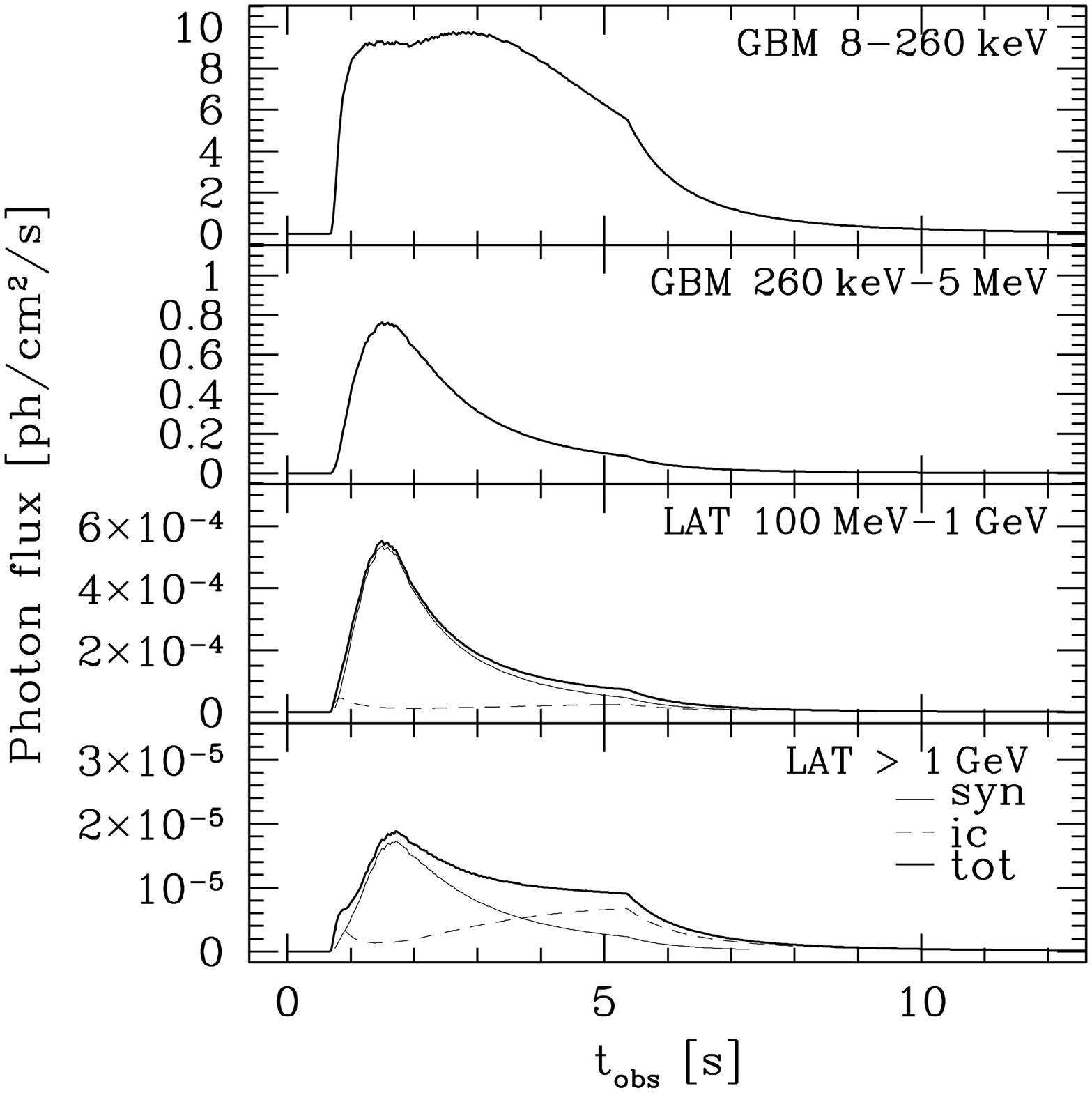} &
\includegraphics[width=0.45\textwidth]{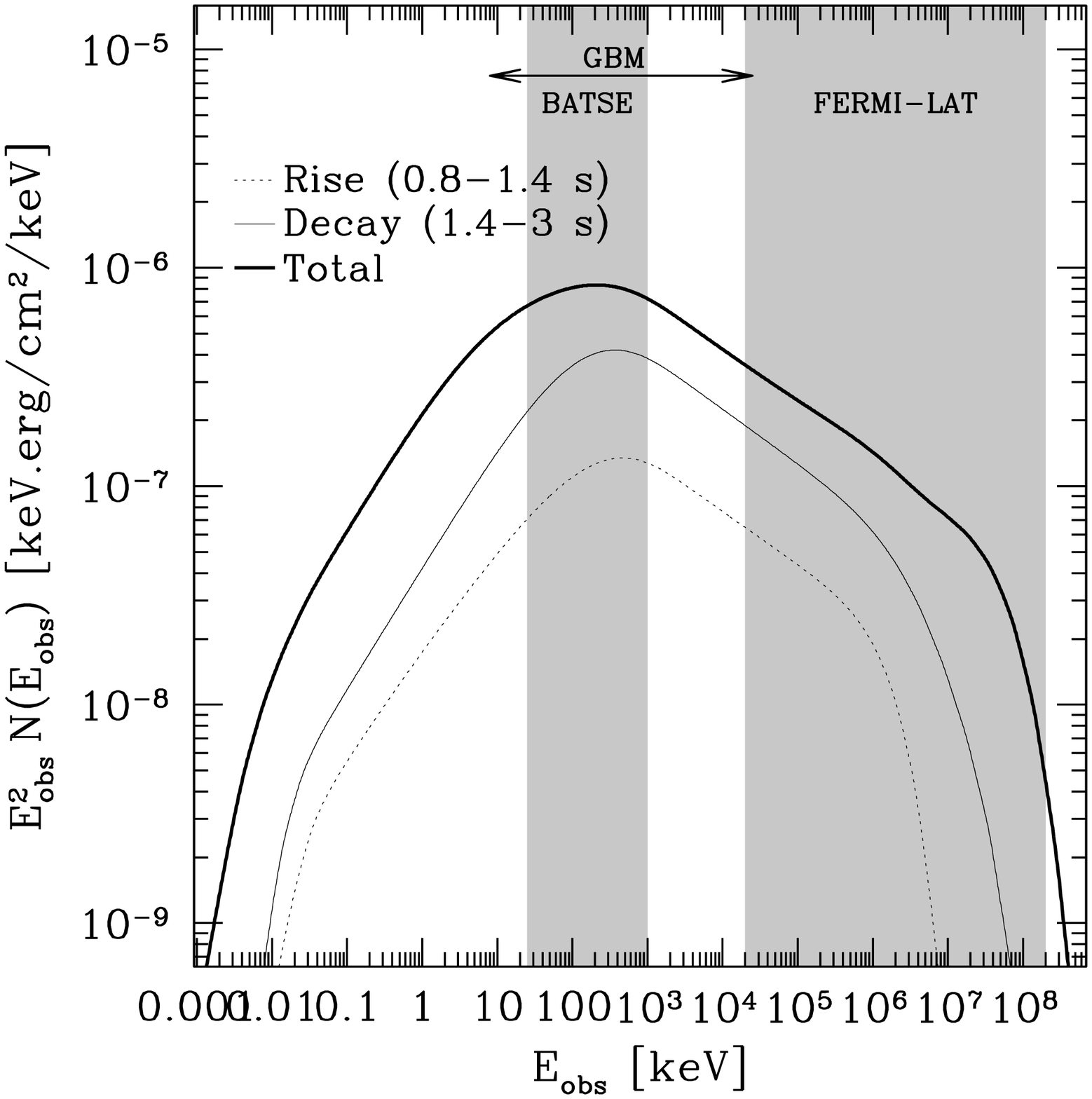}\\
\end{tabular}
\end{center}
\caption{\textbf{A single pulse burst in the ``synchrotron case'' with a
 high magnetic field.} The dynamics is the same as in
 \reffig{fig:exampledyn} except for $\dot{E}=5\times 10^{53}\ \mathrm{erg~s^{-1}}$. The assumed microphysics parameters are
 $\epsilon_\mathrm{B}=\epsilon_\mathrm{e}=1/3$, $\zeta=3\times 10^{-3}$
 and $p=2.5$. 
All radiative processes are included in the calculation. \textit{Left:} observed lightcurves in \textit{Fermi}-{GBM}
 and {LAT} range. 
The synchrotron (thin solid line) and inverse Compton
 (thin dashed line) components are also shown.
\textit{Right:} observed time-integrated spectrum during the
 rise, the early decay and the whole duration of the pulse.}
\label{fig:synHB}
\end{figure*}
\begin{figure*}[t!]
\begin{center}
\begin{tabular}{cc}
\includegraphics[width=0.45\textwidth]{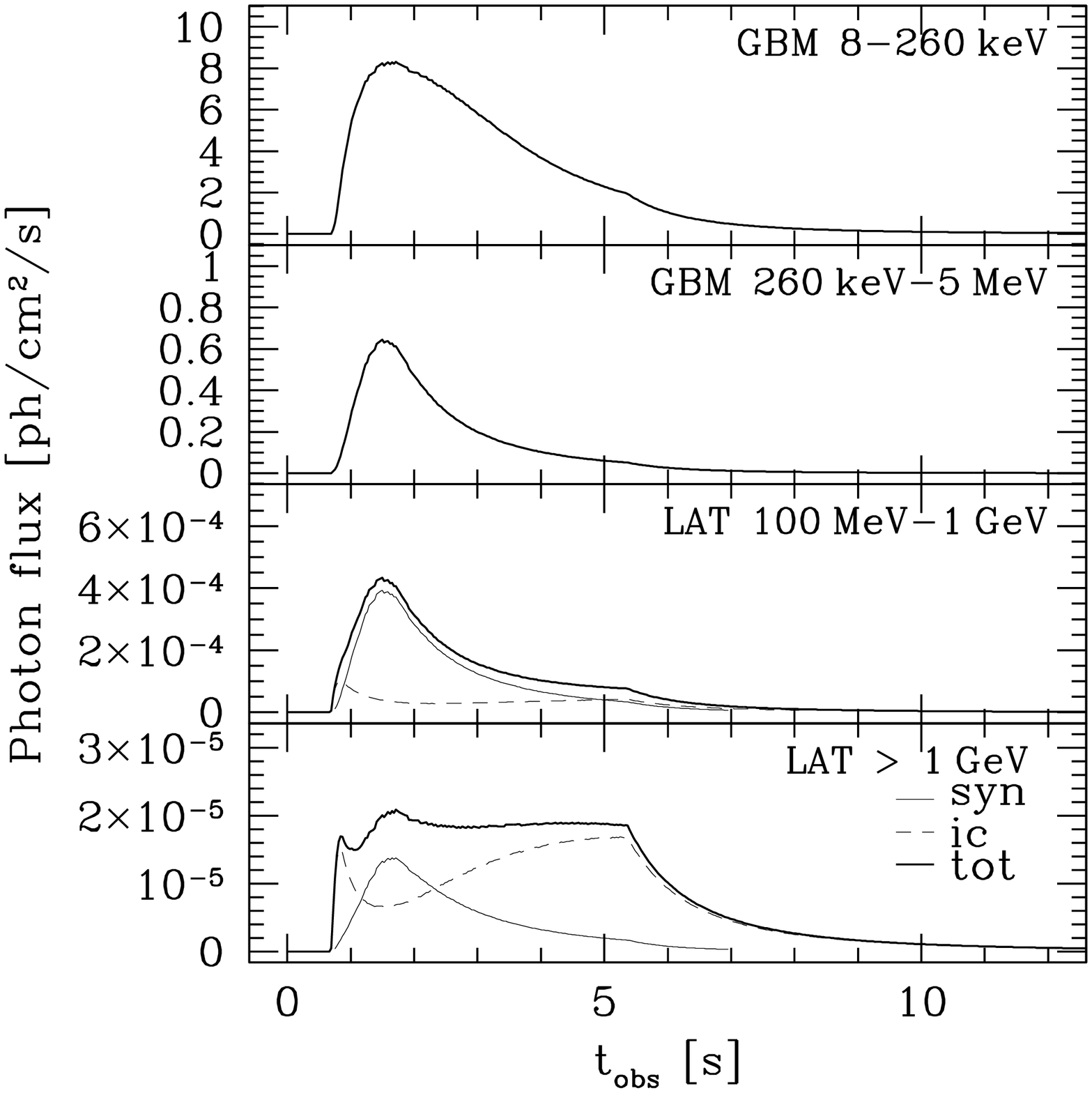} &
\includegraphics[width=0.45\textwidth]{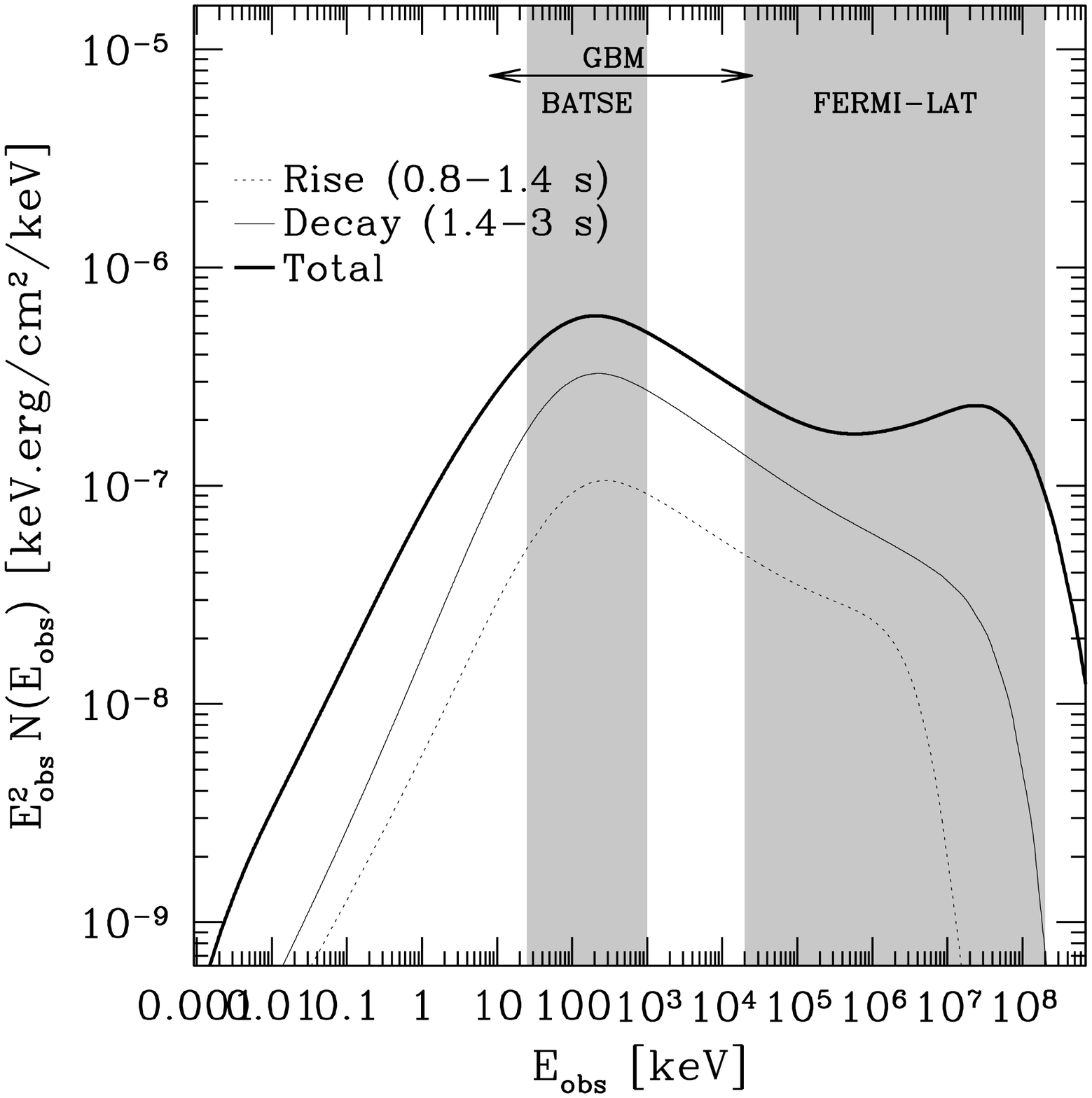}\\
\end{tabular}
\end{center}
\caption{\textbf{A single pulse burst in the ``synchrotron case'' with a
 low magnetic field.} Same as in \reffig{fig:synHB} except for the
 microphysics parameters: 
 $\epsilon_\mathrm{B}=5\times 10^{-3}$, $\epsilon_\mathrm{e}=1/3$,
 $\zeta=2\times 10^{-3}$
 and $p=2.5$.}
\label{fig:synLB}
\end{figure*}

\begin{figure*}[!t]
\begin{center}
\begin{tabular}{cc}
\includegraphics[width=0.45\textwidth]{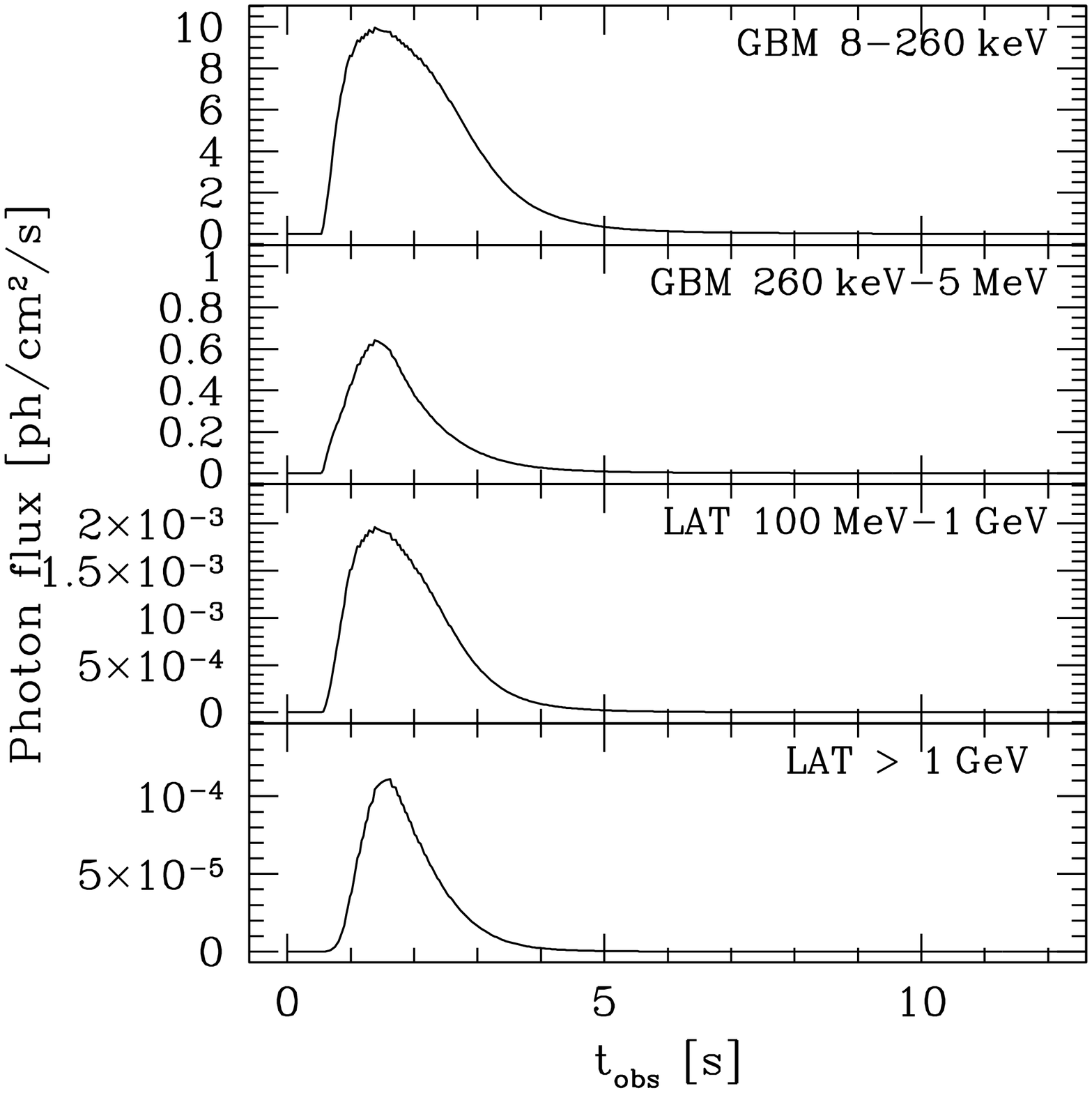} &
\includegraphics[width=0.45\textwidth]{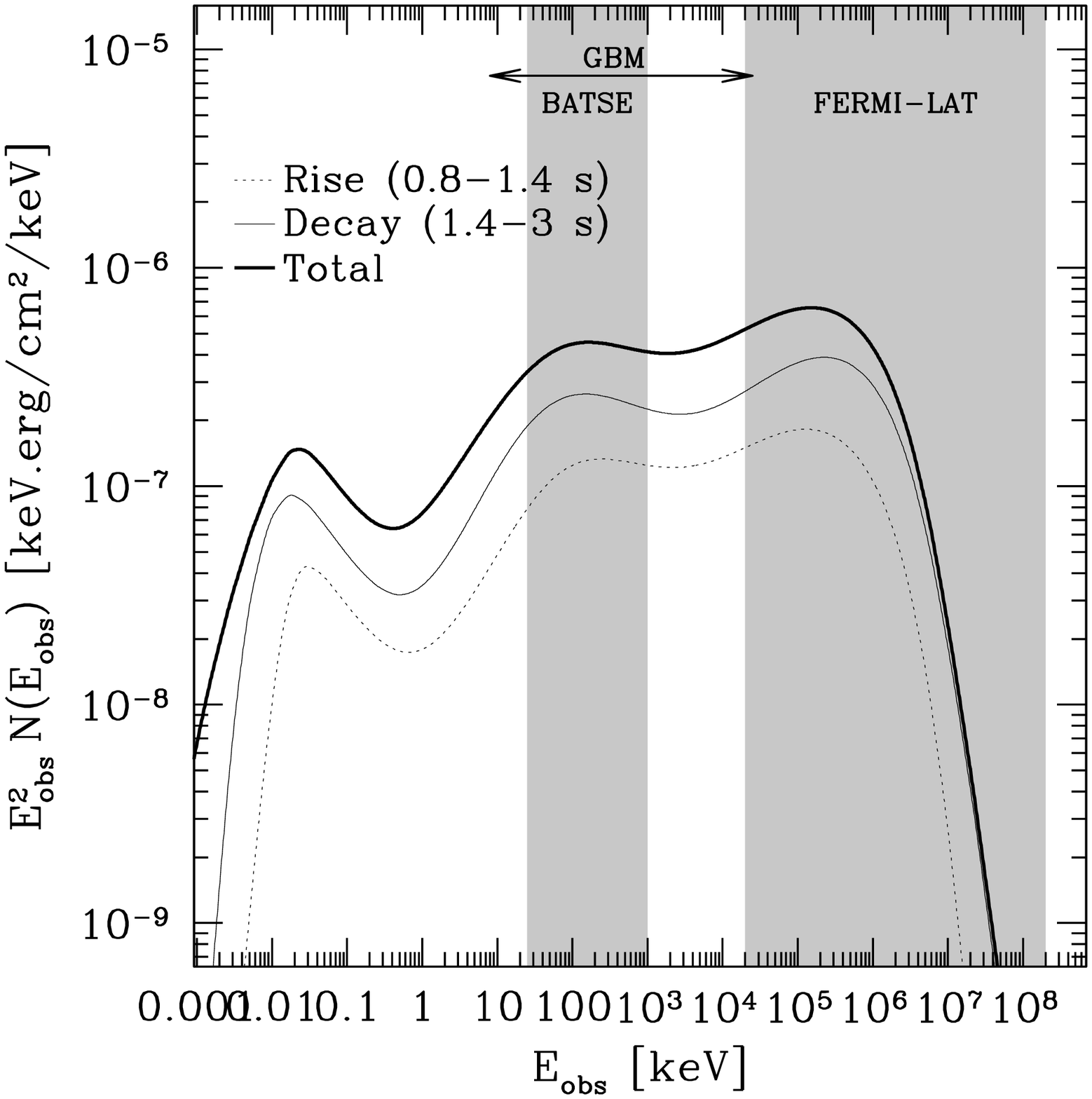}\\
\end{tabular}
\end{center}
\caption{\textbf{A single pulse burst in the ``inverse Compton case''.}
 Same as in \reffig{fig:synHB} except for the initial distribution
 of the Lorentz factor that varies from 100 to 600 and 
for the
 microphysics parameters: 
 $\epsilon_\mathrm{B}=10^{-2}$, $\epsilon_\mathrm{e}=1/3$, $\zeta=1$
 and $p=3.5$. In the left panel, the lightcurves observed both in the \textit{Fermi}-{GBM}+{LAT} energy range are entirely dominated by inverse Compton emission. }
\label{fig:ic}
\end{figure*}

From this study, it appears that using estimate of the Lorentz factor
from the observed cutoff energy such as \refeq{eq:cutoff} is acceptable
when the high-energy spectrum does not show any new bright component in addition
to the low-energy spectrum. 
To confirm any value of the Lorentz factor measured by this method, one
should
do a detailed modelling of the spectrum to check whether the 
process responsible for the observed cutoff has been correctly identified.
Note that the emission from
electron--positron pairs created by
 $\gamma\gamma$ annihilation is not included here and that it could
 provide additional diagnostics \citep{murase:08}. Note also that the
 final shape of the spectrum above the cutoff energy can become
 complicated when considering the spectrum integrated over the
 whole duration of a pulse, as we show in the next section \citep[see also
 the detailed discussion by ][]{granot:08}. It  can make the diagnostics
 more difficult and emphasizes again the necessity of a detailed modelling.\\

\section{Observed time profiles and spectra}
\label{sec:completegrbs}
All the spectra shown in the previous section are computed for a
``typical collision'', using the two shell version of the internal
shock model. However even for
a single pulse, the emission radius, and the physical conditions in the
shocked medium, can span several orders of magnitude during the
propagation of the ``internal shock'' waves (see \reffig{fig:exampledyn}). This leads to a spectral
evolution during the observed pulse that is entirely missed by the
two shell model. In the present section, we show examples of
synthetic bursts computed by coupling the detailed model for the dynamics of
the relativistic outflow (\S~\refsec{sec:dynamics}) 
with the radiative code (\S~\refsec{sec:radiation}), and we discuss the
predicted spectral evolution, as well as the high energy emission (\textit{Fermi}-{LAT}
range). In all examples, a redshift $z=1$ is assumed.

\subsection{A single pulse burst}
We present three synthetic single pulse bursts corresponding to the same relativistic
outflow\,: a total duration of the
relativistic ejection phase $t_\mathrm{w}=2\ \mathrm{s}$, a Lorentz
factor varying from 100 to 400 during the ejection as in
\reffig{fig:exampledyn} and $\dot{E}=5\times 10^{53}\
\mathrm{erg~s^{-1}}$. The dynamics is computed using a
discretization of the outflow in $1000$ shells. The three cases differ by different sets of microphysics
parameters\,: (i) ``synchrotron case'' with a high magnetic field (see \reffig{fig:synHB}), 
$\epsilon_\mathrm{B}=\epsilon_\mathrm{e}=1/3$, $\zeta=3\times 10^{-3}$ and $p=2.5$ ;
(ii) ``synchrotron case'' with a low
magnetic field (see \reffig{fig:synLB}), 
$\epsilon_\mathrm{B}=5\times 10^{-3}$, $\epsilon_\mathrm{e}=1/3$, $\zeta=2\times 10^{-3}$ and $p=2.5$ ;
(iii) ``inverse Compton case'' (see
\reffig{fig:ic}),
$\epsilon_\mathrm{B}=10^{-2}$, $\epsilon_\mathrm{e}=1/3$, $\zeta=1$ and $p=3.5$. 
In this last case, the contrast
$\kappa=\Gamma_\mathrm{max}/\Gamma_\mathrm{min}$ of the initial distribution of
the Lorentz factor has been increased ($\Gamma$ varies from 100 to 600
instead of 400) to increase the dynamical efficiency, and thus compensate for
a lower radiative
efficiency (as well as a lower fraction of the
emission that is radiated in the \textit{Fermi}-{GBM} range). The three pulses
have comparable isotropic equivalent energies radiated in the {GBM} range.
The lightcurves in the
{GBM}+{LAT} range are plotted
for each case, as well as the
time-integrated spectrum during the rise, the decay and the whole
duration of the
pulse. GRB lightcurves in the keV-MeV range usually show a
hard-to-soft evolution \citep[see e.g.][]{bhat:94,ford:95,norris:96}. This spectral evolution is found in
these three examples of synthetic GRBs, as the spectrum
during the rise peaks at higher energy than during the decay phase. \\

In the {GBM} range, the three lightcurves are quite similar, except
for a faster pulse decay in the inverse Compton case. 
We
checked in the three cases that the expected spectral evolution
\citep[see e.g.][]{norris:96} 
in the keV-MeV range
is reproduced, in agreement with the previous results of
\citet{daigne:98}: the photon flux peaks earlier at higher energy and
the duration of the pulse increases at lower energies.\\

In the {LAT} range on the other hand, the spectral
evolution and the corresponding behavior at high energy are
different in the three considered cases. In the ``synchrotron case'', the physical process responsible
for the radiation is not the same in the {GBM} range
(synchrotron) and in the {LAT} range (synchrotron+possible
additional inverse
Compton component, depending on the intensity of the magnetic field). 
Therefore, the lightcurves at low and high-energy do not look
similar. In particular, the inverse Compton component at high energy
emerges later than the synchrotron component, increasing the duration of the pulse in the
{LAT} range. It is due to an evolving
Compton parameter
during the pulse duration. This effect is  more important when
inverse Compton scatterings become more efficient (compare
Figs.~\ref{fig:synHB} and~\ref{fig:synLB}).
For the lowest values of $\epsilon_\mathrm{B}$,
the lightcurve at high energy could 
even peak with a delay with respect to the lightcurve in the {GBM} range
if the inverse Compton component becomes more intense that the
synchrotron component in the high energy range.
This behavior of the {LAT} lightcurves
is due to the evolution of the
physical condition in the shocked medium along the propagation of the
shock wave. It is illustrated in \reffig{fig:dyn_synLB}, where the
ratio of the inverse Compton over the synchrotron component is plotted
as a function of the observer time, as well as the dynamical timescale
$t'_\mathrm{ex}$ and the synchrotron timescale
$t'_\mathrm{syn}$\,:
\begin{figure}[!b]
\begin{center}
\includegraphics[width=0.45\textwidth]{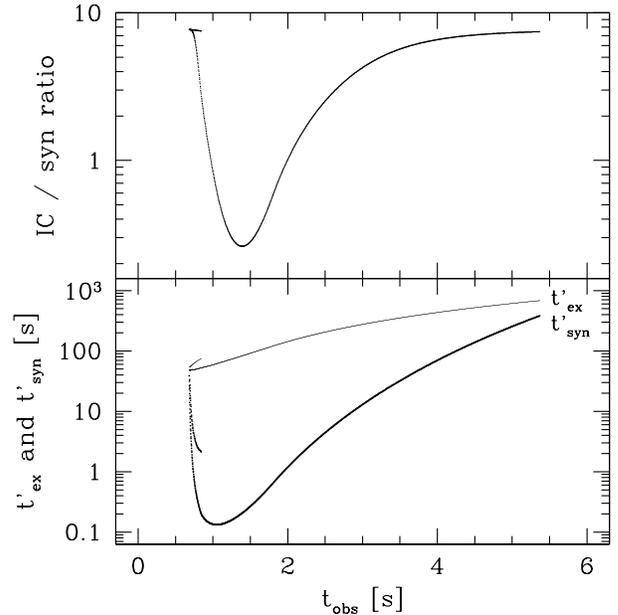}
\end{center}
\caption{\textbf{A single pulse burst in the ``synchrotron case'' with a
 low magnetic field: dynamical evolution during the pulse.} 
The
 evolution of the synchrotron timescale, the dynamical timescale and the
 ratio of the energy radiated in the inverse Compton over the energy
 radiated in the synchrotron component is plotted
 as a function of the observer time for the synthetic single pulse burst shown in \reffig{fig:synLB}.}
\label{fig:dyn_synLB}
\end{figure}
\begin{itemize}
\item
Initially, due to the shape adopted for the initial
distribution of the Lorentz factor in the outflow, the shock is weak
and     the dissipated 
energy per particle is low. This results in moderate electron Lorentz
factors $\Gamma_\mathrm{m}$, and therefore large synchrotron
timescales. On the other hand, these early times correspond to small
radii so the dynamical timescale is still small. In this first phase,
$t'_\mathrm{syn}\la t'_\mathrm{ex}$ and the efficiency of inverse
Compton scatterings is large, as a large fraction of the shocked region
is populated by relativistic electrons (see \refsec{sec:spec_com}). It
results in a weak precursor in the GeV lightcurve. This precursor
   can  disappear if a different  initial distribution of the Lorentz factor in the
outflow is adopted, especially if it  leads to an immediate violent shock (for instance
     with an initial
     discontinuity).
\item 
In a second phase (around the peak of the pulse in the {GBM} range), the
      shock becomes stronger, $\Gamma_\mathrm{m}$ increases and the
      synchrotron timescale decreases. On the other hand, as the radius
      increases, the dynamical timescale increases. This results in
      $t'_\mathrm{syn} \ll t'_\mathrm{ex}$ and a low efficiency for
      inverse Compton scatterings. The emission at high energy is
      dominated by the synchrotron component.
\item A late times (tail of the pulse), the synchrotron timescale
      increases again (mainly due to the decrease of the magnetic field
      as the radius increases) and the efficiency of inverse Compton
      scatterings increases again. The inverse Compton component becomes
      dominant again in the GeV range, which results in a prominent GeV tail of
      the pulse.
\end{itemize}

On the other hand, in the ``inverse Compton case'', inverse Compton
scatterings are the dominant radiative
process both in the {GBM} and the {LAT}
range. Therefore, the lightcurves are much more similar in the different
energy bands.\\

It appears clearly from Figs.~\ref{fig:synHB}, \ref{fig:synLB}
and~\ref{fig:ic} that in addition to the time-integrated spectrum, the
observed spectral evolution and the comparison of the {GBM}
and {LAT} lightcurves are also powerful tools to diagnose the
dominant radiative process and the physical conditions in the shocked
medium (electron distribution and magnetic field).

\subsection{A multi-pulses GRB}
\label{sec:multipulses}
\begin{figure}
\centerline{\includegraphics[width=0.45\textwidth]{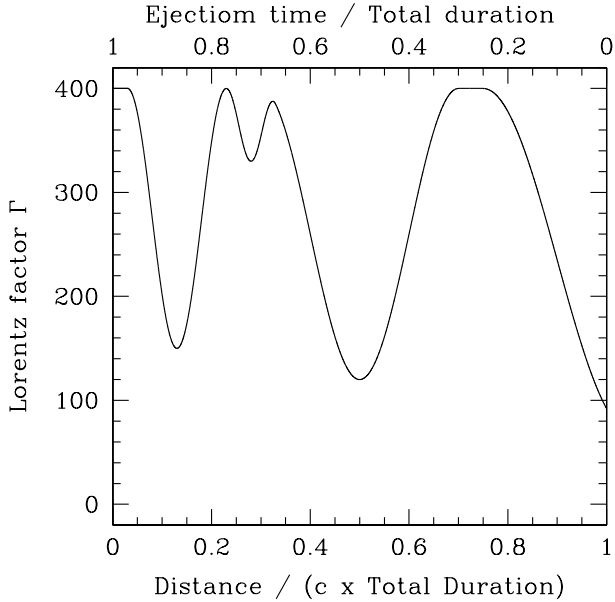}}
\caption{\textbf{An example of a  multi-pulses burst\,:
 initial distribution of the Lorentz factor in the relativistic outflow.}
The initial Lorentz factor in the outflow is plotted as a function of
 the ejection time (top axis) or equivalently the distance from the
 source at the end of the ejection phase (bottom axis).
}
\label{fig:gammamulti}
\end{figure}

\begin{figure*}[!ht]
\begin{center}
\begin{tabular}{cc}
\includegraphics[width=0.45\textwidth]{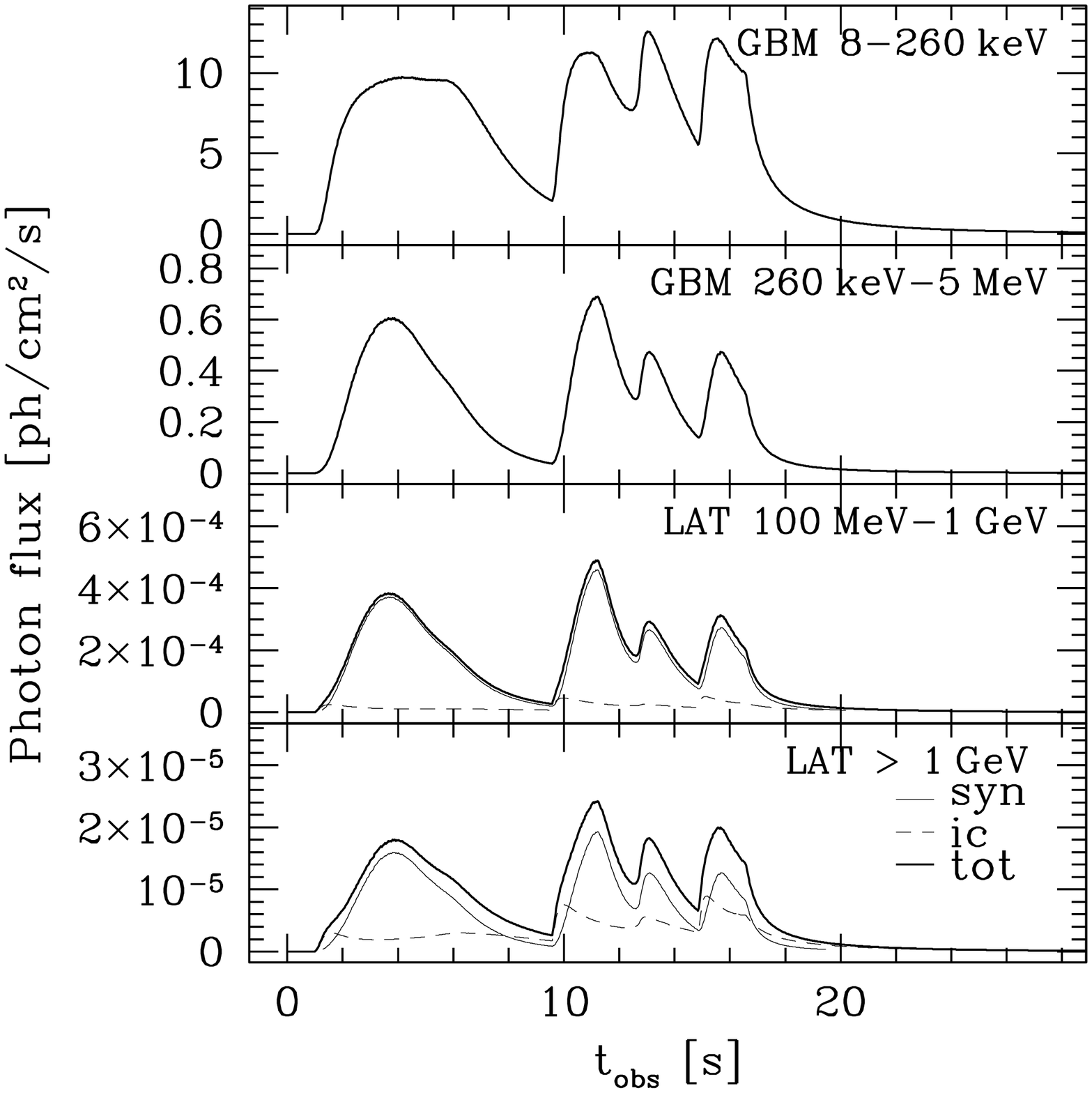} &
\includegraphics[width=0.45\textwidth]{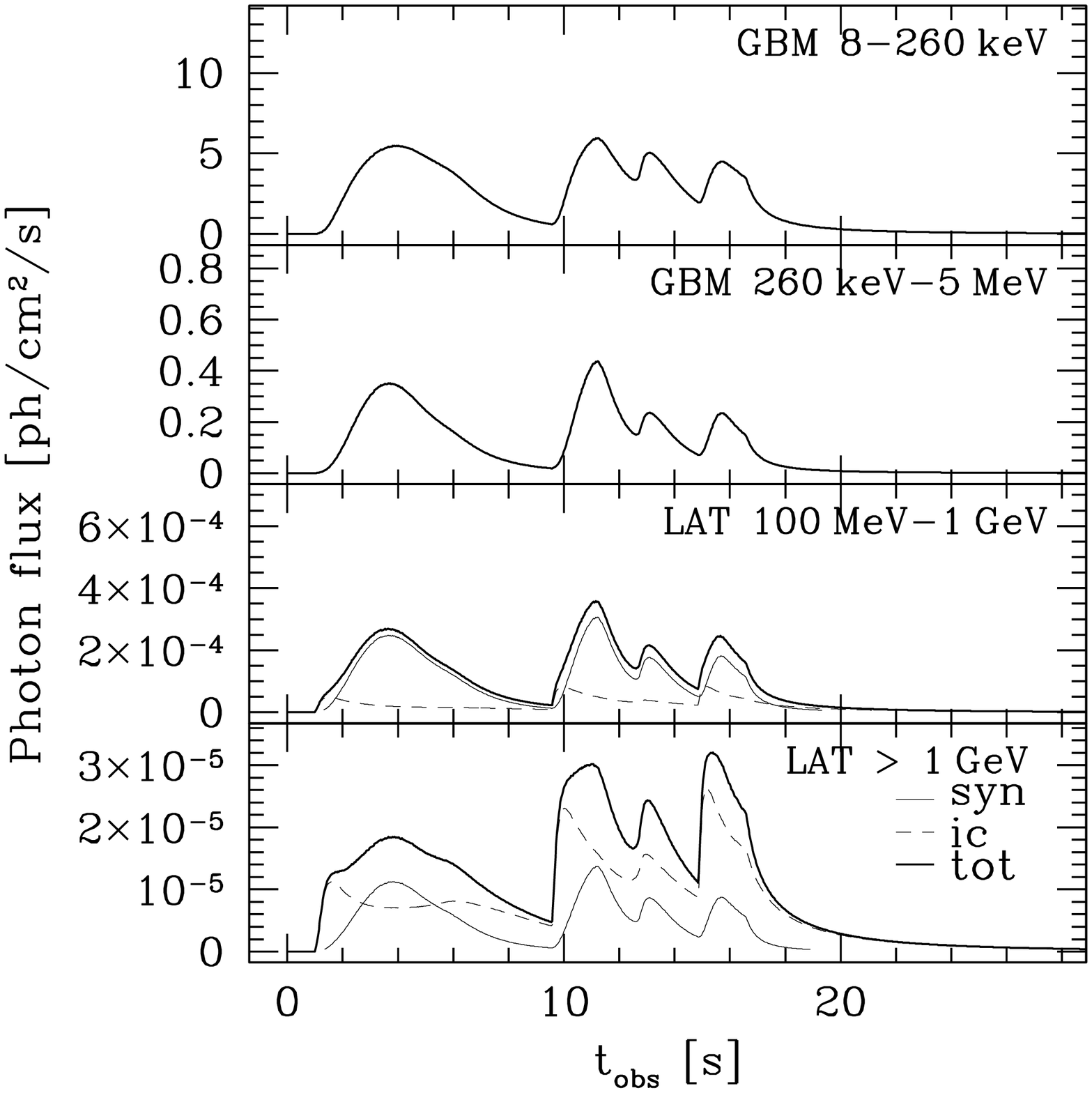}\\
\end{tabular}
\end{center}
\caption{\textbf{An example of a  multi-pulses burst\,:
 lightcurves.}
The dynamics is computed for the initial distribution of the Lorentz
 factor plotted in \reffig{fig:gammamulti}, assuming a total duration of the
 relativistic ejection phase $t_\mathrm{w}=8\ \mathrm{s}$ and an
 injected kinetic power $\dot{E}=5\times 10^{53}\ \mathrm{erg\,s^{-1}}$.
 The two examples differ by the assumptions for the microphysics parameters.\textit{Left:} ``synchrotron case'' with a high magnetic field. The microphysics parameters are $\epsilon_\mathrm{B}=\epsilon_\mathrm{e}=1/3$, $\zeta=0.002$ and $p=2.5$. The observed peak energy of the spectrum integrated over the whole duration of the burst is $\sim 180\ \mathrm{keV}$. \textit{Right:} ``synchrotron case'' with a low magnetic field. The microphysics parameters are $\epsilon_\mathrm{B}=8\times 10^{-4}$, $\epsilon_\mathrm{e}=1/3$, $\zeta=0.001$ and $p=2.5$. The observed peak energy is $\sim 250\ \mathrm{keV}$. In both cases the lightcurves are plotted in different energy channels (\textit{Fermi}-{GBM}+{LAT}) assuming a redshift $z=1$. 
}
\label{fig:multi}
\end{figure*}

\reffig{fig:multi} shows two examples of more complex synthetic GRBs.
The dynamical evolution is the same in both cases, assuming the initial
distribution of the Lorentz factor plotted in \reffig{fig:gammamulti},
which leads to 4 main pulses in the lightcurve. It is computed using a
discretization of the outflow in $4000$ shells. The two examples
correspond to two different sets of microphysics parameters
(``synchrotron case'' with a high or a low magnetic field).
These examples illustrate that in a complex burst, each
pulse exhibits a hard to soft evolution in the main spectral component
and that -- when possible -- the spectral analysis should be made by
integrating the spectrum over a pulse 
rather than over the whole duration of the GRB. The spectral evolution
and the behaviour at high energy that were identified for single pulse
burst are also observed in these multi-pulse GRBs. In particular, the
lightcurve above 1 GeV in the ``synchrotron case'' with a low magnetic
field shows 
more flat topped pulses and prolonged emission in the pulse decays.

\section{Conclusions}
\label{sec:conclusions}
We have developed a detailed model for the prompt emission of gamma-ray
bursts in the framework of the internal shock model. It combines
dynamical simulations that follow the evolution of the physical
conditions (Lorentz factor, density and energy density) in the shocked
regions along the propagation of ``internal'' shock waves in the
relativistic outflow, and a time-dependant radiative code to compute the
emission from shock-accelerated electrons, including the most relevant
processes (adiabatic cooling, synchrotron radiation and self-absorption,
inverse Compton scatterings and photon--photon annihilation). 
We have
used our model to explore the parameter space of the internal shock
model and identify physical diagnostics for \textit{Fermi} data.\\

\noindent We list here our main results:
\begin{enumerate}
\item The comparison of the results of the radiative code with standard
      analytical or semi-analytical estimates of the emitted spectrum
      shows that the synchrotron component is usually well predicted by
      the analytical spectrum from \citet{sari:98}, except when inverse
      Compton scatterings become efficient.
\item We show that the prediction of the high-energy component based on
      the time-averaged electron distribution and the time-averaged
      synchrotron photon spectrum \citep[e.g.][]{sari:01} is less precise, especially above the
      peak of the inverse Compton component. The accuracy of this
      time-averaged prediction decreases as the efficiency of inverse
      Compton scatterings increases. This is mainly due to
      time-dependent effects\,: time-averaged distributions do not take
      into account the time needed to build the photon field.
\begin{table*}[t!]
\caption{\textbf{Physical diagnostics for \textit{Fermi} data:}}
\begin{center}
\begin{tabular}{p{0.36\textwidth}|ccc}
\textbf{Observational constraints\,:} \\
--Is there an additional component in the {LAT} spectrum ? & no (or very weak) & yes &
 yes (possibly very intense)\\
--Is the {LAT} lightcurve in GeV range
 prolonged or delayed compared to {GBM} lightcurve ? & yes (tail) & yes
	 (delayed peak) & no (or very short delay)\\
\hline
 & \\
\textbf{Consequences:}\\
\textit{Radiative processes:}\\
Dominant process in {GBM} range & synchrotron & synchrotron &
	     IC (first peak)\\
Dominant process in {LAT} range & synchrotron + very weak IC & synchrotron + IC &
	     IC (second peak)\\
\textit{Conditions in shocked regions:}\\
Electron distribution & high $\Gamma_\mathrm{m}$, low $\zeta$ & high
	 $\Gamma_\mathrm{m}$, low $\zeta$ & moderate
	     $\Gamma_\mathrm{m}$, $\zeta\to 1$\\
Magnetic field        & high ($\epsilon_\mathrm{B}\to 1/3$) & low
	 ($\epsilon_\mathrm{B}\ll 1$) & low ($\epsilon_\mathrm{B}\ll 1$)\\
\textit{Properties of the relativistic outflow:}\\
Can the Lorentz factor in the outflow be measured from the
 high-energy cutoff (assuming it is due to $\gamma\gamma$ annihilation)
 ? & yes & yes & ? (see text)
\\
\end{tabular}
\end{center}
These diagnostics are based on the results shown in \refsec{sec:parameterspace}
 and \ref{sec:completegrbs}.
\label{tab:diag}
\end{table*}
\item An important consequence of the previous result is that the
      electron cooling rate, and therefore the time-averaged electron
      distribution, is not correctly predicted by the standard
      analytical estimate from \citet{sari:98} when inverse Compton
      scatterings become important. This affects the spectral shape of the
      synchrotron component as well. We find that
 the resulting spectral slope (photon spectrum)
      below the
      peak energy is steeper than the usual value $-3/2$ in synchrotron fast
      cooling regime, in agreement with \citet{derishev:01} who shows
      that the slope can become as steep as $-1$ when Klein-Nishina
      corrections are important.
This may 
reconcile the synchrotron radiation with the observed
      distribution of the low-energy slope $\alpha$ in {BATSE}
      \citep{preece:00,kaneko:06} and \textit{HETE-2} \citep{sakamoto:05} 
      bursts. We will investigate this question  in a
      forthcoming paper.
\item When exploring the parameter space of internal shocks, we find
      that two classes of broad-band spectra can be expected, which 
      correspond to different physical conditions in the shocked
      region :
\begin{itemize}
\item ''Synchrotron case'', where the dominant process in the
      \textit{Fermi}-{GBM} range is synchrotron radiation. It requires
      high electron Lorentz factors and therefore implies that only a
      fraction of the electrons are shock-accelerated. The intensity of
      the inverse Compton component in the {LAT} range depends
      on the intensity of the magnetic field but remains always limited
      due to Klein-Nishina corrections. A high-energy cutoff is present
      due to photon-photon annihilation, which allows us to estimate the
      Lorentz factor of the emitting material. As the high-energy lightcurve
      is made of two different components (synchrotron + inverse
      Compton) whose ratio evolves during a pulse, the lightcurves can
      differ in the {GBM} and the {LAT} range. In
      particular, for low $\epsilon_\mathrm{B}/\epsilon_\mathrm{e}$ ratios that favor inverse
      Compton scatterings, a pulse in the {GBM} can be followed by a tail
      of GeV photon in the {LAT}. When there is a significant
      inverse Compton component in the {LAT},
      the pulse above 1 GeV can even peak with a delay compared to
      the {GBM}.
\item ''Inverse Compton case'', where the synchrotron component peaks at
      low energy and the dominant process in the {GBM} range is
      inverse Compton. A second inverse Compton peak is usually present
      at higher energy but further scatterings are suppressed by
      Klein-Nishina corrections. The cutoff at high energy can be either
      due
       to $\gamma\gamma$ annihilation or to the Klein-Nishina
      suppression of inverse Compton scatterings, which makes more
      difficult to estimate the Lorentz factor of the emitting
      material in this case.
A steep slope for the electron
      distribution is necessary to have two well distinct peaks in the
      inverse Compton component.  The relative intensity of the
      synchrotron and the two inverse Compton peaks in the spectrum
      depends on the ratio $\epsilon_\mathrm{e}/\epsilon_\mathrm{B}$. It
      requires some fine-tuning to have a dominant first inverse Compton
      peak in the {GBM} range, which can lead to a energy crisis
      \citep{piran:08}. As the emission detected in the
      {GBM} and {LAT} lightcurves is due to the
      same process (inverse Compton), the lightcurves in the two
      instruments are much more similar. 
\end{itemize}
\item This study allows us to define physical diagnostics for
      \textit{Fermi} data, based on the spectral shape and the spectral
      evolution, that are summarized in \reftab{tab:diag}. We plan to
      apply these diagnostics as soon as \textit{Fermi} GRB data will be
      made public. Our study emphasizes however that a detailed broad-band
      spectral modelling is always necessary to reach firm conclusions
      regarding the properties of the outflow and the
      physical conditions in the shocked regions.
\end{enumerate}
Several arguments (energetics, spectral evolution) already favor the
``synchrotron case''. Bursts detected by \textit{Fermi} both by the
{GBM} and the {LAT} should allow us to firmly distinguish
these two possibilities and show if all bursts are in the ''synchrotron
case'' or if the ''inverse Compton case'' can also be found. \\

In the ``synchrotron case'', it is assumed that only a fraction of the
electrons is accelerated to very high Lorentz factors
($\Gamma_\mathrm{m}\sim 10^{4}$). If a small fraction of the dissipated
energy in the shock is injected in the remaining electrons, they will
have a Maxwellian distribution with a mean Lorentz factor of a few. We
will investigate in a future work what could be the contribution of
these electrons to the emission and more generally discuss the prompt
optical emission of GRBs in the framework of the internal shock model.

\acknowledgements
The authors thank Prof. P.~Kumar for his prompt and supportive report on
this paper and
Dr. R.~Mochkovitch for many valuable discussions on this
work, and a careful reading of the manuscript. This work is part of the
project \texttt{JETS\_GAMMA} which is funded by the French
National Research Agency (ANR). The work of F.D. was partially
supported by the French Spatial Agency (CNES).

\appendix
\section{Radiative processes}
\label{sec:radproc}
We list here the source and loss terms appearing in
the equations governing the evolution of the electron distribution
(\refeq{eq:electrons}) and of the photon spectrum (\refeq{eq:photons}).
\subsection{Adiabatic cooling}
\begin{equation}
\left.\frac{d\gamma}{dt'}\right|_\mathrm{ad} = -\frac{\gamma}{t'_\mathrm{ex}}\ .
\end{equation}

\subsection{Synchrotron emission}
We assume an isotropic distribution for the pitch angle $\alpha$ between
the electron velocity and the magnetic field, so that
$\left\langle\sin^{2}{\alpha}\right\rangle=2/3$. The synchrotron
power is given by \citet{rybicki:79} and leads to
\begin{eqnarray}
\left.\frac{d\gamma}{dt'}\right|_\mathrm{syn} & = & -\frac{\sigma_\mathrm{T}}{6\pi m_\mathrm{e}c} B^{2}\gamma^{2}\ ,\\
P^\mathrm{syn}_{\nu'}\left(\gamma\right) & = & \frac{\sigma_\mathrm{T} m_\mathrm{e}c^{2}}{3e} B\ \Phi\left(\frac{\nu'}{\nu'_\mathrm{syn}\left(\gamma\right)}\right)\ ,
\end{eqnarray}
where the synchrotron frequency is defined by
\begin{equation}
\nu_\mathrm{syn}\left(\gamma\right) = \frac{3e}{4\pi m_\mathrm{e}c} B \gamma^{2}
\label{eq:nusyn}
\end{equation}
and the function $\Phi$ is defined by
\begin{equation}
\Phi\left(x\right) = \frac{9\sqrt{3}}{8\pi} x \int_{x}^{+\infty}\mathrm{d}u\, K_{5/3}(u)\ .
\end{equation}
This definition has been chosen so that
$\int_{0}^{+\infty}\mathrm{d}x\,\Phi(x)=1$. 
\subsection{Synchrotron self-absorption}
The cross-section is given by \citet{rybicki:79}:
\begin{equation}
\sigma_\mathrm{sa}\left(\gamma,\nu'\right) = \frac{1}{8\pi m_\mathrm{e}\left.\nu'\right.^{2}}\frac{P^\mathrm{syn}_{\nu'}\left(\gamma\right)}{\gamma}\left[2-\frac{\partial\ln{n}}{\partial\ln{\gamma}}\left(\gamma,t'\right)\right]\ .
\end{equation}
In the present version of our radiative code, the corresponding source term $\left.d\gamma/dt'\right|_\mathrm{sa}$ is
not included in the equation for the evolution of electrons (\refeq{eq:electrons}).

\subsection{Inverse Compton scatterings}
We use the kernel derived by \citet{jones:68}, which has an excellent
accuracy, even in the Klein-Nishina regime.
\begin{eqnarray}
\left.\frac{d\gamma}{dt'}\right|_\mathrm{ic} & = & -\frac{3}{4}\frac{h \sigma_\mathrm{T}}{m_\mathrm{e}c} \frac{1}{\gamma^{2}} \int\mathrm{d}\nu'\,\nu' \int\frac{\mathrm{d}\tilde{\nu}'}{\tilde{\nu}'} n_{\tilde{\nu}'}(t') K\left(\gamma,\nu',\tilde{\nu}'\right)\ ,\\
P^\mathrm{ic}_{\nu'} & = & \frac{3}{4}h \sigma_\mathrm{T}c \frac{\nu'}{\gamma^{2}} \int\frac{\mathrm{d}\tilde{\nu}'}{\tilde{\nu}'} n_{\tilde{\nu}'}(t') K\left(\gamma,\nu',\tilde{\nu}'\right)
\end{eqnarray}
with
\begin{eqnarray}
K\left(\gamma,\nu',\tilde{\nu}'\right) & = & \frac{\epsilon}{\tilde{\epsilon}}-\frac{1}{4\gamma^{2}}\ \mathrm{if}\ \frac{\tilde{\epsilon}}{4\gamma^{2}} < \epsilon < \tilde{\epsilon}\ ,\nonumber\\
& = & 2q \ln{q} +\left(1+2q\right)\left(1-q\right)
 +\frac{1}{2}\left(1-q\right)\frac{\left(4\gamma \tilde{\epsilon} q\right)^{2}}{1+4\gamma \tilde{\epsilon} q}\nonumber\\
& & \mathrm{if}\ \tilde{\epsilon} < \epsilon < \frac{4\gamma^{2}\tilde{\epsilon}}{1+4\gamma \tilde{\epsilon}}\ ,
\end{eqnarray}
where $\epsilon=h\nu'/m_\mathrm{e}c^{2}$,
$\tilde{\epsilon}=h\tilde{\nu}'/m_\mathrm{e}c^{2}$ and
$
q = \epsilon / 4\gamma
\tilde{\epsilon} / \left(\gamma-\epsilon\right)
$
.
In the present version of the code, the corresponding loss term
$$
\left.\frac{\partial n_{\nu'}}{\partial t'}\right|_\mathrm{ic,loss}=-c n_{\nu'}(t') \int\mathrm{d}\gamma\,n\left(\gamma,t'\right) \sigma_\mathrm{ic}\left(\gamma,\nu'\right)
$$
is not included in the right-hand part of the equation for the evolution
of photons (\refeq{eq:photons}). For this reason, we can not
compute the emitted spectrum when $\tau_\mathrm{T}^\mathrm{acc}\ga 1$
and can not reproduce a Comptonized spectrum.
\subsection{Photon--photon annihilation}
We use the exact cross section given by \citet{gould:67} for an isotropic photon field, which is a
good approximation as long as the radiative timescale is small compared
to the dynamical timescale (see \refsec{sec:photonfield}).
\begin{eqnarray}
\sigma_\mathrm{\gamma\gamma}\left(\nu',\tilde{\nu}'\right) & = &
\left[\frac{1+\beta^{2}}{1-\beta^{2}}-\beta^{2}-\ln{\frac{1+\beta}{1-\beta}}+4\ln{\frac{2}{1-\beta}}\right]\ln{\frac{1+\beta}{1-\beta}}\nonumber\\
& & -\frac{4\beta}{1-\beta^{2}}+2\beta-\int_{1}^{(1+\beta)/(1-\beta)}\frac{\mathrm{d}x}{x}\,\ln{\left(1+x\right)}
\ ,
\end{eqnarray}
where 
$$
\beta =
\sqrt{\frac{h\nu'h\tilde{\nu}'-\left(m_\mathrm{e}c^{2}\right)^{2}}{h\nu'h\tilde{\nu}'}}\ .
$$
In the present version of the code, the source and loss terms due to
pair production and pair annihilation are not included in
\refeq{eq:electrons}, and the associated radiation is not included in \refeq{eq:photons}.
\section{Numerical method}
\label{sec:mnum}
To solve 
the system of the two Eqs.~(\ref{eq:electrons})
and~(\ref{eq:photons}) for the evolution of the electron distribution $n\left(\gamma,t'\right)$
and the photon spectrum $n_{\nu'}(t')$ in the comoving frame of the
shocked material, we have developed a numerical scheme that is a good
compromise between accuracy and computing speed. We use normalized
variables $\tilde{t}=t'/t'_\mathrm{ex}$,
$\epsilon=h\nu'/m_\mathrm{e}c^{2}$,
$\tilde{n}\left(\gamma,\tilde{t}\right)=n\left(\gamma,t\right)/n_\mathrm{e}^\mathrm{acc}$
and $\tilde{n}_{\epsilon}(\tilde{t})=\left(m_\mathrm{e}c^{2}/h\right)n_{\nu'}(t')/n_\mathrm{e}^\mathrm{acc}$.
The photon
spectrum at time $\tilde{t}_\mathrm{k}$ after the
$\mathrm{k}^\mathrm{th}$ step is stored in a fixed grid
$\epsilon_\mathrm{j}$ for $j=1\to M$, i.e. 
$\tilde{n}_\mathrm{j}^{\mathrm{(k)}}=\tilde{n}_{\epsilon_\mathrm{j}}\left(\tilde{t}_\mathrm{k}\right)$.
At $\tilde{t}=0$, no photons are present so that $\tilde{n}_\mathrm{j}(0)=0$.
To take advantage of the short radiative timescale, the electron
distribution at time $t_\mathrm{k}$ is stored in a moving
(``Lagrangian'') grid
$\gamma_\mathrm{i}^{\mathrm{(k)}}=\gamma_\mathrm{i}\left(\tilde{t}_\mathrm{k}\right)$
for $i=0\to N$, i.e.
$$
\Delta\tilde{n}_\mathrm{i}^\mathrm{(k)} =
\int_{\gamma_\mathrm{i-1}^\mathrm{(k)}}^{\gamma_\mathrm{i}^\mathrm{(k)}}\mathrm{d}\gamma\,\tilde{n}\left(\gamma,\tilde{t}\right)\
.
$$
This insures that the number of electrons is exactly conserved.
At $\tilde{t}=0$, this grid is adjusted to the distribution
of accelerated electrons given by \refeq{eq:epowerlaw}, i.e.
$\gamma_{0}^\mathrm{(0)}=\Gamma_\mathrm{m}$,
$\gamma_{N}^\mathrm{(0)}=\Gamma_\mathrm{M}$ and $\Delta\tilde{n}_\mathrm{i}^\mathrm{(0)}=\left(\gamma_\mathrm{i-1}^\mathrm{(0)}/\Gamma_\mathrm{m}\right)^{1-p}-\left(\gamma_\mathrm{i}^\mathrm{(0)}/\Gamma_\mathrm{m}\right)^{1-p}$.
In practice, all results presented in this paper are obtained using
logarithmic grids for $\epsilon_\mathrm{j}$ and
$\gamma_\mathrm{i}^\mathrm{(0)}$ with $N=M=100$. The time-evolution of
$\tilde{n}_\mathrm{j}^\mathrm{(k)}$ is computed using a discretized
version of \refeq{eq:photons}. On the other hand, to compute the
time-evolution of $\gamma_\mathrm{i}^\mathrm{(k)}$, we replace
\refeq{eq:electrons} by
\begin{equation}
\frac{d\gamma}{d\tilde{t}}\left(\tilde{t}\right)=t'_\mathrm{ex}\left(
\left.\frac{d\gamma}{dt'}\right|_\mathrm{ad}
+\left.\frac{d\gamma}{dt'}\right|_\mathrm{syn}
+\left.\frac{d\gamma}{dt'}\right|_\mathrm{ic}
\right)\ .
\end{equation}
In fast cooling regime the values of
$\gamma_\mathrm{i}^\mathrm{(k)}$ decrease rapidly, which is a valuable
advantage, as the timestep of the simulation is given by the 
radiative timescale of the most energetic electrons still present in the
medium. 
We observe a rapid increase of the
timestep following the cooling of the electrons, and reach therefore
$\tilde{t}=1$ in a small number of steps. In addition, as
$-d\gamma/d\tilde{t}$ is an increasing  function of $\gamma$, the width
of each cell in the electron grid
($\gamma_\mathrm{i}^\mathrm{(k)}-\gamma_\mathrm{i-1}^\mathrm{(k)}$) is
decreasing with time, which allows at the end of the simulation to merge
cells of cooled electrons and therefore gain an additional reduction of
the computing time. Energy is not necessarily conserved in our code due to the
finite size of the electron and photon grids: typically, the error is a few percents
and is never larger than 10 \%.
\section{Formation of the radiation field: the Compton parameter $Y(t')$.}
\label{sec:Y}
When only adiabatic cooling and synchrotron radiation are considered,
the equation for the evolution of electrons has a simple analytic
solution, given by
\begin{equation}
n\left(\gamma,t'\right) = \left(\frac{\gamma_{0}\left(\gamma,t'\right)}{\gamma}\right)^{2} n\left(\gamma_{0}\left(\gamma,t'\right),t'=0\right)\ ,
\end{equation}
where $\gamma_{0}\left(\gamma,t'\right)$ is the initial Lorentz factor
at $t'=0$ of an electron having cooled to the Lorentz factor $\gamma$ at
time $t'$, i.e.
\begin{equation}
\gamma_{0}\left(\gamma,t'\right) = \frac{\Gamma_\mathrm{c}}{\left(1+\frac{\Gamma_\mathrm{c}}{\gamma}\right)e^{-t'/t'_\mathrm{ex}}-1}\ .
\end{equation}
At time $t'$, electrons have Lorentz factors in the interval 
\begin{equation}
\Gamma_\mathrm{m}(t')= \frac{\Gamma_\mathrm{c}}{\left(1+\frac{\Gamma_\mathrm{c}}{\Gamma_\mathrm{m}}\right)e^{t'/t'_\mathrm{ex}}+1} \le \gamma \le \Gamma_\mathrm{M}(t')=\frac{\Gamma_\mathrm{c}}{\left(1+\frac{\Gamma_\mathrm{c}}{\Gamma_\mathrm{M}}\right)e^{t'/t'_\mathrm{ex}}+1}\ .
\end{equation}
\begin{figure}[!t]
\centerline{\psfig{file=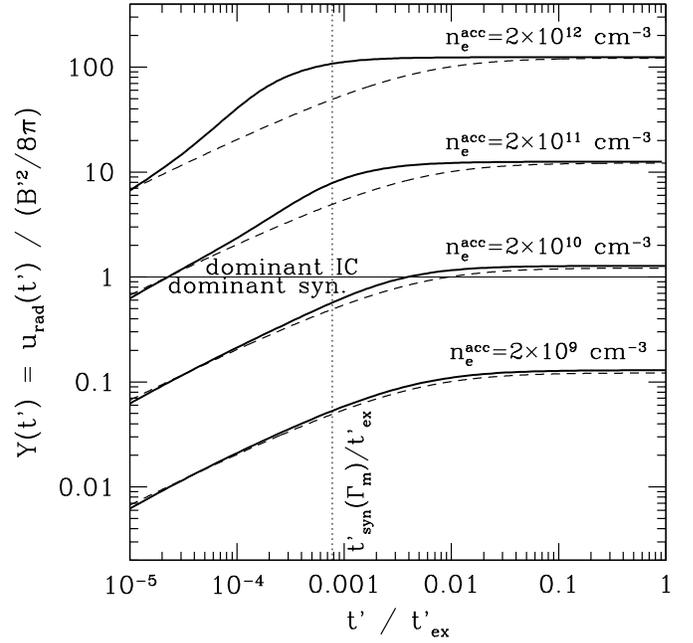,width=\linewidth}}
\caption{\textbf{Time evolution of the Compton parameter.} 
The Compton parameter $Y(t')$ is plotted for $\Gamma_\mathrm{m}=100$,
 $B'=10^{4}\ \mathrm{G}$, $t'_\mathrm{ex}=10\ \mathrm{s}$ and four
 values of the initial density of relativistic electrons which are
 labelled in the Figure. The numerical calculation (thick solid line) is performed including adiabatic cooling,
 synchrotron radiation (without self-absorption) and inverse Compton
 scatterings. The analytical expression of $Y(t')$ described in
 appendix~\ref{sec:Y} is plotted with a dashed line. A horizontal thin line
 indicates the limit $Y=1$ between a regime where electron cooling is
 dominated by synchrotron radiation or inverse Compton scatterings. A
 vertical dotted line indicates the synchrotron timescale for electrons
 with Lorentz factor $\Gamma_\mathrm{m}$.
}
\label{fig:testY}
\end{figure}
The time-averaged electron distribution 
$$
\bar{n}\left(\gamma\right)=\frac{1}{t'_\mathrm{ex}}\int_{0}^{t'_\mathrm{ex}}\mathrm{d}t'\,n\left(\gamma,t'\right)
$$
can be computed exactly from these expressions. An accurate approximate
expression is given by \refeq{eq:nbarge}. When inverse Compton
scatterings are included, this exact solution for the electron evolution
is not valid any more, but remains very accurate as long as the
synchrotron process is still dominant, i.e. as long as $Y(t')\ll 1$. It
is therefore interesting to compute the evolution of the Compton
parameter in this case. It is defined as
$Y(t')=u_\mathrm{rad}(t')/\left(\left.B'\right.^{2}/8\pi\right)$.
After some algebra, it can be written as
\begin{eqnarray}
Y(t') & = & \frac{4}{3}(p-1) \left(\sigma_\mathrm{T}n_\mathrm{e}^\mathrm{acc}c t'_\mathrm{syn}\left(\Gamma_\mathrm{m}\right)\right)\Gamma_\mathrm{m}^{2}\left(\frac{\Gamma_\mathrm{m}}{\Gamma_\mathrm{c}}\right)^{p-2}\nonumber\\
& & \times \int_{0}^{\Gamma_\mathrm{c}/\Gamma_\mathrm{m}}\mathrm{d}y\,y^{p-2}\left[
\ln{\frac{y e^{t'/t'_\mathrm{ex}}}{(1+y)e^{t'/t'_\mathrm{ex}}-1}}
+\frac{(1+y)\left(e^{t'/t'_\mathrm{ex}}-1\right)}{y\left((1+y)e^{t'/t'_\mathrm{ex}}-1\right)}
\right]\ .
\label{eq:Yt}
\end{eqnarray}
The final value at $Y_\mathrm{ex}=Y\left(t'=t'_\mathrm{ex}\right)$ is
given in \refsec{sec:Yex}. We plot in \reffig{fig:testY} the
evolution of $Y(t')$ for increasing values of $Y_\mathrm{ex}$. The agreement between the numerical
result and the analytical solution is excellent as long as 
the synchrotron radiation is still the dominant process, i.e. for
$Y(t')<1$. The discrepancy is larger when the transition $Y'=1$
occurs in a time shorter than the synchrotron timescale. In any case, in
fast cooling regime the energy density contained in the photon field
at $t'=t'_\mathrm{ex}$ equals $u_\mathrm{e}^\mathrm{acc}$ (initial
energy density injected in
relativistic electrons) and therefore the asymptotic value
is always equal to the theoretical prediction
$Y_\mathrm{ex}=u_\mathrm{e}^\mathrm{acc}/\left(\left.B'\right.^{2}/8\pi\right)$
if the Thomson regime is valid (which is the case in all examples
plotted in \reffig{fig:testY}).

\bibliographystyle{aa}
\bibliography{paper}
\end{document}